\documentclass[12pt]{article}

\usepackage{url}
\usepackage{scicite}

\usepackage{times}

\usepackage{times}
\usepackage{graphicx}
\usepackage{bm}
\usepackage[figurename=Fig.]{caption}

\usepackage{longtable}

\newenvironment{sciabstract}{%
\begin{quote} \bf}
{\end{quote}}

\def\simless{\lower2pt\hbox{$\buildrel {\scriptstyle <}
   \over {\scriptstyle\sim}$}}
\def\gtrsim{\lower2pt\hbox{$\buildrel {\scriptstyle >}
   \over {\scriptstyle\sim}$}}
\def\fss{\hbox{$.\!\!^s$}}
\def\farcs{%
 \mbox{%
  \kern  0.13ex.%
  \kern -0.55ex\raisebox{.9ex}{\scriptsize$\prime\prime$}%
  \kern -0.1ex%
 }%
}%

\newcounter{lastnote}
\newenvironment{scilastnote}{%
\setcounter{lastnote}{\value{enumiv}}%
\addtocounter{lastnote}{+1}%
\begin{list}%
{\arabic{lastnote}.}
{\setlength{\leftmargin}{.22in}}
{\setlength{\labelsep}{.5em}}}
{\end{list}}

\title{{\Large\bf Kepler-47: A Transiting Circumbinary Multi-Planet System}}

\author
{{\rm Jerome A. Orosz,$^{1\ast}$
William F.\ Welsh,$^{1}$ 
Joshua A. Carter,$^{2,\dagger}$ 
Daniel C. Fabrycky,$^{3,\dagger}$ } \\
{\rm William D. Cochran,$^{4}$
Michael  Endl,$^{4}$
Eric B. Ford,$^{5}$  
Nader  Haghighipour,$^{6}$} \\
{\rm Phillip J. MacQueen,$^{4}$
Tsevi Mazeh,$^{7}$
Roberto Sanchis-Ojeda,$^{8}$ 
Donald R. Short,$^{1}$} \\
{\rm Guillermo Torres,$^{2}$
Eric Agol,$^{9}$
Lars A. Buchhave,$^{10,11}$ 
Laurance R.  Doyle,$^{12}$ } \\
{\rm Howard Isaacson,$^{13}$
Jack J. Lissauer,$^{14}$
Geoffrey W.  Marcy,$^{13}$ 
Avi Shporer,$^{15,16,17}$ }  \\
{\rm Gur Windmiller,$^{1}$  
Thomas   Barclay,$^{14,18}$ 
Alan P. Boss,$^{19}$
Bruce D. Clarke,$^{12,14}$}  \\
{\rm Jonathan Fortney,$^{3}$
John C. Geary,$^{2}$ 
Matthew J. Holman,$^{2}$ 
Daniel Huber,$^{14}$  }   \\
{\rm Jon M.\ Jenkins,$^{12,14}$
Karen Kinemuchi,$^{14,18}$
Ethan Kruse,$^{2}$ 
Darin Ragozzine,$^{2}$ } \\
{\rm Dimitar Sasselov,$^{2}$ 
Martin Still,$^{14,18}$
Peter  Tenenbaum,$^{12,14}$ 
Kamal Uddin,$^{14,20}$} \\
{\rm Joshua N. Winn,$^{8}$ 
David G. Koch,$^{14}$
\& William J. Borucki$^{14}$}\\
\\
\tiny{$^{1}$Astronomy Department, San 
  Diego State University, 5500 Campanile Drive, San Diego, CA 92182,}
\tiny{USA $^{2}$Harvard-Smithsonian Center for Astrophysics,
   60 Garden Street, Cambridge, }\\
\tiny{MA 02138, USA $^{3}$Department of Astronomy 
  \& Astrophysics, University of 
  California, Santa Cruz,  CA 95064, USA}
\tiny{$^{4}$McDonald Observatory, The University of Texas at 
   Austin, Austin,  TX } \\
\tiny{78712-0259, USA $^{5}$Astronomy } 
   \tiny{Department, University of Florida, 211
   Bryant Space Sciences Center,}
  \tiny{Gainesville, FL 32111, USA}
\tiny{$^{6}$Institute for Astronomy and NASA Astrobiology} \\
\tiny{Institute
   University of Hawaii-Manoa,}
   \tiny{2680 Woodlawn} 
 \tiny{ Dr., Honolulu, HI 96822, USA}
\tiny{$^{7}$School of Physics and Astronomy, Tel Aviv University,
    Tel Aviv 69978,}
\tiny{ Israel $^{8}$Department of Physics, }\\ 
\tiny{and Kavli Institute for Astrophysics and 
    Space Research,}
   \tiny{Massachusetts  Institute of Technology, Cambridge, } 
   \tiny{MA 02139, USA}
\tiny{$^{9}$Department of Astronomy, BOX 351580,} 
  \tiny{University
   of Washington, }\\ 
\tiny{Seattle, WA 98195, USA}
\tiny{$^{10}$Niels Bohr Institute, University of} 
  \tiny{Copenhagen, 
  Juliane Maries vej 30, 2100 Copenhagen,  Denmark}
\tiny{$^{11}$Centre for Star \& Planet  Formation, Natural History} \\
   \tiny{ Museum of Denmark, University of Copenhagen,} 
   \tiny{\O ster Voldgade 5-7, }
 \tiny{  1350 Copenhagen, Denmark}
\tiny{$^{12}$SETI Institute, 189 Bernardo Avenue, Mountain View, 
  CA 94043, USA $^{13}$Astronomy } \\
\tiny{Department, University of California, Berkeley,
CA 94720, USA} 
\tiny{$^{14}$NASA Ames Research } 
  \tiny{Center, Moffett Field, CA 94035}
\tiny{$^{15}$Las Cumbres Observatory Global Telescope Network, 
   6740 }  \\
   \tiny{Cortona Drive, Suite 102,}
   \tiny{Santa Barbara, CA 93117, USA}
\tiny{$^{16}$Department of Physics, Broida Hall,
    University of} 
   \tiny{California, Santa
   Barbara, CA 93106, USA}
  \tiny{$^{17}$Division of Geological and Planetary } \\
  \tiny{Sciences,  California Institute of 
 Technology, Pasadena, CA 91125, USA} 
\tiny{$^{18}$Bay Area Environmental Research Institute, Inc.,}
 \tiny{ 560 Third Street West, Sonoma, CA 95476, USA $^{19}$Department } \\
\tiny{of Terrestrial Magnetism, Carnegie }
   \tiny{Institution for Science,}
\tiny{5241 Broad Branch Road, N.W.
    Washington, D.C. 20015-1305  U.S.A.}
\tiny{$^{20}$Orbital Sciences Corporation,
     45101 Warp Drive, Dulles, }\\
\tiny{VA 20166, USA $^\ast$To whom }
  \tiny{correspondence should be addressed; E-mail:  
orosz@sciences.sdsu.edu.}\tiny{$^{\dagger}$Hubble Fellow}
}

\date{}

\begin{document} 




\thispagestyle{empty}
\maketitle



\begin{sciabstract}
We report the detection of Kepler-47, a system consisting of
two planets orbiting around an eclipsing pair of stars. The inner
and outer planets have radii  3.0 and 4.6 times that of
the Earth,
respectively.
The binary star consists of a Sun-like star and a companion
roughly one-third its size, 
orbiting each other every 7.45 days.
With an orbital period of 49.5 days, eighteen transits of the inner
planet have been observed, allowing a detailed characterization of
its orbit and those of the stars.
The outer planet's orbital period is 303.2 days, and although the
planet is not Earth-like, it resides  within  the 
classical
``habitable zone'',
where liquid
water could exist on an Earth-like planet.
With its two known planets, Kepler-47 establishes that close binary
stars can host complete planetary systems.
\\
\end{sciabstract}



The extremely precise and nearly continuous observations provided by
the Kepler spacecraft \cite{Koch_2010} has enabled the detection
of over 2300 planet candidates \cite{Borucki_2010,Batalha_2012}, and
over 2100 eclipsing binary stars \cite{Prsa_2011,Slawson_2011}.  A
synergy of these efforts has helped  establish the
class of circumbinary planets, which are planets that orbit around
a pair of stars \cite{Doyle_2011,Welsh_2012,Orosz_2012}.  
Their detection
has led to a revitalized effort to
understand planet formation around binary stars
\cite{Meschiari_2012,Paardekooper_2012}. 
A circumbinary planet
can reveal itself in two ways.  If the planet's orbital plane is
favorably aligned, the planet may transit across one or both of the
stars, causing a small decrease in the amount of light from the
system.  If the planet is sufficiently massive and close, the planet
can perturb the stellar orbits \cite{Borkovits_2003}.  The most
readily observable manifestation of this perturbation is a change in
the times when the eclipses occur.

In contrast to a single planet orbiting a single star, a planet 
in a circumbinary system must transit a ``moving target.''  
As a consequence, the time intervals between the transits as well as
their duration can vary substantially. The transits can deviate 
from having a constant period
by up to several days and can vary in duration by 
several hours. 
These transit timing and duration variations can
be taken to be the signature of a circumbinary planet because
no other known mechanism can cause such effects.
Modeling of the timing and duration changes can be used to
precisely determine the orbits of the planet
and stars 
\cite{Doyle_2011,Welsh_2012,Orosz_2012}.

Kepler observations of the binary star 
system Kepler-47 (KIC 10020423, also KOI-3154) 
show primary eclipses 
(smaller and fainter star blocking the brighter, more massive
``primary'' star)
every 7.45 days with a depth of 13\%. Also present are secondary
eclipses with a depth of 0.8\% (Fig.\ 1) and a quasi-periodic modulation 
in the out-of-eclipse regions
of $\sim 2-4$\% 
caused by star-spots on the primary star \cite{SOM}.
The Kepler data span 1050.5 days and 
visual inspections of the light curve
revealed the signals of two candidate 
circumbinary planets, with periods of $\sim 50$ and 
$\sim 303$ days.
Three transits of the longer-period candidate 
(hereafter the outer planet) were readily apparent,
but those of the shorter-period candidate (hereafter the inner
planet) were more difficult to find because of their shallower
depth. Using the predictions of a preliminary model of the system 
as a guide, a total of 18 transits of the 
inner planet were detected.
The transits have timings that
can deviate from strict periodicity by
up to several hours and their 
durations vary significantly, strongly suggesting their origins are
from
circumbinary planets (Fig. 2).
All of these events are transits over the primary star.

To characterize the stellar orbit, 
we obtained Doppler spectroscopy of the system
\cite[{\rm Fig.\ 1}]{SOM}.  Radial velocity variations of the primary star were readily
detected, but the secondary star is too faint to have been measured.
Usually when the radial velocity measurements of
only one component in a spectroscopic binary are available, the masses
of the stars cannot be uniquely determined.  However, 
the
transit times and 
durations provide constraints on the geometric
configuration of the stellar orbits and specify the stellar mass
ratio, which in combination with the primary's radial velocities,
allow both stellar masses and the physical scale to be determined.


To determine the system parameters, we used a photometric-dynamical model 
\cite{Carter_2011}
similar to that used for the four previously known transiting
circumbinary planets \cite{Doyle_2011,Welsh_2012,Orosz_2012}. 
This model assumes spherical bodies interacting
via Newtonian gravity \cite{SOM}, and
is used to fit the radial velocity data and
the Kepler time-series photometry.
We determined
the stellar masses 
as described above, and the relative sizes of the bodies
from the eclipses and transits in
the light curve. Information
on the inclination, eccentricity, and mutual inclination of the
planetary orbits is also implicit in the combination of photometric
and radial-velocity data. 
Gravitational perturbations
caused by the planets on the stars and on each other could, in principle, 
also constrain the masses, but for Kepler-47 the expected
masses of the planets 
are too small to create a measurable effect over the time span of our data.
The small radii of the transiting objects strongly suggests
they are of planetary mass
(Table 1); dynamical considerations described below
make this conclusion secure.

The inner planet, Kepler-47 b,
is the smallest transiting circumbinary planet yet detected, 
with a radius of
$3.0 \pm 0.1$ Earth radii. 
Its mass is too small to  be directly measured, but a
$3\sigma$ 
upper limit of 2 Jupiter masses has been determined based on the 
nondetection of timing variations of the stellar orbit \cite{SOM}.  
Because the planet's mass is
unknown, its density is also unknown and it is not possible to
distinguish between a rocky composition and a more volatile-enriched
composition. 
We can make a plausible mass etimate by using both an empirical
mass-radius relation based on transiting exoplanets \cite{Kane_2012}
and a
limited empirical mass-radius relation for planets in the
Solar System \cite{Lissauer_2011} yielding 
$\sim7-10$ Earth masses or
$\sim 0.4-0.6$ Neptune
masses.
The planet's 49.5-day orbital period is 6.6 times the period of
the stellar binary. This is $\sim 77\%$ longer than the critical
period (28~d) within which the planet would be susceptible to
dynamical instability due to interactions with the stars
\cite{Holman_1999}.
While this 77\% margin is notably larger than for the other known 
transiting circumbinary planets, 
i.e., 14\%, 21\%, 24\%, 42\% for Kepler 16, 34, 35, and 38, respectively, 
the planet is still somewhat
close to the instability limit, a feature shared 
by all known transiting circumbinary planet systems.

The outer planet, Kepler-47 c,
has a radius of $4.6 \pm 0.2$ Earth radii, making it slightly larger
than the planet Uranus.  As before, the planet's mass is too small
to be measured directly,
and we derived a $3\sigma$ upper limit of 
28 Jupiter masses  \cite{SOM}.
Based on its radius, we find a plausible mass of 
$\sim 16-23$ Earth masses or $\sim 0.9-1.4$ Neptune masses,
using these empirical  mass-radius relations
\cite{Kane_2012,Lissauer_2011}.
With only 3 transits currently available, the outer planet's orbital
eccentricity is poorly constrained.  A perfectly circular orbit would
fit the data, and a low-eccentricity orbit seems plausible given the
low eccentricity of the stellar binary ($e=0.023$) and of planet b
($e<0.035$).  The photometric-dynamical model provides only an upper
limit on the eccentricity, $e<0.4$ with 95\% confidence, and the
requirement of long-term stability only rules out eccentricities
larger than 0.6 \cite{SOM}.

Due to the orbital motion of the stars, the outer planet
is subject to variations in the incident
stellar flux (i.e.\ insolation), even if the planet's orbit
is circular (Fig. 3).
The average insolation is similar to the amount the Earth 
receives from the Sun: for a circular orbit it is 87.5\% of 
the Sun-Earth insolation, and varies by $\sim9$\%. 
This places Kepler-47 c well within the classical ``habitable zone'', 
defined as the range of distances 
from the host star(s) where liquid water could persist on the
surface of an Earth-like planet \cite{Kasting_1993}.
While Kepler-47 c is probably a gas giant and thus not suitable 
for life, its location is notable as it demonstrates that circumbinary 
planets can exist in habitable zones.
Although the definition of the habitable zone assumes 
a terrestrial planet atmosphere which does not apply for Kepler-47 c,
large moons, if present, would be interesting worlds to investigate.

A 0.2\% deep transit-like event is present at time 2,455,977.363 
(BJD)
that is not caused by either of the two planets. A search for additional
transits has revealed several more tentative transit
events \cite{SOM}, but we caution 
that the star is faint (the Kepler
magnitude is 15.178), there are large modulations due to star-spots, and 
the data contain correlated ``red'' noise, making small, non-periodic 
transit detection challenging.
The marginal evidence at the present time is insufficient to place
confidence on any additional candidate planet(s).

The primary star is similar to the Sun in both mass and radius, 
and dominates the luminosity of the binary system, having
60 times the bolometric luminosity of the secondary star
(or 176 times the brightness in the Kepler bandpass).
A spectroscopic analysis  
gives an effective temperature of
$5640 \pm 100~K$ for the primary star
(Table S2), with a metallicity slightly
less than solar ([M/H]$=-0.25\pm 0.08$ dex).
The star's rotation period as determined from
the star-spot modulation in the light curve \cite{SOM}
is only 4\% longer than the orbital
period, suggesting that the spin and orbital angular momenta have been
synchronized by tidal interactions.  Supporting this interpretation,
the obliquity of the primary star (the angle between the spin and
orbital axes) must be smaller than about $20^{\circ}$, based on the
observable effects of the secondary star eclipsing  star-spots on the
primary star \cite{SOM,Sanchis_2011,Nutzman_2011,Desert_2011,Sanchis_2012}.
Star-spot crossings also perturb the shape and depth of the primary
eclipses, leading to systematic trends in the eclipse times, and 
limit the precision with which one
can infer the planets' masses.  In addition, the loss
of light due to star-spots causes eclipses to appear 
slightly deeper than they would for an unspotted star,
biasing the determination of the stellar and planetary
radii too high by a few percent.

With Kepler-47 b and c, there are six confirmed transiting circumbinary
planets currently known. Their orbital periods relative to their host binary
stars show no tendency to be in resonance, and their radii are Saturn-size
and smaller. Given that Jupiter-size planets are easier to detect, their
absence in the Kepler data
suggests that the formation and migration history of circumbinary
planets may disfavor
Jupiter-mass planets orbiting close to the stars, in accord with
\cite{Pierens_2008}.

The planets in Kepler-47 are expected to have formed much farther out than
their present orbits, at locations where the conditions for the formation of
giant planets are more favorable \cite{Meschiari_2012,Paardekooper_2012}.
The planets have likely migrated to their current orbits as a result of
interactions with the circumbinary disk. The multiplicity and coplanarity of
the orbits strengthens the argument for a single-disk formation and a
migration scenario for circumbinary planetary systems. However, unlike
orbits around a single star, the environment around a binary star is much
more dynamic and tends to augment planet-planet interactions. The relatively
large distance between the orbits of the inner and outer planets in the
Kepler-47 system is consistent with requirements for dynamical stability
\cite{Youdin_2012}.

%

The previously detected transiting circumbinary planet systems show no
evidence for more than a single planet. The multi-planet nature of the
Kepler-47 system establishes that despite the chaotic environment around
binary stars, planetary systems can form and persist close to the binary,
and invites a broader investigation into how circumbinary planets compare to
planets and planetary systems around single stars.

\bibliography{scibib}

\begin{thebibliography}{10}

\bibitem{Koch_2010}
D. G. Koch, W. J. Borucki, G. Basri, 
N. M. Batalha, T. M.  Brown, et al.,
Kepler mission design, realized photometric performance, and early science.
{\em Astrophys. J.}  {\bf 713}, L79-L86 (2010).

\bibitem{Borucki_2010}
W. J. Borucki,  D. Koch, G. Basri, N. Batalha, T. Brown,
et al., Kepler Planet-Detection Mission: Introduction and First Results.
{\em Science} {\bf 327}, 977 (2010).

\bibitem{Batalha_2012}
N. M. Batalha, J. F. Rowe, S. T. Bryson, T. Barclay, C. J. Burke,
et al., 
Planetary candidates observed by Kepler, III: analysis of the first 
16 months of data.
{\em Astrophys. J. Suppl.\ Ser.} submitted, arXiv:1202.5852, (2012).

\bibitem{Prsa_2011}
A. Pr{\v s}a, A., N. Batalha, R. W. Slawson, L. R. Doyle, W. F. Welsh,
et al.,
Kepler eclipsing binary stars. I. 
catalog and principal characterization of 1879 eclipsing 
binaries in the first data release.
{\em Astron. J.} {\bf 141}, 83 (2011).

\bibitem{Slawson_2011}
R. W. Slawson, A. Pr{\v s}a, W. F. Welsh, J. A. Orosz, M. Rucker,
et al., Kepler eclipsing binary stars. II. 2165 
eclipsing binaries in the second data release.
{\em Astron. J.} {\bf 142}, 160 (2011).

\bibitem{Doyle_2011}
L. R. Doyle, J. A. Carter, D. C. Fabrycky, R. W. Slawson,
S. B. Howell, et al., Kepler-16: a transiting circumbinary planet.
{\em Science} {\bf 333}, 1602 (2011).

\bibitem{Welsh_2012}
W. F. Welsh, J. A. Orosz, J. A. Carter, D. C. Fabrycky, E. B.
Ford, et al., Transiting circumbinary planets Kepler-34 b and Kepler-35 b.
{\em Nature} {\bf 481}, 475-479 (2012).

\bibitem{Orosz_2012}
J. A. Orosz, W. F. Welsh, J. A.  Carter, E. Brugamyer, L. A.,
Buchhave, et al.,
The Neptune-sized circumbinary planet Kepler-38b.
{\em Astrophys. J.} in press, arXiv:submit/0534300 (2012).



\bibitem{Meschiari_2012}
S. Meschiari, 
Circumbinary planet formation in the Kepler-16 system. 
I. N-body simulations.
{\em Astrophys. J.} {\bf 752}, 71 (2012).

\bibitem{Paardekooper_2012}
S.-J. Paardekooper,  Z. M. Leinhardt, P. Th\'ebault,  \& C. Baruteau,  
How not to build Tatooine: 
the difficulty of in situ formation of circumbinary planets 
Kepler 16b, Kepler 34b, and Kepler 35b.
{\em Astrophys. J.}
{\bf 754}, L16 (2012).

\bibitem{Borkovits_2003}
T. Borkovits, B. \'Erdi, E. Forg\'acs-Dajka,  \&
T. Kov\'acs, 
On the detectability of long period perturbations in close 
hierarchical triple stellar systems.
 {\em Astron.\ Astroph.} {\bf 398}, 1091-1102 (2003).

\bibitem{SOM}
See the Supplementary Materials on {\em Science} Online. 


\bibitem{Carter_2011}
J. A. Carter, D. C. Fabrycky, D. Ragozzine, M. J. Holman, S. N. Quinn,
et al., KOI-126: A triply eclipsing 
hierarchical triple with two low-mass stars.
{\em Science} {\bf 331}, 562 (2011).

\bibitem{Kane_2012}
S. R. Kane,  \& D. M. Gelino, 
The habitable zone gallery.
{\em Pubs. Astron. Soc. Pac.} {\bf 124}, 323-328 (2012).

\bibitem{Lissauer_2011}
J. J. Lissauer, D. Ragozzine, D. C.  Fabrycky, J. H. Steffen,
E. B. Ford, 
et al., Architecture and dynamics of Kepler's 
candidate multiple transiting planet systems.
{\em Astrophys. J. Suppl. Ser.} 
{\bf 197}, 8 (2011).


\bibitem{Holman_1999}
M. J. Holman,  \& P. A. Wiegert,
Long-term stability of planets in binary systems.
{\em Astron. J.} {\bf 117}, 621-628 (1999).

\bibitem{Kasting_1993}
J. F. Kasting, D. P. Whitmire, \& R. T. Reynolds,
Habitable zones around main sequence stars.
{\em Icarus} {\bf 101}, 108-128 (1993).


\bibitem{Sanchis_2011}
R. Sanchis-Ojeda, J. N.  Winn, M. J.  Holman, J. A.  Carter, D. J. Osip, 
\& C. I. Fuentes,  
Starspots and spin-orbit alignment in the WASP-4 exoplanetary system.
{\em Astrophys. J.} {\bf 733}, 127 (2011). 

\bibitem{Nutzman_2011}
P. A. Nutzman, D. C.  Fabrycky,  \& J. J. Fortney, 
Using star spots to measure the spin-orbit 
alignment of transiting planets.
{\em Astrophys. J.} {\bf 740}, L10 (2011).

\bibitem{Desert_2011}
J. M. D\'esert, D. Charbonneau, B.-O. Demory, S. Ballard,
J. A. Carter, et al.,
The Hot-Jupiter Kepler-17b: discovery, obliquity from 
stroboscopic starspots, and atmospheric characterization.
et al., {\em Astrophys. J. Suppl. Ser.} {\bf 197}, 14 (2011).

\bibitem{Sanchis_2012}
R. Sanchis-Ojeda, D. C.  Fabrycky, 
J. N.  Winn, T.  Barclay, B. D. Clarke,
et al.,  Alignment of the stellar spin with the 
orbits of a three-planet system.
{\em Nature} {\bf 487}, 449 (2012).




\bibitem{Pierens_2008}
A. Pierens,  \& R. P. Nelson, 
On the formation and migration of 
giant planets in circumbinary discs.
{\em Astron.\ Astroph.} {\bf 483}, 633-642 (2008).

\bibitem{Youdin_2012}
A.~N. Youdin, K.~M. Krautter, \& S.~J. Kenyon,
Circumbinary Chaos: Using Pluto's Newest Moon to Constrain the Masses of Nix
and Hydra.
{\em Astrophys. J.} {\bf 755}, 17 (2012).

\bibitem{Underwood_2003}
D. R. Underwood, B. W.  Jones,  \& P. N. Sleep,
The evolution of habitable zones during stellar lifetimes
and its implications on the search for extraterrestrial life.
{\em Int. J. Ast. Bio.}
{\bf 2}, 289 (2003).


\end{thebibliography}

\clearpage

\begin{thebibliography}{10}

\bibitem[1]{Koch_2010S}
D. G. Koch, W. J. Borucki, G. Basri, 
N. M. Batalha, T. M.  Brown, et al.,
Kepler mission design, realized photometric performance, and early science.
{\em Astrophys. J.}  {\bf 713}, L79-L86 (2010).

\bibitem[2]{Borucki_2010S}
W. J. Borucki, D. Koch, G. Basri, N. Batalha, T. Brown,
et al., Kepler Planet-Detection Mission: Introduction and First Results.
{\em Science} {\bf 327}, 977-980 (2010).

\bibitem[3]{Batalha_2012S}
N. M. Batalha, J. F. Rowe, S. T. Bryson, T. Barclay, C. J. Burke,
et al., 
Planetary candidates 
observed by Kepler, III: analysis of the first 16 months of data.
{\em Astrophys. J. Suppl.\ Ser.} submitted, arXiv:1202.5852, (2012).

\bibitem[4]{Prsa_2011S}
A. Pr{\v s}a, A., N. Batalha, R. W. Slawson, L. R. Doyle, W. F. Welsh,
et al.,
Kepler eclipsing binary stars. I. 
catalog and principal characterization of 1879 eclipsing 
binaries in the first data release.
{\em Astron. J.} {\bf 141}, 83 (2011).

\bibitem[5]{Slawson_2011S}
R. W. Slawson, A. Pr{\v s}a, W. F. Welsh, J. A. Orosz, M. Rucker,
et al., Kepler eclipsing binary stars. II. 2165 
eclipsing binaries in the second data release.
{\em Astron. J.} {\bf 142}, 160 (2011).

\bibitem[6]{Doyle_2011S}
L. R. Doyle, J. A. Carter, D. C. Fabrycky, R. W. Slawson,
S. B. Howell, et al., Kepler-16: a transiting circumbinary planet.
{\em Science} {\bf 333}, 1602-1606 (2011).

\bibitem[7]{Welsh_2012S}
W. F. Welsh, J. A. Orosz, J. A. Carter, D. C. Fabrycky, E. B.
Ford, et al., Transiting circumbinary planets Kepler-34 b and Kepler-35 b.
{\em Nature} {\bf 481}, 475-479 (2012).

\bibitem[8]{Orosz_2012S}
J. A. Orosz, W. F. Welsh, J. A.  Carter, E. Brugamyer, L. A.,
Buchhave, et al.,
The Neptune-sized circumbinary planet Kepler-38b.
{\em Astrophys. J.} in press,  arXiv:1208.3712 (2012).




\bibitem[9]{Meschiari_2012S}
S. Meschiari, 
Circumbinary planet formation in the Kepler-16 system. 
I. N-body simulations.
{\em Astrophys. J.} {\bf 752}, 71 (2012).

\bibitem[10]{Paardekooper_2012S}
S.-J. Paardekooper,  Z. M. Leinhardt, P. Th\'ebault,  \& C. Baruteau,  
How not to build Tatooine: 
the difficulty of in situ formation of circumbinary planets 
Kepler 16b, Kepler 34b, and Kepler 35b.
{\em Astrophys. J.}
{\bf 754}, L16 (2012).

\bibitem[11]{Borkovits_2003S}
T. Borkovits, B. \'Erdi, E. Forg\'acs-Dajka,  \&
T. Kov\'acs, 
On the detectability of long period perturbations in close 
hierarchical triple stellar systems.
 {\em Astron.\ Astroph.} {\bf 398}, 1091-1102 (2003).

\bibitem[12]{SOMS}
See the Supplementary Materials on {\em Science} Online. 


\bibitem[13]{Carter_2011S}
J. A. Carter, D. C. Fabrycky, D. Ragozzine, M. J. Holman, S. N. Quinn,
et al., KOI-126: A triply eclipsing 
hierarchical triple with two low-mass stars.
{\em Science} {\bf 331}, 562-565 (2011).

\bibitem[14]{Kane_2012S}
S. R. Kane,  \& D. M. Gelino, 
The habitable zone gallery.
{\em Pubs. Astron. Soc. Pac.} {\bf 124}, 323-328 (2012).

\bibitem[15]{Lissauer_2011S}
J. J. Lissauer, D. Ragozzine, D. C.  Fabrycky, J. H. Steffen,
E. B. Ford, 
et al., Architecture and dynamics of Kepler's 
candidate multiple transiting planet systems.
{\em Astrophys. J. Suppl. Ser.} 
{\bf 197}, 8 (2011).


\bibitem[16]{Holman_1999S}
M. J. Holman,  \& P. A. Wiegert,
Long-term stability of planets in binary systems.
{\em Astron. J.} {\bf 117}, 621-628 (1999).

\bibitem[17]{Kasting_1993S}
J. F. Kasting, D. P. Whitmire,  \& R. T. Reynolds,
Habitable zones around main sequence stars.
{\em Icarus} {\bf 101}, 108-128 (1993).


\bibitem[18]{Sanchis_2011S}
R. Sanchis-Ojeda, J. N.  Winn, M. J.  Holman, J. A.  Carter, D. J. Osip, 
\& C. I. Fuentes,  
Starspots and spin-orbit alignment in the WASP-4 exoplanetary system.
{\em Astrophys. J.} {\bf 733}, 127 (2011). 

\bibitem[19]{Nutzman_2011S}
P. A. Nutzman, D. C.  Fabrycky,  \& J. J. Fortney, 
Using star spots to measure the spin-orbit 
alignment of transiting planets.
{\em Astrophys. J.} {\bf 740}, L10 (2011).

\bibitem[20]{Desert_2011S}
J. M. D\'esert, D. Charbonneau, B.-O. Demory, S. Ballard,
J. A. Carter, et al.,
The Hot-Jupiter Kepler-17b: discovery, obliquity from 
stroboscopic starspots, and atmospheric characterization.
et al., {\em Astrophys. J. Suppl. Ser.} {\bf 197}, 14 (2011).

\bibitem[21]{Sanchis_2012S}
R. Sanchis-Ojeda, D. C.  Fabrycky, 
J. N.  Winn, T.  Barclay, B. D. Clarke,
et al.,  Alignment of the stellar spin with the 
orbits of a three-planet system.
{\em Nature} {\bf 487}, 449-453 (2012).



\bibitem[22]{Pierens_2008S}
A. Pierens,  \& R. P. Nelson, 
On the formation and migration of 
giant planets in circumbinary discs.
{\em Astron.\ Astroph.} {\bf 483}, 633-642 (2008).

\bibitem[23]{Youdin_2012S}
A.~N. Youdin, K.~M. Krautter, \& S.~J. Kenyon,
Circumbinary Chaos: Using Pluto's Newest Moon to Constrain the Masses of Nix
and Hydra.
{\em Astrophys. J.} {\bf 755}, 17 (2012).


\bibitem[24]{Underwood_2003S}
D. R. Underwood, B. W.  Jones,  \& P. N. Sleep,  
The evolution of habitable zones during stellar lifetimes 
and its implications on the search for extraterrestrial life.
{\em Int. J. Ast. Bio.}
{\bf 2}, 289-299 (2003).

\bibitem[25]{Brown_2011}
T. M.  Brown, D. W.  Latham, 
M. E. Everett,  \& G. A. Esquerdo,
Kepler input catalog: photometric calibration and stellar classification.
{\em Astron. J.}
{\bf 142}, 112
(2011)


\bibitem[26]{Kinemuchi_2012}
K. Kinemuchi, T. Barclay, M. Fanelli, J. Pepper, M. Still, 
\& S. B. Howell, 
Demystifying Kepler data: a primer for systematic artifact mitigation.
{\em Publ. Astron. Soc. Pac.} in press, arXiv:1207.3093, (2012).

\bibitem[27]{Mazeh_2008}
T. Mazeh,  in Observational Evidence for Tidal Interaction
in Close Binary Systems, eds.\ M.-J. Goupil \& J.-P. Zahn, EAS
Publications Series, vol.\ 29, p. 29 (2008).

\bibitem[28]{tull1998}
R. G. Tull,
High-resolution fiber-coupled spectrograph of the Hobby-Eberly Telescope.
{\em Proc.\ Soc.\ Photo-opt.\ Inst.\ Eng.} {\bf 3355}, 
387-398 (1998).

\bibitem[29]{tull1995}
R. G. Tull,
P. J. MacQueen,
C. Sneden, C. \& D. L. Lambert,
The high-resolution cross-dispersed echelle white-pupil 
spectrometer of the McDonald Observatory 2.7-m telescope.
{\em Publ. Astron. Soc. Pacif.} {\bf 107}, 251-264 (1995).


\bibitem[30]{vogt1994}
S. S. {Vogt},
S. L. {Allen},
B. C. {Bigelow},
L. {Bresee},
B. {Brown},  et al.,
HIRES: the high-resolution echelle spectrometer on the Keck 10-m Telescope.
{\em Proc.\ Soc.\ Photo-opt.\ Inst.\ Eng.} 
{\bf 2198}, 362 (1994).

\bibitem[31]{marcy2008}
G. W. {Marcy}, 
R. P. {Butler},
S. S. {Vogt}, 
D. A. {Fischer}, 
J. T. {Wright},  et al.,
Exoplanet properties from Lick, Keck and AAT.
{\em Physica Scripta Volume T} {\bf 130}, 014001 (2008).


\bibitem[32]{rucinski1992}
S. M. Rucinski, 
Spectral-line broadening functions of WUMa-type binaries. I - AW UMa.
{\em AJ} {\bf 104}, 1968-1981 (1992).

\bibitem[33]{Nidever_2002}
D. L. Nidever, G. W. Marcy, R. P. Butler, D. A. Fischer, \& S. S. Vogt,
Radial velocities for 889 late-type stars.
{\em Astrophys. J. Suppl. Ser.} {\bf 141}, 502-522 (2002).

\bibitem[34]{Chubak_2012}
C. Chubak, G. W. Marcy, D. A. Fischer, 
A. W. Howard, H. Isaacson, H., et al.,
Precise radial velocities of 2046 nearby FGKM stars and 131 standards.
{\em Astrophys. J. Supp. Ser.} submitted, arXiv:1207.6212v1, (2012).

\bibitem[35]{Buchhave_2012}
L A. Buchhave, D. W. Latham, A. Johansen,
M. Bizzarro, G. Torres,  et al.,
An abundance of small exoplanets around stars with a 
wide range of metallicities.
{\em Nature} {\bf 486}, 375-377 (2012).


\bibitem[36]{Czesla_2009}
S. Czesla, K. F. Huber, U.  Wolter, S.  Schr\"oter, 
J. H. M. M. Schmitt,
How stellar activity affects the size estimates of extrasolar
planets. 
{\em Astron. Astroph.} 
{\bf 505}, 1277-1282 (2009).

\bibitem[37]{Carter_2011b}
J. A. Carter, J. N. Winn, M. J. Holman, D. C. Fabrycky,
Z. K. Berta, 
et al., The transit light curve project. XIII. sixteen
transits of the Super-Earth GJ 1214b. 
{\em Astrophys. J.} {\bf 730}, 82 (2011).

\bibitem[38]{Mandel_2002}
K. Mandel,  \& E. Agol, 
Analytic light curves for planetary transit searches. 
{\em Astrophys. J.} 
{\bf 580}, L171-L175 (2002).


\bibitem[39]{Hirano_2012}
T. Hirano, R. Sanchis-Ojeda, Y. Takeda, N. Narita, J. N. Winn, 
et al. 
Measurements of stellar inclinations for Kepler planet candidates. 
{\em Astrophys. J.} in press, arXiv:1205.3233, (2012).

\bibitem[40]{Winn_2005}
J. N. Winn, R. W. Noyes, M. J. Holman,
D. Charbonneau, Y. Oha, 
et al. 
Measurement of spin-orbit alignment in an
extrasolar planetary system. 
{\em Astrophys. J.} 
{\bf 631}, 1215-1226 (2005).


\bibitem[41]{Carter_2012}
J. A. Carter, E. Agol, W. J. Chaplin, S. Basu, T. R. Bedding, 
et al.,  Kepler-36: a pair of planets 
with neighboring orbits and dissimilar densities.   
{\em Science} {\bf 337}, 556-559  (2012). 

\bibitem[42]{Soderhjelm1411984}
S. Soderhjelm, Third-order and tidal effects 
in the stellar three-body problem.   {\em Astron. Astrophy.} {\bf 
141}, 232--240 (1984). 

\bibitem[43]{Mardling5732002}
R. A. Mardling, 
\& D. N. C. Lin,  Calculating the tidal, spin, and dynamical evolution of 
extrasolar planetary systems.   
{\em The Astroph. J.} {\bf 573}, 829--844 (2002). 

\bibitem[44]{Press2002}
W. H. Press, S. A. Teukolsky, W. T. Vetterling, 
\& B. P. Flannery,  Numerical recipes in C++ : the art of scientific computing.  
(Cambridge University Press, Cambridge,
2002). 

\bibitem[45]{terBraak} 
C. J. F. ter Braak, \& J. A. Vrugt, J.A., 
A Markov Chain Monte Carlo version of 
the genetic algorithm Differential Evolution: easy Bayesian computing for 
real parameter spaces. {\it Statistics
and Computing} {\bf 16}, 239--249 (2006).

\bibitem[46]{Orosz_2000}
J. A. Orosz, \& P. H. Hauschildt,
The use of the NextGen model atmospheres for cool giants 
in a light curve synthesis code.
{\em Astron.\ Astrophys.} {\bf 364}, 265-281 (2000).

\bibitem[47]{Avni_1976} 
Y. Avni,  
The eclipse duration of the X-ray pulsar 3U 0900-40.
{\em Astrophys. J.} {\bf 209}, 574-577 (1976).

\bibitem[48]{Wilson_1979}
R. E. Wilson, 
Eccentric orbit generalization and simultaneous solution of 
binary star light and velocity curves.
{\em Astrophys. J.} {\bf  234}, 
1054-1066 (1979).

\bibitem[49]{Gimenez_2006}
A. Gim\'enez, 
Equations for the analysis of the light curves of extra-solar 
planetary transits.
{\em Astron. Astrophys.}  {\bf 450}, 1231-1237 (2006).


\bibitem[50]{Fabrycky_2012}
D. Fabrycky, Non-Keplerian dynamics of exoplanets.
in Exoplanets. ed.\ S. Seager, University of Arizona Press, p. 217 (2011).


\bibitem[51]{Lopez-Morales:07} 
M. L\'opez-Morales,  
On the correlation between the magnetic activity levels, 
metallicities, and radii of low-mass stars.
{\em Astrophys. J.} {\bf 660}, 732-739 (2007).

\bibitem[52]{Torres:10} 
G. Torres, J. Andersen, \&
A. Gim\'enez, 
Accurate masses and radii of normal stars: modern results and applications.
{\em Astron. \& Astrophys. Rev} {\bf 18}, 67-126 (2010).


\bibitem[53]{Yi:01} 
S. Yi, P. Demarque, Y. -C. Kim,
Y. -W. Lee, C. H. Ree, T. Lejeune,  \& S. Barnes, 
Toward better age estimates for stellar populations: 
the Y$^2$ isochrones for solar mixture.
{\em Astrophys. J. Suppl. Ser.} {\bf 136},
417-437 (2001).

\bibitem[54]{Demarque:04} 
P. Demarque, J. H.  Woo, 
Y. -C. Kim, \& S. K. Yi, 
Y$^2$ Isochrones with an improved core overshoot treatment.
{\em Astrophys J. Suppl. Ser.} {\bf 155}, 667-674 (2004).

\bibitem[55]{Pietrinferni_2004}
A. Pietrinferni, S. Cassisi, M.  Salaris,  \& F. Castelli, 
A large stellar evolution database for population synthesis studies. I. 
scaled solar models and isochrones.
{\em Astrophys. J.} {\bf 612}, 168-190 (2004).

\bibitem[56]{Dotter:08} 
A. Dotter, B.  Chaboyer, 
D. Jevremovi\'c, V.  Kostov, E. Baron,  \& J. W. Ferguson, 
The Dartmouth stellar evolution database.
{\em Astrophys. J. Suppl. Ser.} 
{\bf 178}, 89-101 (2008).


\bibitem[57]{Selsis_2007}
F. Selsis, J. F. Kasting, B. Levrad, J.  Paillet, 
I. Ribas,  \& X. Defosse,
Habitable planets around the star Gliese 581?.  
{\em Astron.\ Astroph.} {\bf 475},
1373-1387 (2007).


\bibitem[58]{Williams_2002}
D. M. Williams, \& D. Pollard,  
Earth-like worlds on eccentric orbits: excursions beyond the habitable zone.
{\em Int. J. Ast. Bio.} {\bf 1}, 61-69
(2002).

\bibitem[59]{Cahoy_2010}
K. L. Cahoy, M. S. Marley,  \& J. J. Fortney, 
Exoplanet albedo spectra and colors as a 
function of planet phase, separation, and metallicity.
{\em Astrophys. J.} {\bf 724}, 189-214 (2010).
\end{thebibliography}

\bibliographystyle{Science}


\begin{scilastnote}
\item[]
Kepler was selected as the 10th mission of the Discovery Program.
Funding for this mission is provided by NASA, Science Mission
Directorate. 
The Kepler data presented in this paper were obtained from the Mikulski
Archive for Space Telescopes (MAST). The Space Telescope Science Institute
(STScI) is operated by the Association of Universities for Research
in Astronomy, Inc., under NASA contract NAS5-26555.
Support for MAST for non-HTS data is provided by the NASA Office
of Space Science via grant NXX09AF08G and by other grants and
contracts.  
This work is based in part on 
observations obtained with the Hobby-Eberly Telescope, 
which is a joint project of the 
University of Texas at Austin, 
the Pennsylvania State University, 
Stanford University, Ludwig-Maximilians-Universit\"at M\"unchen, and 
Georg-August-Universit\"at G\"ottingen.
JAO and WFW acknowledge support from the 
Kepler Participating Scientist 
Program via NASA grant NNX12AD23G; JAO, WFW and GW also gratefully
acknowledge support from the NSF via grant AST-1109928.
G.T. acknowledges partial support for this work from NSF grant
AST-1007992.
J.A.C. and D.C.F. acknowledge NASA support through Hubble
Fellowship grants HF-51267.01-A and HF-51272.01-A, respectively,
awarded by
STScI.
\end{scilastnote}

{\bf Supplementary Materials}

Materials and Methods

Supplementary Text 

Figs.\ S1 to S25

Tables S1 to S9

Full reference list


\clearpage

\begin{table}
\begin{centering}
\begin{tabular}{lrr}
\hline
\hline
parameter & best fit & $1\sigma$ uncertainty \\
\hline
\multicolumn{3}{c}{Bulk Properties} \\
\hline
~~Mass of Star A, $M_A$ ($M_\odot$) & $1.043$ & $0.055$ \\
~~Mass of Star B, $M_B$ ($M_\odot$) & $0.362$ & $0.013$ \\
~~Radius of Star A, $R_A$ ($R_\odot$) & $0.964$ & $0.017$ \\
~~Radius of Star B, $R_B$ ($R_\odot$) & $0.3506$ & $0.0063$ \\
~~Temperature of Star A, $T_{{\rm eff},A}$ (K) & 5636 & 100 \\
~~Temperature of Star B, $T_{{\rm eff},B}$ (K) & 3357 & 100 \\
~~Luminosity of Star A, $L_A$ ($L_{\odot}$)  & 0.840 & 0.067 \\
~~Luminosity of Star B, $L_B$ ($L_{\odot}$)  & 0.014 & 0.002 \\
~~Radius of Planet b, $R_b$ ($R_\oplus$) & $2.98$ & $0.12 $ \\
~~Radius of Planet c, $R_c$ ($R_\oplus$) & $4.61$ & $0.20 $ \\
\hline
\multicolumn{3}{c}{Stellar Orbit} \\
\hline
~~Semimajor Axis, $a_{AB}$ (AU) & $0.0836$ & $0.0014 $ \\
~~Orbital Period, $P_{AB}$ (day) & $  7.44837695$  & $0.00000021$\\
~~Eccentricity, $e_{AB}$ & $0.0234$ & $0.0010$\\
~~Argument of Periapse, $\omega_{AB}$ (Degrees) & $   212.3$ & $   4.4 $ \\
~~Orbital Inclination, $i_1$ (deg) & $ 89.34$ & $ 0.12 $ \\
\hline
\multicolumn{3}{c}{Planet b Orbit} \\
\hline
~~Semimajor Axis, $a_b$ (AU) & $0.2956$ & $0.0047 $ \\
~~Orbital Period, $P_b$ (day) & $  49.514$ & $ 0.040 $ \\
~~Eccentricity (95\% conf.), $e_b$ & $ < 0.035$ &  \\
~~Orbital Inclination, $i_b$ (deg) & $ 89.59$ & $ 0.50 $ \\
~~Mutual Orbital Inclination, $I_b$ (deg) & $ 0.27$ & $ 0.24 $ \\
\hline
\multicolumn{3}{c}{Planet c Orbit}  \\
\hline
~~Semimajor Axis, $a_c$ (AU) & $ 0.989$ & $ 0.016 $ \\
~~Orbital Period, $P_c$ (day) & $ 303.158$ & $ 0.072 $ \\
~~Eccentricity (95\% conf.), $e_c$ & $ < 0.411$ &  \\
~~Orbital Inclination, $i_c$ (deg) & $89.826$ & $ 0.010 $ \\
~~Mutual Orbital Inclination, $I_c$ (deg) & $ 1.16$ & $ 0.46 $ \\
\hline
\hline
\end{tabular}
\caption{{\bf A summary of the results for the photometric-dynamical model.}
For brevity some of the fitting parameters are not listed here. See 
Table S5
for a complete listing of fitting parameters.}
\end{centering}
\end{table}

\clearpage
\thispagestyle{empty}
\begin{figure}
\begin{center}
\includegraphics[scale=0.75,angle=-00]{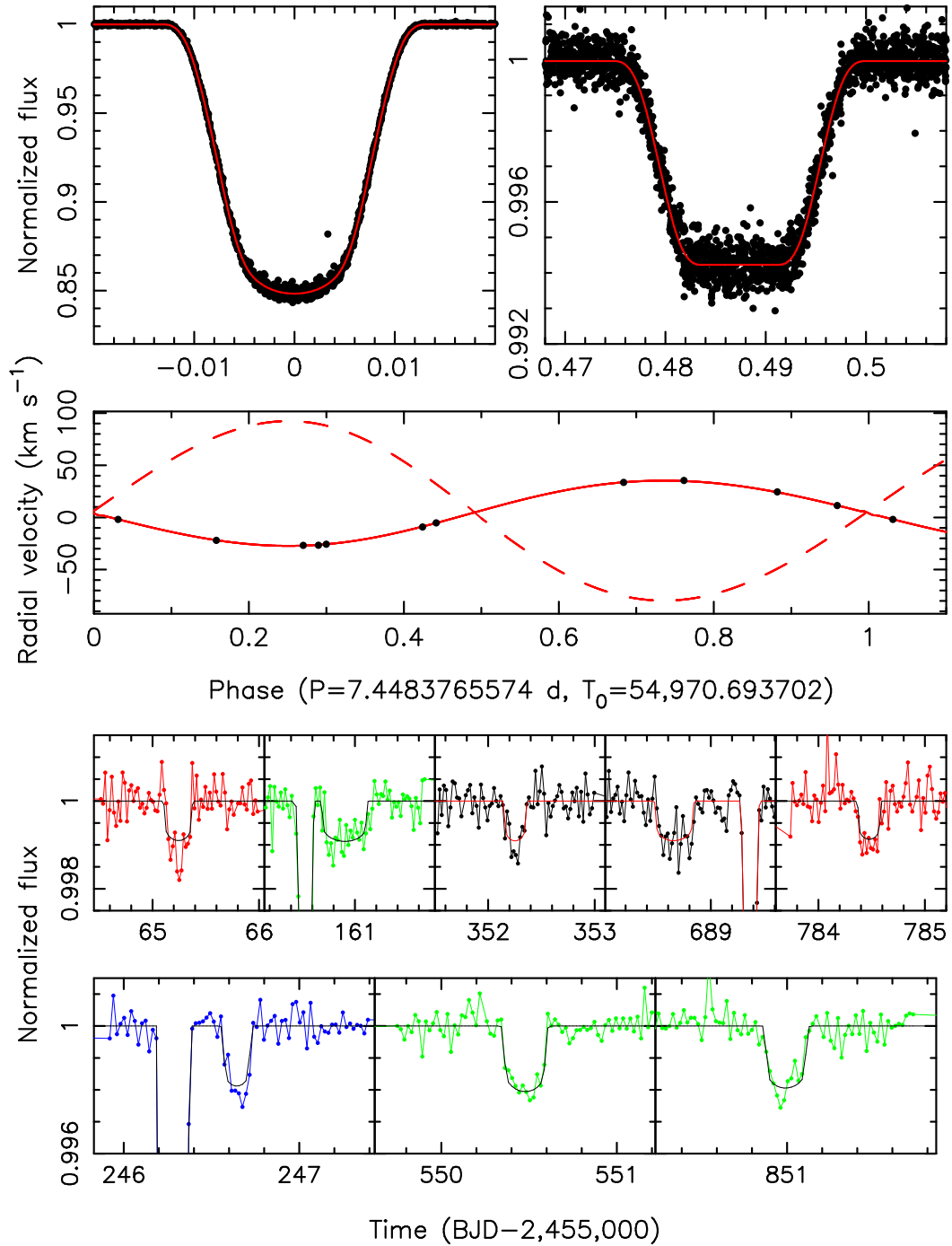}
\caption{{\bf Light curves and velocity curve data with model fits.}
Top: Normalized and detrended flux 
is plotted versus orbital phase for the primary and secondary
eclipses, along with the binary star model.  Middle: The radial
velocities of the primary star and the best-fitting model
are plotted versus the orbital phase.  The expected radial velocity
curve of the secondary star is shown with the dashed line.  Bottom:
The normalized and detrended flux near five representative transits of
the inner planet and all three transits of the outer planet are shown.
See Figs.\ S13, S14, and S15
for plots of all 18 transits of the inner planet and
plots of the residuals of the various model fits.
}
\end{center}
\end{figure}
\clearpage

\thispagestyle{empty}
\begin{figure}
\begin{center}
\includegraphics[scale=0.65,angle=-90]{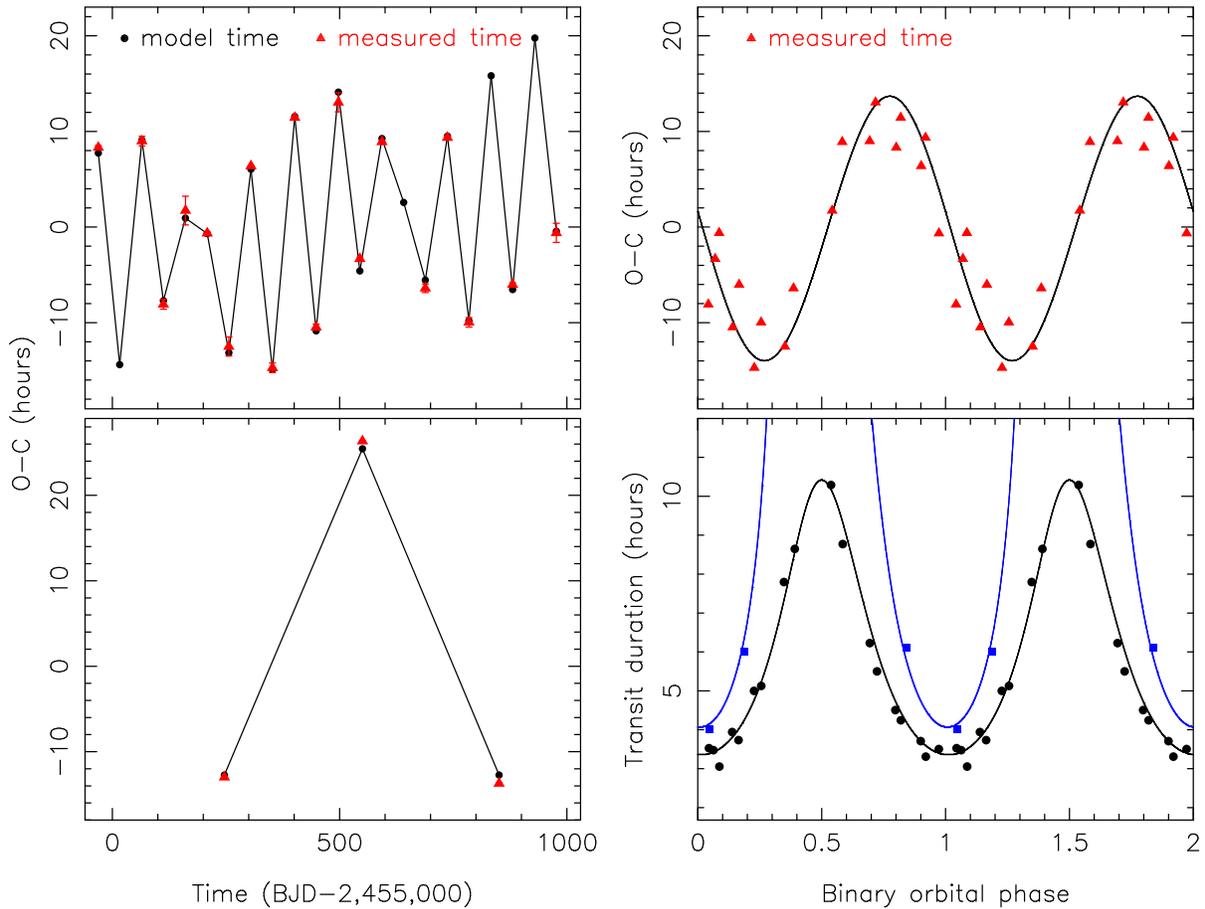}  
\caption{{\bf Planetary transit time and duration variations.}
Left:
The
observed minus expected times of transit computed from a linear
ephemeris are shown versus  time (an ``O--C'' curve). The triangles
show the measured deviations, and the filled circles are the predictions from
the photometric-dynamical model.  Four transits of the inner planet
occurred in data gaps or regions of corrupted data.  Top
right: The O-C values of the inner planet are shown as a function
of the binary phase, where the primary eclipse occurs at phase 0.0
and the secondary eclipse is at phase 0.487.  
Two cycles have been shown for clarity.
The solid curve is
the predicted deviation assuming a circular, edge-on orbit for the planet.
The lateral displacement
of the primary near the eclipse phases
is minimal and therefore the deviation of the transit time
from a linear ephemeris is near zero.  The primary is maximally displaced 
near the quadrature phases, so transits
near those phases 
show the most offset in time.
Bottom right:
The
durations 
of the transits for the inner planet (filled circles) and the
outer planet (filled squares) as a function of the orbital phase of
the binary.  
The solid curves are the predicted durations assuming a circular,
edge-on orbit for the planet.
At phases near the primary eclipse, the planet and the primary star
are moving in opposite directions, resulting in a narrower
transit.  At
phases near the secondary eclipse, the planet and the primary star are
moving in the same direction, resulting in a longer transit.   
The outer planet is moving slower than the
inner planet, resulting in longer transits at the same binary phase.
\label{plotmultiplanetOC}}
\end{center}
\end{figure}

\clearpage
\begin{figure}
\includegraphics[scale=0.65,angle=-90]{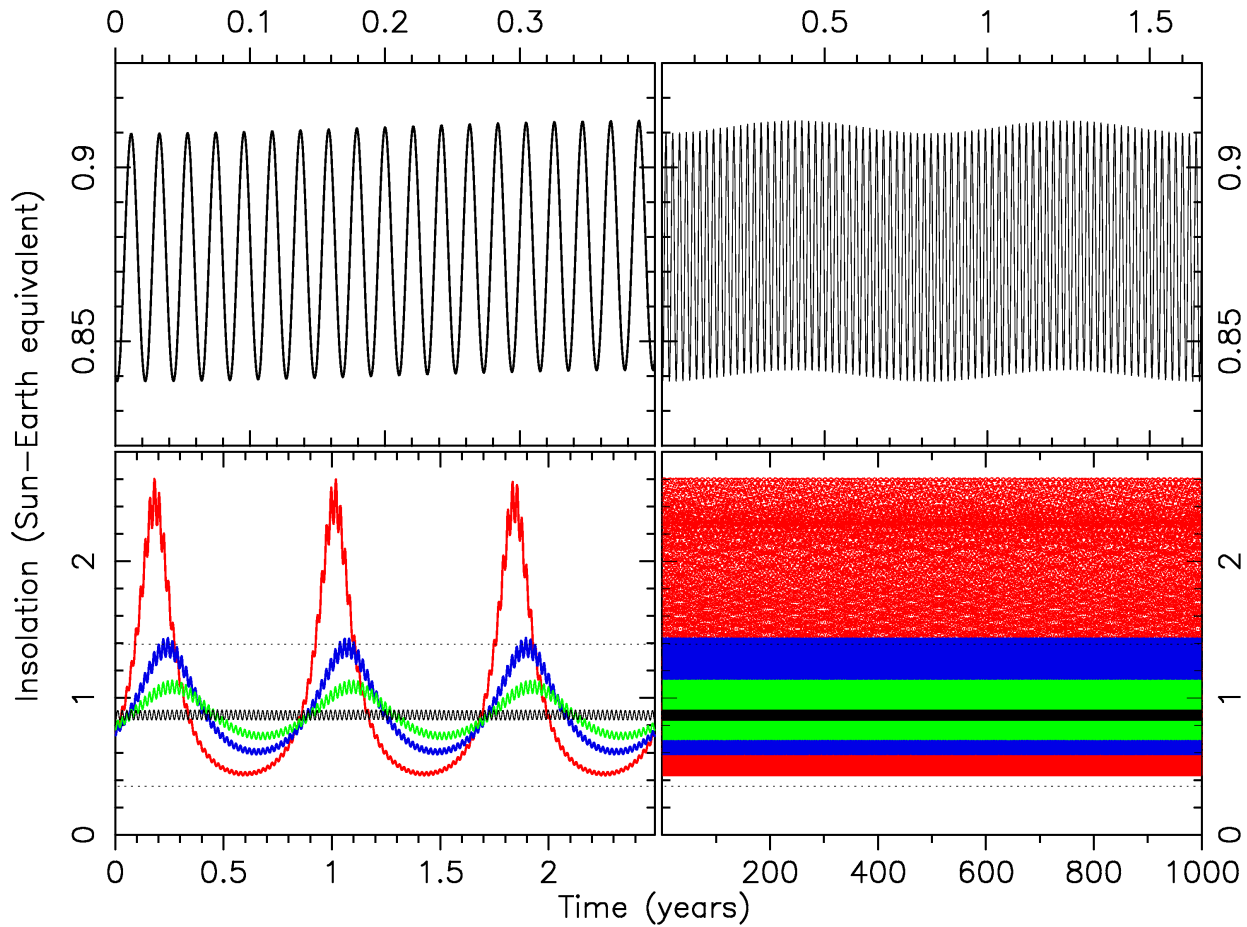}
\caption{{\bf    
The time-varying insolation S received by Kepler-47 c, for different
assumed eccentricities.} The insolation is in units of the Solar 
luminosity at a distance of 1 AU ($S_{\rm Sun}=1368$ W m$^{-2}$). 
The upper panels are for a zero eccentricity orbit and highlight the 
insolation variations caused by the 7.4-day orbit of the binary.
The lower panels show eccentricities of $e = 0.0$, 0.1, 0.2, and 0.4
(colored black, green, blue, and red, respectively), and illustrate
the longer time-scale variations. 
The dotted lines mark the limits for the inner and outer edges of 
the habitable zone, following the prescription in 
\protect\cite{Underwood_2003} for the onset of 
a runaway greenhouse effect and the maximum greenhouse effect.} 
\end{figure}


\clearpage

\newpage
\setcounter{page}{1}

\section*{Supporting Online Material (SOM)}

{\bf
We provide additional details regarding the
detection and characterization of Kepler-47 in 
this supplement.  \S\ref{altdesig} gives alternate
designations and other information for Kepler-47.
\S\ref{dataprep} discusses the Kepler data preparation
and detrending.
\S\ref{rotperiod} discusses how the rotational period
of the primary star is derived.
\S\ref{specobs} discusses the ground-based spectroscopic
observations.
\S\ref{SPC} describes how the effective temperature,
gravity, and metallicity of the primary were measured.
\S\ref{eclipsetimes} gives an overview of how the times of
mid-eclipse for the primary and secondary eclipses
were measured.
\S\ref{starspot} discusses the effects of star-spots
on the measurement of the eclipse times and
other parameters.
\S\ref{transittimes} presents measurements of the transit times
and the detection of a transit event possibly due to a third
planet.
\S\ref{sec:photo} gives a full discussion of the photometric-dynamical
model.
\S\ref{ELC} presents a discussion of independent light curve
modeling done with the ELC code.
\S\ref{upperlimit} discusses how upper limits on the masses of
the planets were derived.
\S\ref{stability} considers the long-term stability of the planetary
orbits.
\S\ref{evol} gives a comparison of the stellar properties with
evolutionary models.
\S\ref{habit} presents details of the habitable zone
in Kepler-47.

}

\section{Materials and Methods}

\subsection{Alternate designations, celestial coordinates,
and apparent magnitudes}\label{altdesig}

Kepler-47 appears in the
Kepler Input Catalog 
\cite[{\rm KIC}]{Brown_2011} as KIC 10020423.  Other designations
include Kepler Object of Interest KOI-3154 and
2MASS J19411149+4655136.  The J2000 celestial coordinates given in
the KIC are 
$\alpha=19^{\rm h}41^{\rm m}11\fss 501$,      
$\delta=+46^{\circ}55^{\prime}13\farcs 69$, and 
the apparent magnitudes are $r=15.126$ and $ {\rm Kp}=15.178$.

\subsection{Kepler data preparation and detrending}\label{dataprep}

In this study we make use of data from Kepler Quarters Q1 through
Q12 (May, 2009 through late March, 2012).
We used the ``simple aperture photometry'' (SAP) provided by
the Kepler pipeline and available at the
Mikulski Archive for Space Telescopes (MAST).
The Kepler SAP light curves show instrumental trends
\cite{Kinemuchi_2012}, 
so further
processing is necessary.  The detrending must be done for
each Quarter separately since the object appears on a different
detector module.  The amount of detrending needed depends on
the specific task.  When modeling
the eclipses and transits, a fairly aggressive detrending is used where
both the instrumental trends and the spot modulations are removed. 
In this case, the eclipses and transits are masked out, and a high
order cubic spline is fit to short segments whose end points are
usually defined by gaps in the data collection due to monthly
data downloads, rolls between Quarters, or spacecraft safe modes.
The segments are normalized to the spline fits, and the segments
are reassembled.
The SAP light curves and the detrended light curve with the spot
modulation removed are shown
in Fig.\ \ref{plotrawmulti}.
Other tasks such as spot modeling require much less aggressive
detrending, in which case low-order polynomials are used
to stitch together different segments across the Quarters.

The time difference between the last Q12 observation and the first Q1 observation
is 1050.51 days.  During that interval, Kepler was collecting data
92.55\% of the time, and 44389 cadences out of the 47580 in
total were flagged
as good (SAP\_QUALITY=0), for a duty cycle of 86.34\%.  Not all observations
with SAP\_QUALITY$>$0 are necessarily useless, depending on the purpose,
so the 86.34\% duty cycle is a lower limit.

\subsection{Rotational period from star-spot induced stellar 
variations}\label{rotperiod}

Fig.\ \ref{plotmultiQ159} shows closer-in views of the light curves
from Q1, Q5, and Q9.  A modulation of up to 3\%
in the out-of-eclipse regions due to
star spots rotating into and out of view is evident. This modulation has
a period that is close to, but not exactly equal to the eclipse
(e.g.\ orbital) period.  
Fig.\ \ref{autocorr} depicts the autocorrelation of the cleaned 
detrended
light curve, after the primary and secondary eclipses were removed and 
replaced by the value of the mean light curve with a typical random noise. 
The autocorrelation reveals clear modulation with a period of about 
7.8 days. Presumably, the clock behind the modulation is the stellar rotation
of the primary, 
which has brightness variation due to inhomogeneous distribution of 
stellar spots 
(as the primary star dominates the light in the Kepler bandpass, we
assume it is the source of the modulation).
To obtain a more precise value of the 
stellar rotation we measured the lags of 
the first 12 peaks of the autocorrelation and fitted 
them with a straight line as shown in Fig.\ \ref{lagline}.
From the slope of the fitted line
we derived a value of 
$7.775 \pm 0.022$ days as our best value for the stellar rotation period.
This period is slightly longer than the orbital period of 7.448 days.
It is interesting to note that the transition between synchronized 
and unsynchronized binaries for pre-main sequence  and young stars 
appears between 
7 and 8 days, as depicted by   \cite{Mazeh_2008}.

\subsection{Spectroscopic observations}\label{specobs}

We observed Kepler-47 four times with the High-Resolution-Spectrograph 
\cite[{\rm HRS}]{tull1998} 
at the Hobby-Eberly Telescope (HET).
Spectra with a resolving power of $R=30,000$ were obtained on UT 2012, 
April 23, May 18 \& 20  and  June 5.
We used the ``600g5822'' setting of HRS that delivers a spectrum from 4814
to 6793\,\AA. 
The data were reduced with our own HRS reduction
script using standard IRAF routines.
We selected a total exposure of 3600 seconds per spectrum (divided into three
sub-exposures of 1200 seconds each
to facilitate cosmic-ray removal). The
signal-to-noise (S/N) levels of the HRS spectra range from 30:1 to 55:1 at
5500\,\AA, 
depending on seeing conditions.
Adjacent to every visit to Kepler-47 we also observed the Kepler field
standard star HD~182488 to be used for the radial velocity determination.

In addition to the HET observations, we observed Kepler-47 six times using
the Tull Coud\'e spectrograph \cite{tull1995} 
at the Harlan J. Smith 2.7m telescope
(HJST). The
data were obtained with our standard instrumental setup that covers the
a wavelength range of 3760-10,200~\AA\
and uses a 1.2 arcsecond slit
that yields a resolving power of $R=60,000$. We obtained data 
during the nights
of UT 2012, May 1, 2, 4-6  and on June 26.
Exposure times ranged from 3600 to 4800 seconds (again divided in 1200
second sub-exposures) and the S/N is typically around 14:1
at 5500\AA. Each of these nights we also observed
HD~182488 to serve as a radial velocity  standard.
The data were reduced with our own reduction
scripts using standard IRAF routines.  
After some experimentation, it was discovered
that better measurements of the radial velocities were obtained from
spectra that did not have the sky background subtracted.

An additional spectrum of Kepler-47 was obtained using the
10 m Keck 1 telescope   
and the HIRES spectrograph\cite{vogt1994}.  
The spectra   were collected using the standard planet
search setup and reduction\cite{marcy2008}. 
The resolving power is $R=60,000$ at
5500~\AA. Sky subtraction, using the ``C2 decker'' was implemented with
a slit that projects to $0.87 \times 14.0$ arcsec on the sky. 
The wavelength calibrations
were made  using
Thorium-Argon lamp spectra.

The radial velocities of Kepler-47 were measured using
the ``broadening function'' technique \cite{rucinski1992}.
The broadening functions (BFs) are rotational
broadening kernels, where the centroid of the peak yields
the Doppler shift and where the width of the peak is
a measure of the rotational broadening.  The BF analysis
is often better suited for measuring radial velocities of
binary stars in cases where the velocity difference between the 
two stars is small compared to the spectral resolution.
A high quality spectrum of a slowly rotating star is needed for the BF
analysis, and for this purpose we used 
observations of HD182488 (spectral type G8V)
taken with each telescope+instrument combination.  The derived radial velocities
are insensitive to the precise spectral type of the template, as
similar radial velocities are found when using templates
of early G to late K.
The adopted template radial velocity  was 
$-21.508$
km s$^{-1}$ \cite{Nidever_2002}.

We prepared the spectra  for the BF analysis by normalizing each
echelle order to the local continuum using cubic splines, trimming
the low signal-to-noise ends of each order, and merging the orders
by interpolating to a log-linear wavelength scale.  
The wavelength ranges used for the final BF analysis was
4830-5770~\AA\ for the HET spectra and 5138-5509~\AA\ for the HJST spectra.
Fig.\ \ref{plotBF} shows example BFs from HET and HJST spectra. 
The spectrum is single-lined, as only one peak is evident in the
BFs from the HET.  
Some simple simulations were performed, and non-detection of a second
star in the HET spectra indicates the secondary star is $\gtrsim 10$ times
fainter than the primary star, consistent with the expectations based
on the eclipse depths, where a flux ratio of $\sim 1/176$ is expected.
In the case of the HJST, two peaks are apparent.
However, one of the peaks is due to the sky background since it 
is stationary in velocity, and changes strength relative
to the other.  
The FWHM of the BF peaks were consistent with the instrumental
broadening, which indicates the rotational velocity of the primary is
at best only marginally
resolved.  

Gaussian functions were fit to the BF peaks to determine the relative 
Doppler shifts.
The appropriate barycentric velocity corrections were applied and the
contribution of the template radial velocity was removed, thereby placing
the radial velocities on the standard IAU radial velocity scale
defined by  \cite{Nidever_2002} and
 \cite{Chubak_2012}.  The Keck HIRES pipeline automatically
produces radial measurements for single stars on the IAU scale, accurate to
0.1 km s$^{-1}$ or better \cite{Chubak_2012}.
Having established Kepler-47 as a single-lined binary, we simply adopted 
the pipeline measurement.  The radial velocity measurements for all
11 observations are given in Table S1.

\subsection{Spectroscopic parameters}\label{SPC}

The effective temperature $T_{\rm eff}$, 
surface gravity $\log g$, the metallicity
[m/H], and the rotational velocity $V_{\rm rot}\sin i$
of the primary were measured using the Stellar Parameter
Classification (SPC) code \cite{Buchhave_2012}.  SPC 
uses a cross-correlation analysis against a large grid of 
model spectra in the wavelength region 5050 to 5360\,\AA.  Since all
of the absorption lines in this region are used, the SPC analysis
is ideal for spectra with low signal-to-noise.  The first
three HET observations
were combined to yield a spectrum with a signal-to-noise
ratio of $\approx 116$ in the order containing the Mg b features
near 5169\,\AA\  (the fourth HET observation had relatively high sky
contamination and was not used).   
The derived spectroscopic parameters are given in Table
S2.

\subsection{Stellar eclipse times and corrections}\label{eclipsetimes}

The times of mid-eclipse for the primary and secondary eclipses
in Kepler-47 were measured using the technique described in
\cite{Welsh_2012}.  
For completeness we give most of the details
here as well.  Given an initial linear ephemeris and an initial
estimate of the eclipse widths, the data near the eclipses were
isolated and locally detrended using a cubic polynomial with
the eclipses masked out.  The detrended data were then folded on
the linear ephemeris and an eclipse template was made
by fitting a cubic Hermite spline.  The Piecewise Cubic Hermite Spline
(PCHS)
model template was then
iteratively cross-correlated with each individual eclipse event
to produce a measurement of the time at mid-eclipse.   After each
iteration, a new PCHS model was produced by using the latest
measured times to fold the data.  Fig.\ \ref{plot3154PCHS} shows
the folded eclipse profiles and the final PCHS models.  The fits
are generally good, although there is increased scatter near the middle
of the primary eclipse, presumably due to the effects of star spots.
Table S3 gives the eclipse
times.  The cycle numbers for the secondary eclipse are not exactly half
integers because  the orbit is eccentric.

The eclipse times were fitted with a linear ephemeris and
the Observed minus Computed (O-C) residual times were calculated and
are shown in Fig.\ \ref{plotmultiOC}.  For the primary, there are
coherent deviations of up to two minutes.  While not strictly periodic,
there is a quasiperiod of $\approx 178$ days seen in a periodogram
(Fig.\ \ref{plot3154lomb}).  This modulation is most likely 
a beat frequency between the stellar rotation and the binary motion,
similar to what is observed for Kepler-17 \cite{Desert_2011}.
If the secondary passes in front of a big spot during the primary
eclipse, the spot anomaly will
introduce a shift on the eclipse timing since the projected stellar disk
of the primary on the sky will no longer have a symmetric surface brightness
distribution.
The shift of the eclipse time
will depend on the size and position of
the spot and the position on the eclipse chord. A long-lasting spot will
introduce shifts in consecutive eclipses, but the shift will change with
time since the spot will be at a different position on the eclipse chord at
each eclipse. More specifically for this system, a spot with a period of
rotation of 7.775 days
will effectively recede on the transit chord $360^{\circ}(7.4484~d -
7.775~d)/7.4484~d = -15.79^{\circ}$  each eclipse. In order to come back to 
the exact
same position, and hence complete a full cycle in the O-C diagram,
$(360^{\circ}/15.79^{\circ})P_{\rm orb} = 22.8P_{\rm orb} = 170$ days
will be needed, which is close
to the period of the observed signal.  In reality, the spots also change
with time and may also drift in latitude over time, so the signal near the beat
frequency is blurred somewhat.

There is a correlation between the O-C residual time of the primary eclipse and
the local slope of the out-of-eclipse portions of
the SAP light curve during the eclipse, as shown in
Fig.\ \ref{slopevsoc}.  A large negative slope in the light curve surrounding
an eclipse indicates a dark
spot is rotating into view.  
The ``center of light'' of the primary will
be shifted to the opposite side of the stellar disk, resulting in a slightly
later time of mid-eclipse.  Likewise, a large positive slope surrounding
an eclipse indicates a dark spot is rotating out of view, which results
in a slightly earlier time of mid-eclipse.  Finally, when the slope is near zero,
the spots are centered on the stellar disk, and no change in the eclipse time is
seen.  
A linear function was fitted to the data in Fig.\ \ref{slopevsoc}, 
and the times of primary eclipse were corrected.  The O-C diagram resulting
from these corrected times (Fig.\ \ref{plotmultiOC}) has much less scatter.  No
periodicities  are evident
(Fig.\ \ref{plot3154lomb}).

The best-fitting ephemerides for the corrected primary eclipse times and
the secondary eclipse times are 


\vspace{1em}
\begin{tabular}{rcr@{\,$\pm$\,}ll}
$P_A$ & = & $7.44837605$ & $0.00000050$ d    & Kepler-47 primary \\
$P_B$ & = & $7.44838227$ & $0.00000342$ d   & Kepler-47 secondary \\
$T_0(A)$ &=& $2,454,963.24539$ &    $0.000041$ & Kepler-47 primary \\
$T_0(B)$ &=& $2,454,959.426986$  &  $0.000277$ & Kepler-47 secondary \\
\end{tabular}

\vspace{1em}
\noindent The difference between the primary and secondary periods is 
$0.52\pm 0.30$ seconds, with the secondary period being longer.

\subsection{The effect of star-spots on the eclipses: 
possible biases and spin-orbit alignment.}\label{starspot}

Star-spots cause the light curve to exhibit modulations that can
be used to measure the rotation period of the primary star.
Star-spots can also affect the determination
of certain system parameters. It has been shown that there
is a correlation between the eclipse timing variations and the local slope 
of the stellar flux variations at the times of 
the eclipses. In order to confirm that star-spots are the main cause
of the eclipse timing variations, and to evaluate their impact on our
ability to measure the size of the secondary star, we attempt to
model the effect of spots on individual eclipses \cite{Sanchis_2012,
Czesla_2009,Carter_2011b}.

The data from each primary eclipse are isolated by keeping only 3 
hours of observations before and after the eclipse. The out-of-eclipse 
part of each dataset is then fitted with a linear function. 
The fit is
subtracted from the data, then the data are normalized so the 
out-of-eclipse flux is equal to unity. The detrended eclipse
light curves are folded with a linear ephemeris, 
and this folded light
curve is fitted with a standard model for light loss due to a
dark body passing in front of a limb darkened star \cite{Mandel_2002}. 
This no-spot model has only four free parameters: squared radius ratio
$(R_B/R_A)^2$, impact parameter $b$,
normalized semimajor axis for the secondary orbit 
$R_A/a_B$, and a linear limb darkening coefficient $u_1$.

The effect spots have on individual primary eclipses is manifest
in two ways: the depth of each eclipse changes since it is 
measured relative to the changing stellar flux,
and the shape of each eclipse is distorted which leads to a shift
in the measured mid-eclipse time.
Visual inspection of the eclipse residuals shows
that this last effect can be well-modeled in most cases by adding 
just one large star-spot on the surface of the primary star. Since 
the rotation of the star happens on a longer time-scale than the eclipse 
itself, we held the position of the star-spot fixed during each 
individual eclipse. 
The latitude of the spots cannot be well constrained with single eclipse 
events, so we fix the position of the spot so that the center of the 
secondary star trajectory intersects the center of the spot. Our spot
model adds five additional parameters: three parameters describe the
spot itself -- its angular radius, the flux contrast (related to
the spot temperature), and the position along the eclipse chord. The
fourth parameter is the out-of-eclipse flux, which corrects for the
depth variations. Finally, the time of mid-eclipse is free.
We set up a pixilated model of the star with a circular spot, in which 
the flux is calculated as the surface integral of the intensity of the 
visible hemisphere of the star. 

The best-fitting model for several consecutive eclipses is compared
with a no-spot model in Fig.\ \ref{roberto1}, showing how the
model captures the essential effect of spots on the eclipses. For every
eclipse we obtained a new eclipse time, and fitted a linear
ephemeris to these times. The scatter was found to be substantially 
reduced from the initial timings (see Fig.\ \ref{plotmultiOC}, 
upper panel), by a factor of 30\%,
which shows that indeed the scatter is due to spots.  The improvement
on the scatter is similar to the one obtained through the local
slope correction, so this serves as a good consistency check.

Our model also estimates the fraction of the star covered by spots 
at the time of each eclipse. This quantity is not very precise, but can 
help us estimate the effect of spots on the measurement of the 
eclipse depth and hence the radius ratio $R_A/R_B$.
We divide the square of the radius ratio from the spot model 
by each observed local out-of-eclipse flux to mimic the apparent
depth that one would obtain by  
fitting each eclipse individually. The results are plotted in 
Fig.\ \ref{roberto2},
where one can clearly see how the depth of each eclipse changes
with time. The variations seem to have a time-scale similar to the 
uncorrected eclipse timing
variations, which is expected since the scatter in both are due to spots. 
A variation with the observing season is also apparent, which is a clear
indication that there are different levels of Quarterly contamination.
With these eclipse depths we can estimate the inferred secondary star 
radius $R_B$ from each eclipse, using a fixed $R_A$ from Table 1
(see Fig.\ \ref{roberto2}).
The values obtained do not have a large scatter, and they all agree within 
$1\sigma$ with the value obtained from the photometric-dynamical model 
fit. Thus the correction to the secondary star radius because of the 
presence of spots is not significant, although a slightly smaller radius 
is favored. 

We can also use these spot models to gain information 
about the obliquity of the system
\cite{Sanchis_2011,Nutzman_2011,Desert_2011,Sanchis_2012,Hirano_2012}.
In Fig.\ \ref{roberto1}
we can clearly see how the spot model shows that a spot is moving
backwards with each eclipse, which means that the spot trajectory is
contained within the boundaries of the eclipse chord. This backwards
movement makes it seem as if the star is rotating backwards 
(retrograde) very slowly, but this is simply a stroboscopic alias effect.
The spot appears to move backwards because the star's rotation period is 
slightly longer than the orbital period. 

If we assume that the entire trajectory of a spot is contained 
on the part of the primary star eclipsed by the secondary star 
then we can estimate the obliquity of the system 
\cite{Sanchis_2011} to be
smaller than $\arctan (R_B/R_A)\approx 20^{\circ}$. 
The obliquity is likely to be smaller
since we have detected more than 10 spots receding with different
velocities, and these different velocities could be due to spots at
different latitudes exhibiting differential rotation. 
We note that the obliquity of this target will be very hard to measure 
with the Rossiter-McLaughlin effect \cite{Winn_2005} due to its faintness, 
so additional investigation of its spots might be the preferred method
to further constrain the obliquity.

In principle, we can use the spectroscopic 
$V_{\rm rot}\sin i$ together with an 
estimated rotation period and size of the primary star to obtain 
information on the inclination of the primary star. 
The spectroscopic observed
$V_{\rm rot}\sin i=4.1\pm 0.5$ km s$^{-1}$,
while the inferred 
$V_{\rm rot} =2\pi R_A/P_{\rm rot}=6.3\pm 0.2$ km s$^{-1}$.
This would imply a highly inclined star 
($i_s\approx 40^{\circ}$).  Note, however, that
the measured value of the rotational velocity
is below the resolution of the spectra, so its value
should be treated with caution.
In addition, differential rotation can make it 
harder to compare the surface integrated 
projected rotational velocity 
$V_{\rm rot}\sin i$
to the equatorial
rotational velocity
$2\pi R_A/P_{\rm rot}$ \cite{Hirano_2012}.

\subsection{Transit times for the inner and outer 
planet and the search for additional
transits}\label{transittimes}

All of the transits of the outer planet and about half of the
transits of the inner planet are evident in the SAP light
curves before any detrending.   The rest became 
visible when the data are carefully detrended.  
A symmetric polynomial
``U-function'' template with an adjustable width and
depth was used to estimate
the times of mid-transit and their durations.  
A cubic polynomial was used to detrend each transit using five different
duration windows around the transit, and the best-fitting
one was adopted.
The fits for each transit were iterated to determine
the best-fitting time of mid-transit and the duration.  This method worked
well in some cases and failed to converge in other cases.  In cases
where the convergence failed, the time was estimated using an interactive plotting
program, and an uncertainty of 30, 60 or 90 minutes was assigned
based on the judged quality of the transit.  Table S4 gives the
measured times and durations 
and their uncertainties, and the corresponding model times and durations.
We note that the measured times
and the durations were only used to establish starting models for the
photometric-dynamical models described below.  The actual (detrended) light
curve was modeled directly.

One ``orphan'' transit occurring about 12 hours after a transit of planet b
was noticed in the Q12 data (Fig.\ \ref{plotorphan1}).
This transit cannot be accounted for by
either the inner or the outer planet (the intervals between the nearest
transits are 0.5 days and 127 days, respectively).  To estimate the
significance of the orphan event, a model consisting of two Gaussians
was fit to the segment of the detrended light curve shown in
Fig.\ \ref{plotorphan1}, which contains 103 data points.  The uncertainties
on each point were scaled to give $\chi^2_{\nu}=1$ for 96 degrees of
freedom.  The Gaussian in the model
at the location of the orphan was replaced by
the background level of 1.0 and the resulting $\chi^2$ value increased to
205.5, giving a formal significance of $\sim 10.5 \sigma$.

No other
orphan transits with a significance of $>3\sigma$ were found using visual
searches.  An automated search algorithm, dubbed the ``Quasi-periodic 
Automated Transit 
Search'' 
(QATS) was also used to search for additional transits.  The QATS algorithm
can allow for unequal time intervals between the transit events.  For a given
trial period for a potential planet, the expected transit duration at each time
in the light curve is computed using
a circular orbit for the planet.  The data are
corrected for the different transit durations and shifted to a common phase to
increase the signal-to-noise (the correction for the different widths is
quite good, provided the planet's orbit is nearly circular).
A ``periodogram'' is constructed by plotting
the significance versus the trial period.
QATS detected the inner planet at high significance.
Unfortunately QATS is very sensitive to detrending errors for longer periods, and in
fact did not detect the outer planet.  No additional planets with periods shorter
than 150 days were detected at the significance level of the inner planet.

Although the overall duty cycle of the data collection
by Kepler is quite high, a non-trivial amount of the light curve is
occupied by the stellar eclipses, which in the case
of Kepler-47 is $\approx 3-6$ times more than it is for Kepler-16, 34, 35, and 38.
Although one could in principle find transits during primary and secondary eclipses,
in Kepler-47 this is extremely difficult owing to the effects of star-spots.
The primary and secondary eclipse durations together are 0.014 in
orbital phase, which is 0.104 days.  A combined
total of 256 primary and secondary
eclipses were observed, giving a total of 26.62 days
lost for the purposes of transit searches, lowering the duty cycle to
83.8\%.  

Finally,
if
the transit is due to another planet in the Kepler-47 system, its radius would be
$\approx 4.5$ Earth radii.   Without more transit events, it is nearly
impossible to determine what the orbital period of such a planet would be.  If its
orbit is more inclined relative to the other planets, it would not necessarily
transit the stars near each conjunction.  In addition, if 
there is precession of the orbit, it is possible
for sequences of transits to come and go over long time scales.   Thus, the 
orphan transit could in principle belong to a planet in
between the inner and outer one, in spite of the lack of other observed
transits.

\subsection{Photometric-dynamical model}\label{sec:photo}

We modeled the Kepler light curve of Kepler-47 using a dynamical
model to predict the motions of the planets and stars, and a
eclipse/transit model to predict the light curve.

\subsubsection{Description of the model}

The ``photometric-dynamical model'' refers to the model that was used
to fit the Kepler photometry.  This model is analogous to that
described in the analyses of KOI-126 \cite{Carter_2011}, Kepler-16
\cite{Doyle_2011}, Kepler-34 and Kepler-35 \cite{Welsh_2012},
Kepler-36 \cite{Carter_2012}, and Kepler-38 \cite{Orosz_2012}.

Four bodies were involved in this problem; however, the planets'
gravitational interaction with the stars and with each other was
determined to be observationally negligible.  We therefore assumed the
planets to be massless in our model.  The motion of the stellar binary
was Keplerian and could be predicted analytically.  The planets were
modeled as orbiting in the two-body potential of the stars.  The
motion of each planet was determined via a three-body numerical
integration. This integration utilized a hierarchical (or Jacobian)
coordinate system. In this system, ${\bf r_{b}}$ (${\bf r_{c}}$) is
the position of Planet b (Planet c) relative to the center-of-mass of
the stellar binary (which corresponds to the barycenter in this
approximation), and ${ \bf r_{\rm EB} }$ is the position of Star B
relative to Star A.  The computations are performed in a Cartesian
system, although it is convenient to express ${\bf r_b}$ (${\bf
  r_{c}}$) and ${\bf r_{\rm EB}}$ and their time derivatives in terms
of osculating Keplerian orbital elements: instantaneous period,
eccentricity, argument of pericenter, inclination, longitude of the
ascending node, and time of barycentric conjunction: $P_{b, c, {\rm
    EB}}$, $e_{b, c, {\rm EB}}$, $i_{b,c, {\rm EB}}$, $\omega_{b,c,
  {\rm EB}}$, $\Omega_{b,c, {\rm EB}}$, $T_{b,c, {\rm EB}}$,
respectively.  We note that these parameters do not necessarily
reflect observables in the light curve; the unique three-body effects
make these parameters functions of time (and we refer to these
coordinates as ``osculating'').

The accelerations of the three bodies are determined from Newton's
equations of motion, which depend on ${\bf r_b}$ (${\bf r_c}$), ${\bf
  r_{\rm EB}}$ and the masses
\cite{Soderhjelm1411984,Mardling5732002}.  For the purpose of
reporting the masses and radii in Solar units, we assumed $G M_{\rm
  Sun} = 2.959122 \times 10^{-4}$ AU$^{3}$ day$^{-2}$ and $R_{\rm Sun}
= 0.00465116$ AU.  We used a Bulirsch-Stoer algorithm \cite{Press2002}
to integrate the coupled first-order differential equations for
$\dot{\bf r}_{b, {\rm EB}}$ and ${\bf r}_{b, {\rm EB}}$.

The spatial coordinates of all four bodies at each observed time are
calculated and used as inputs to model the light curve.  The computed
flux was the sum of the fluxes assigned to Star A, Star B, and a
seasonal (being the four ``seasons'' of the 
Kepler spacecraft
orientation) source of ``third light,'' minus any missing flux due to
eclipses or transits (only planetary transits across Star A were
computed, those across Star B are not significant in the 
Kepler
data). The loss of light due to eclipses was calculated as
follows. All objects were assumed to be spherical. The sum of the
fluxes of Star A and Star B was normalized to unity and the flux of
Star B was specified relative to that of Star A. The radial brightness
profiles of Star A and Star B were modeled with a linear
limb-darkening law, i.e., $I(r)/I(0) = 1-u_1 (1-\sqrt{1-r^2})$ where
$r$ is the projected distance from the center of a given star,
normalized to its radius, and $u$ is the linear limb-darkening
parameter.  The limb darkening coefficient of Star B was fixed (to $u
= 0.5$); letting it vary freely resulted in a negligible change to
final parameter posterior.

The radial velocity of Star A was computed from the time derivative of
the position of Star A along the line of sight (analytically, in this
case) and compared to the radial velocity data.

The continuous model is integrated over a 29.4 minutes interval centered 
on each long cadence sample before being compared to the long cadence 
Kepler data.

\subsubsection{Local detrending of Kepler data} \label{sec:detrend}

The Kepler light curve (``SAP\_FLUX'' from the standard fits
product) for Kepler-47, spanning twelve Quarters, is reduced to only
those data within 0.5--1 day of any primary or secondary eclipse or
any transit of either planet.  As
noted above, some data are missing as a result of
observation breaks during Quarterly data transfers or spacecraft safe
modes.

Each continuous segment of data has a local cubic correction in time
divided into it.  The parameters of this polynomial correction are
found through an iterative process, as described as follows.  In the
first step, we masked the eclipses of the stars and the transits of
the planets and then performed a robust nonlinear least-squares fit
to each continuous segment.  The data, having divided out this
correction, were then ``fit'' with the photo-dynamical model by
determining the highest likelihood solution from a Markov Chain Monte
Carlo simulation.  The best-fit model was then divided into the data
and the local nonlinear fits were recomputed (this time without
masking the eclipses and transits).  This process was repeated until
the corrections converged to a sufficient tolerance.

\subsubsection{Specification of parameters} \label{sec:params}

A reference epoch for the three-body integration was specified for
each planet near a particular transit.  Those epochs were chosen to be
2,454,969.216 BJD and 2,455,246.6545 BJD for planets b and c,
respectively.

The model has 33 adjustable parameters. Two parameters are related to
Star A: the stellar density times the gravitational constant, $G
\rho_A$, and the stellar mass times $G$, $G M_A$. One parameter gives
the mass ratio of the stars, $q \equiv M_B/M_A$. Six parameters encode
the eccentricities and arguments of pericenter for the planetary and
stellar orbits about the barycenter in a way that reduces nonlinear
correlations:
\begin{eqnarray}
        h_{b,c} &\equiv & \sqrt{e_{b,c}}\sin \omega_{b,c} \\
        k_{b,c} & \equiv & \sqrt{e_{b,c}}\cos \omega_{b,c}\\
        H & \equiv & e_{EB}\sin \omega _{EB} \\
        K & \equiv & e_{EB} \cos \omega_{EB} 
\end{eqnarray}

The remaining osculating parameters, 11 in total, are the periods
$P_{b,c}$, $P_{EB}$, the orbital inclinations $i_{b,c}$, $i_{EB}$, the
times of conjunction with barycenter, $T_{b,c}$, $T_{EB}$ and the
difference between the nodal longitudes of the planets to that of the
stellar binary $\Delta\Omega_{b,c}$.  The absolute nodal angle
relative to North of the stellar binary cannot be determined and was
fixed to zero in practice.

Three more parameters are the relative radii of Star B and the planets
to that of Star A: $r_B \equiv R_B/R_A$ and $r_{b,c} \equiv
R_{b,c}/R_A$.  One parameter, $u$, parameterizes the linear limb
darkening law for Star A (described above).  Another parameter gives
the relative flux contribution of Star B, $F_B/F_A$.

A single parameter, $\sigma_{\rm LC}$, describes the width of the
probability distribution for the photometric noise of the long cadence
observations, assumed to be stationary, white and
Gaussian-distributed.

Three parameters characterize the radial velocity measurements: the
constant offset of the radial velocity, $\gamma$, the offset between
the HET and HJST velocities, $\Delta \gamma$, and a ``stellar jitter''
term, $\sigma_{RV}$, which contributes to the measured errors for each
radial velocity observation, in quadrature.
Because only one Keck observation was made, this radial velocity 
could not be offset to match the HET and HJST velocities in a sensible
way, and therefore was omitted in the modeling.

Additionally, we specify 4 more parameters describing the relative
extra flux summed in the aperture.  The four parameters specify the
constant extra flux in each 
Kepler ``season.'' The 
Kepler spacecraft
is in one of four orientations during a year; a constant level of
``third light'' is assumed for all Quarters sharing a common season.

\subsubsection{Priors and likelihood} \label{sec:like}

We assumed uniform priors for all 33 parameters.  For the eccentricity
parameters, this corresponds to uniform priors in $e_{b,c}$ and
$\omega_{b,c}$, but a prior that scales as $e_{EB}$ for the stellar
eccentricity.  This eccentricity is sufficiently determined that this
non-uniform prior  does not dominate the posterior distribution.

The likelihood ${\cal L}$ of a given set of parameters was 
taken to be the product of likelihoods based on the 
photometric data and radial velocity data:
\begin{eqnarray}
        {\cal L} &\propto&  \sigma_{\rm LC}^{-N_{\rm LC}} \exp 
      \left[-\sum_i^{N_{\rm LC}} 
\frac{(\Delta F^{LC}_i)^2}{2 \sigma_{\rm LC}^2} \right]  \\ \nonumber
                        && \times \prod_{j}^{N_{\rm RV}} \left(\sigma_{j}^2+
\sigma_{RV}^2\right)^{-1/2} \exp \left[ -\frac{(\Delta V_j)^2}
{2\left(\sigma_{j}^2+\sigma_{RV}^2\right)} \right]
\end{eqnarray}
where $\Delta F^{\rm LC}_i$ is the $i$th photometric data residual,
$\sigma_{\rm LC}$ is the width parameter describing the photometric
noise of the long cadence data, $\Delta V_j$ is the $j$th radial
velocity residual, $\sigma_j$ is the uncertainty in the $j$th radial
velocity measurement and $\sigma_{RV}$ is the stellar jitter term
added in quadrature with the $\sigma_j$.

\subsubsection{Best-fit model} \label{sec:best}

We determined the best-fit model by maximizing the likelihood.  The
maximum likelihood solution was found by finding the highest
likelihood in a large draw from the posterior as simulated with a
Markov Chain Monte Carlo simulation as described below.
%
Fig.\ \ref{plot3154planet1} shows 18 transits of the inner planet, and Fig.\
\ref{plot3154planet1res} shows the residuals (observed data minus the model).
With a few exceptions, there are no strong patterns in the residuals.
Fig.\ \ref{plot3154planet2res} shows the model fits and the residuals
for the outer planet.  There are no patterns evident in the residuals.
The residuals for the fits to the primary eclipses are shown in Fig.\
\ref{fitsFlux_100_Primaries_resid}
and the residuals for the fits to the secondary
eclipses are shown in Fig.\ \ref{fitsFlux_100_Secondaries_resid}.   Spot crossing
events are evident in many of the primary eclipses.
Fig.\ \ref{show3154RVres} shows the radial 
velocity measurements and the best-fitting
model and the residuals of the fit.  Generally the absolute value of the
radial velocity
residuals is less than about 200 m s$^{-1}$.

The photometric noise parameter, $\sigma_{LC}$, has a best-fit value
of $\sigma_{LC} = 629.5$ ppm.  For comparison, the root-mean-square
deviation of the best-fit residuals is $626.9$ ppm.  
This is similar to the expected noise in the light curve
of 635 ppm as estimated using an on-line tool provided
by the Kepler Guest Observer 
Office\footnote{http://keplergo.arc.nasa.gov/CalibrationSN.shtml},
where we used an apparent magnitude of
Kp=15.18 and 20 pixels in the aperture.
For this
$\sigma_{LC}$, the $\chi^2$-metric for the photometric data is $\chi^2
= 10576$ with 10629 degrees of freedom.  If we fail to include planet
b in our model (by setting its radius to zero), 
the $\chi^2$ increases by $\Delta \chi^2 =
343.4$.  If we ignore planet c, 
the $\chi^2$ increases by $\Delta \chi^2 = 248.2$.

The stellar jitter parameter, $\sigma_{RV}$, has a best-fit value of
$\sigma_{RV} = 0.31$ km s$^{-1}$.  The value of $\chi^2$ for the
radial velocity data alone is $\chi^2 = 8.85$ for the 10 radial
velocity observations.

Fig.\ \ref{schem} shows schematic diagrams of the Kepler-47 orbits.  
The projected orbits of planets b and c cross the projected disk of
the primary, and so transits of both planets across the primary occur,
as do occultations of both planets by star A.  The former events
are observed, whereas the latter events are not observable given
the noise level.  On the other hand, owing to its small radius,
the projected disk of star B does not intersect the projected
orbits of the planets, and as such no transits of star B or occultations
due to star B occur for the best-fitting orbital configuration. 
Due to the uncertainties in the relative nodal angles,
transits of the planets across star B might occur for a subset of
the acceptable solutions.  However,
even if transits across star B did occur,
the expected
transit depth  would be $\sim 30$ times weaker than the transits
across the primary, and would not be observable in the light curve
given the noise level.

\subsubsection{Parameter estimation methodology} \label{sec:mcmc}

We explored the parameter space and estimated the posterior parameter
distribution with a Differential Evolution Markov Chain Monte Carlo
(DE-MCMC) algorithm \cite{terBraak}.

We generated a population of 100 chains and evolved them through
approximately 200,000 generations.  The initial parameter states of
the 100 chains were randomly selected from an over-dispersed region in
parameter space bounding the final posterior distribution.  The first
10\% of the links in each individual Markov chain were clipped, and
the resulting chains were concatenated to form a single Markov chain,
after having confirmed that each chain had converged according to the
standard criteria.

The parameter values and derived values reported in Tables
S5 and S6
 beside the best-fit values (see 
above), were found by computing the 15.8\%, 50\%, 84.2\%
levels of the cumulative distribution of the marginalized posterior
for each parameter.  Figure \ref{corr100} shows two-parameter joint
distributions between all parameters.  This figure is meant to
highlight the qualitative features of the posterior as opposed to
providing quantitative ranges.  The numbers in that figure correspond
to the model parameters in Table \ref{tab:tab1} with the same number
listed as in the first column, if available.

Figures \ref{fig:ecc} and \ref{fig:omega} show the posterior
distribution in the eccentricity and argument of pericenter planes The
distribution of the three-dimensional inclination between the planets'
orbits and the invariable plane is shown in Figure
\ref{fig:inclination_angle}.

\subsubsection{Predicted ephemerides and transit parameters}

Tables S7 and 
S8
provide the predicted times of transit, impact parameters, 
normalized transit velocities and durations over 7 years, 
starting with 
Kepler Quarter Q13.

\subsection{ELC light curve models}\label{ELC}

Although the secondary star is not detected spectroscopically,
its temperature can be estimated using the temperature of
the primary derived using SPC, and the temperature ratio
derived from modeling
the eclipses.  In order to find the temperature ratio,
and to have an independent check on the results from the 
photometric-dynamical model,
we modeled
the light and velocity curves using the 
Eclipsing Light Curve
(ELC) code \cite{Orosz_2000} with 
its genetic algorithm and Monte Carlo Markov Chain optimizers.
The free parameters include
the temperature ratio $T_B/T_A$, the primary's limb darkening
parameters $x_A$ and $y_A$ for the  quadratic
limb darkening law [$I(\mu)=I_0(1-x(1-\mu)-y(1-\mu)^2)$],
the orbital parameters
($e$, $\omega$, $i$), and
the fractional radii $R_A/a$ and $R_B/a$.  The stellar masses
and the orbital period
were held fixed at the values derived from the photometric-dynamical
model discussed above.

In ELC, the shapes of the stars are computed using
a ``Roche'' potential modified to account for nonsynchronous rotation
and eccentric orbits \cite{Avni_1976,Wilson_1979}.  Given the mass ratio
and the fractional radii, the volumes of each star are found by
numerical integration.  The effective radius of each star is taken
to be the radius of a sphere with the same volume as the equipotential
surface.  In the case of Kepler-47, the stars are very nearly
spherical.  For the primary at periastron, 
the ratio of the polar radius 
to its  effective radius is 0.99988, and the ratio of
the radius along the line of centers to the effective radius is
1.00007.  The amplitude of the out-of-eclipse modulation in the
light curve due to ellipsoidal variations, reflection, and Doppler
boosting is on the order of 400 ppm, which is $\approx 75$ times smaller
than the modulation due to star-spots.
Thus the use of spherical stars in the photometric-dynamical
model is a very good approximation.

Since the numerical integrations are very CPU intensive, 
ELC has a fast ``analytic'' mode where the equations given
in \cite{Gimenez_2006} are used.
The normalized light curve was divided into 41 segments containing
two or three pairs of primary and secondary eclipses.  These segments
were modeled separately in order to help assess the systematic errors 
associated with the changing star-spots and the changes in the contamination
from Quarter to Quarter.  For each fitting parameter, we computed the mean
of the best-fitting values and the standard deviation.
Table \ref{tab:ELC} gives the mean values and standard
deviations, which we adopt as $1\sigma$ errors.  

Based on the temperature of the primary derived from the SPC 
analysis, and the temperature ratio found from the ELC models,
we derive a temperature of $3357\pm 100$ K for the secondary.


\subsection{Upper limits on planetary masses}\label{upperlimit}

Upper limits on the masses of the planets can be placed separately as
follows.  The mass of the inner planet is best constrained
by the lack of eclipse timing variations due to gravitational
perturbations from that planet.  The planet will induce short-term
eclipse timing variations with a period equal to the planet's period.
It will also
cause precession of the binary.  
Over the time-scale of a few
years,
the binary precession
will cause a slight change in the phase
difference between the primary and secondary
eclipses, which can be observed as a slight difference between
the primary and secondary eclipse periods.  
Numerical simulations showed that in this case the stronger
upper limit comes from the lack of short-term eclipse timing
variations.  A grid of masses for the inner planet was used,
and equations of motion for a three body system were integrated,
holding the orbital parameters of the binary at their best-fitting values
(the nature of the perturbations on the binary are insensitive
to anything except the planet's mass). 
The period and epoch of the binary was found, and the predicted
times of eclipse were compared to the measured times.
The $\chi^2$ value changes smoothly with the planet's mass, and
going to $\chi^2=\chi^2_{\rm min}+9$ gives a $3\sigma$ upper
limit of 2.7 Jupiter masses for the uncorrected eclipse
times and 2.0 Jupiter masses for the eclipse times corrected for
the effect of star-spots.  We adopt the latter value as the $3\sigma$
upper limit on the mass of the inner planet.

The mass of the outer planet was best constrained by light travel
time (LTT) effects.   We fit an LTT orbit to the corrected 
eclipse times,
using a period of 303.13 days and constraining the eccentricity
to be $e<0.2$.  While
no convincing signal is seen at that period, the best-fitting
orbit formally has a semiamplitude of $3.84\pm 1.84$ seconds.
Given the total mass of the binary and the
period of the outer planet, we find a $3\sigma$ upper mass limit of
28 Jupiter masses.

\subsection{Stability of orbits and limits on eccentricity}\label{stability}

We carried out an extensive study of the dynamics of the system and
its long-term stability. The orbits of the two planets
were integrated, numerically, for different values of their masses and
orbital eccentricities. To determine an upper limit for
the eccentricity of planet c, we held constant all orbital elements at
their best-fit values and integrated the system varying the
eccentricity of this planet. Results indicated that the system
maintained stability for at least 100 Myr and for $e_c<0.6$.
An examination of the semimajor axis, eccentricity, and orbital
inclination of each planet during the course of the integrations
showed that the variations of these quantities were negligibly small,
supporting the idea that the two planets do not disturb each
other's orbits. The results stayed unchanged when the masses of the two
planets were increased to 0.21 and 0.63 Jupiter masses, roughly ten
times their plausible values based on the empirical mass-radius
relations \cite{Kane_2012,Lissauer_2011}.

Both the photometric-dynamical model and stability simulations used a
Newtonian 4-body numerical integrator. A more physical model would
include the precession of the binary due to general relativity (GR)
and the tidal and rotational bulges. Expressions for the rate of
precession due to these effects
\cite{Fabrycky_2012}
show that GR dominates, and it would cause a
full periastron rotation in $\sim6700$ years. In the current
observations, such precession would cause the period of primary and
secondary eclipse to differ fractionally by $\sim 10^{-7}$, whereas
the uncertainty of this quantity is $4.6 \times10^{-7}$. The GR
precession period is much longer than the periastron period of the
planets -- e.g., numerical integrations of planet c showing a
$\sim 560$ year precession cycle due to the effective quadrupolar
gravitational potential of the binary -- so it has little dynamical
importance.  Therefore GR and other precession effects are neither
detectable nor significantly change our assessment of stability,
so GR has little dynamical importance.

%

\subsection{Comparison with stellar evolution models}\label{evol}

The reasonably precise absolute dimensions determined for the stars in
Kepler-47 (4--5\% relative errors for the masses, and 1.8\% for the
radii) offer an opportunity to compare the measurements with models of
stellar evolution. This is of particular interest for the late M-dwarf
secondary, given that low-mass stars have shown discrepancies with
theory in the sense that they are generally larger and cooler than
predicted.  These anomalies are believed to be due to stellar activity
\cite{Lopez-Morales:07,Torres:10}.

In Fig.\ \ref{plottrackg} 
we compare the measurements for the primary star with a
stellar evolution track from the Yonsei-Yale series 
\cite{Yi:01,Demarque:04},
interpolated to the exact mass we measure. The
metallicity of this model is set by our spectroscopic determination of
${\rm [Fe/H]} = -0.25$, where we assume the iron abundance tracks
the metallicity measurement from SPC.
The model is consistent with the observations
to within less than $2\sigma$, and the small difference may be due
either to slightly biased spectroscopic parameters (temperature and
metallicity) or a slightly overestimated mass for the primary star.
As a check, we produced a photometric estimate of the temperature
using available photometry from the KIC
and empirical color-temperature calibrations along
with the reddening listed in the KIC. The result suggests a value
closer to 5900 K than 5600 K, although we consider this evidence to be
somewhat circumstantial and highly dependent on reddening.  We
confirmed that the level of agreement between theory and observation
is independent of the adopted model physics by comparing the primary
star parameters with BaSTI stellar evolutionary tracks 
\cite{Pietrinferni_2004},
which yielded similar results as the Yonsei-Yale
models.

In Fig.\ 
\ref{plotmrt2a} 
we compare the measurements for both components against
models from the Dartmouth series \cite{Dotter:08},
which
incorporate physical ingredients (equation of state, non-grey boundary
conditions) more appropriate for low-mass stars. We find the radius of
the secondary of Kepler-47 to be consistent with these models, which
would be an exception to the general trend mentioned above, although
the mass error is large enough ($\sim$4\%) that the conclusion is not
as strong as in other cases. Its temperature, however, is lower than
predicted for a star of this mass by about 200 K. This deviation is in
the same direction as seen for other low-mass stars. Because the
secondary is so faint, we have no information on its activity level.
Age estimates for the system from this figure and the previous one are
somewhat conflicting, and only allow us to say that Kepler-47 is very
roughly of solar age.  

\subsection{Details of habitable zone}\label{habit}

To determine the insolation limits of the habitable zone for Kepler-47 c, we
follow the relations given by \cite{Underwood_2003} 
that include the stellar
temperature as well as luminosity. The temperature term accounts for the
different relative amount of infrared flux to total flux, which is important
for atmospheric heating. We use the criteria of a runaway greenhouse 
effect for the
inner boundary and the maximum greenhouse effect for a cloud-free carbon
dioxide atmosphere for the outer boundary. This is more conservative than
the ``recent Venus'' and ``early Mars'' criteria, but less conservative than
the ``water loss'' and ``first carbon dioxide condensation'' criteria
\cite{Underwood_2003}. 
The secondary star emits only 1.7\% 
as much
energy as the primary star (and only 0.58\% in the Kepler bandpass),
so its contribution is neglected. The resulting insolation limits are shown
as the dotted lines in Figure 3 (main text). The relations 
given in 
\cite{Selsis_2007}, i.e.\  
a cloud-free atmosphere yield nearly identical limits.

The average insolation for Kepler-47 c for a circular orbit is 87\% of the
Sun-Earth insolation, and varies by $\sim 9\%$ peak-to-peak. For an
eccentricity of 0.2 the mean insolation is 89\% and varies from 59\% to
144\% of the Sun-Earth value; for an eccentricity of 0.4, the mean is 96\%
and varies from 43\% to 261\%. Even in this latter case, which is ruled
out at the 95\% confidence level by the photometric-dynamical model, the
mean is less than the Sun-Earth value, and it is the mean insolation that is
most relevant for habitability \cite{Williams_2002}.
Thus for
all allowed eccentricities, Kepler-47 c lies in the habitable zone.

Because the primary star dominates the system both in luminosity and mass
(so the primary star remains near the barycenter), the variation in
insolation is relatively small for a circular planetary orbit. This is seen
in the upper left panel of Figure 3, where the variations are caused by the
7.4-day orbit of the primary star. For large eccentricities, the variation
in insolation are dominated by the non-circular orbit of the planet.

It must be stressed that the habitable zone is defined such that liquid
water could persist for a biologically significant time period on an
Earth-like planet (i.e.\ with a terrestrial CO$_{2}$/H$_{2}$O/N$_{2}$
atmosphere, plate tectonics, etc.), and the formulations of 
\cite{Kasting_1993,Underwood_2003,Selsis_2007}
explicitly assume such conditions. For
Kepler-47 c these conditions are not met. Nevertheless, the main point is
that Kepler-47 c receives approximately the same amount of energy from its
stars that the Earth receives from the Sun.

While it neglects most atmospheric physics, the equilibrium temperature
$T_{eq}$ of the planet is still a useful characterization.
Assuming that the entire surface of the planet radiates isothermally
(i.e.\ the stellar insolation is efficiently advected around the planet),
and for a Bond albedo of $A_{B}$=0.7, appropriate for a Neptune-size
planet and 1 Sun-Earth insolation 
\cite{Cahoy_2010},
a value of $T_{eq} \sim 200 \ K$ is found for eccentricities
from 0.0 to 0.3.
For $A_{B}$=0.34, corresponding to the albedos of Jupiter and
Saturn, $T_{eq} \sim 243 ~ K$.
For an Earth-like albedo of 0.29,
which is appropriate for a habitable-zone planet,
$T_{eq} \sim 247 ~ K$.
The greenhouse effect will lead to temperatures at the 1-bar pressure 
level that are higher by several tens of degrees.

\newpage

\renewcommand{\thefigure}{S\arabic{figure}}
 \begin{figure} 
\hspace{-.46in}\includegraphics[angle = -90,clip,width=1.15\textwidth]{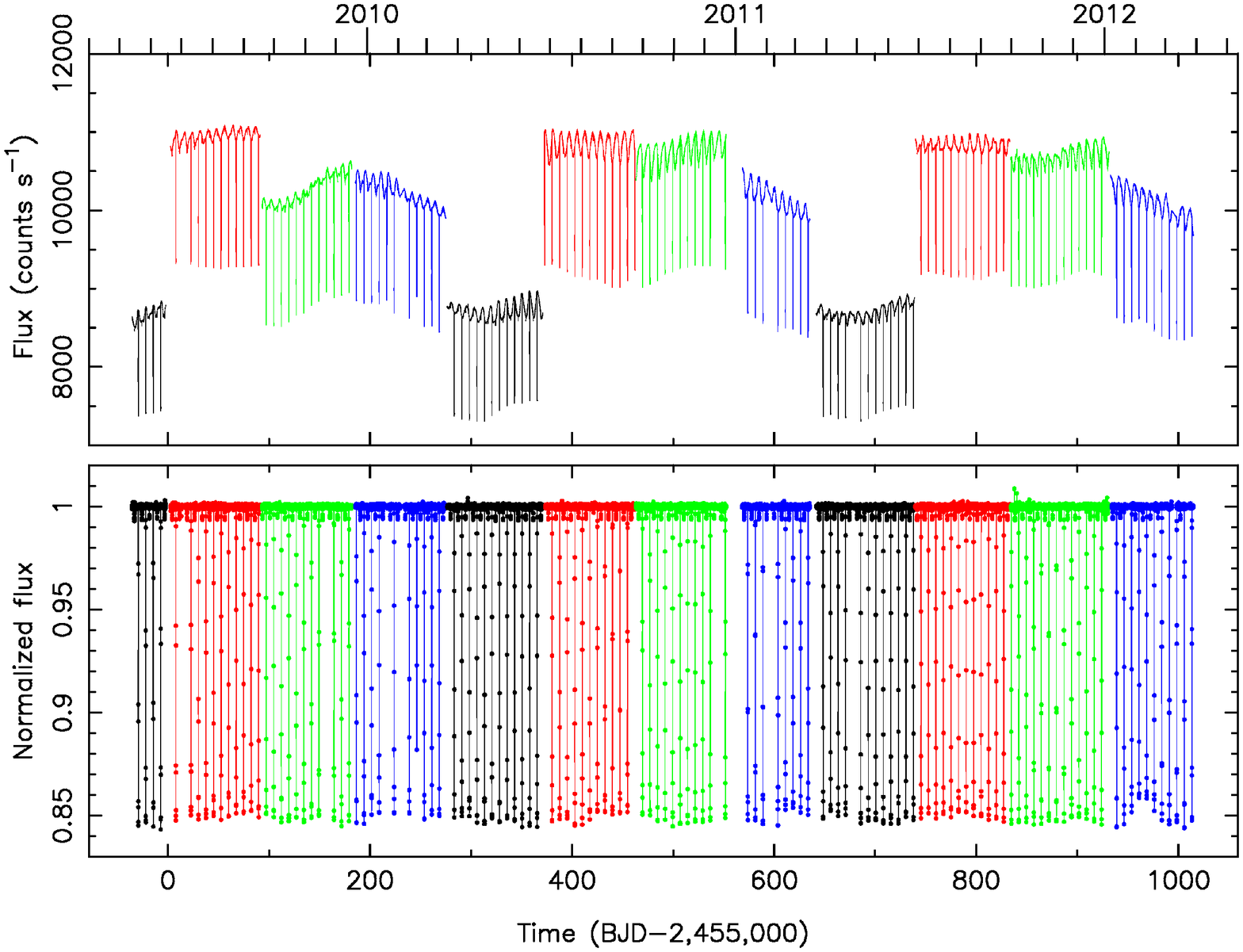}
\caption{{\bf SAP and detrended light curves.}
Top:
The SAP light curves of Kepler-47 are shown.
The colors denote the season and hence the spacecraft
orientation where 
black is for Q1, Q5, and Q9,
red is for Q2, Q6, and Q10, green is for Q3, Q7,
and Q11, and blue is for Q4, Q8, and Q12.  
Bottom:
The
normalized and detrended light curve with the spot modulation
removed is shown.  Fifteen primary eclipses 
and thirteen secondary eclipses were missed
during monthly data downloads, spacecraft rolls between Quarters, 
spacecraft safe modes, and interruptions caused by solar flares.
\label{plotrawmulti}}
\end{figure} 

\renewcommand{\thefigure}{S\arabic{figure}}
 \begin{figure} 
    \begin{center}
      \includegraphics[angle = -90,clip=yes,width=0.98\textwidth]{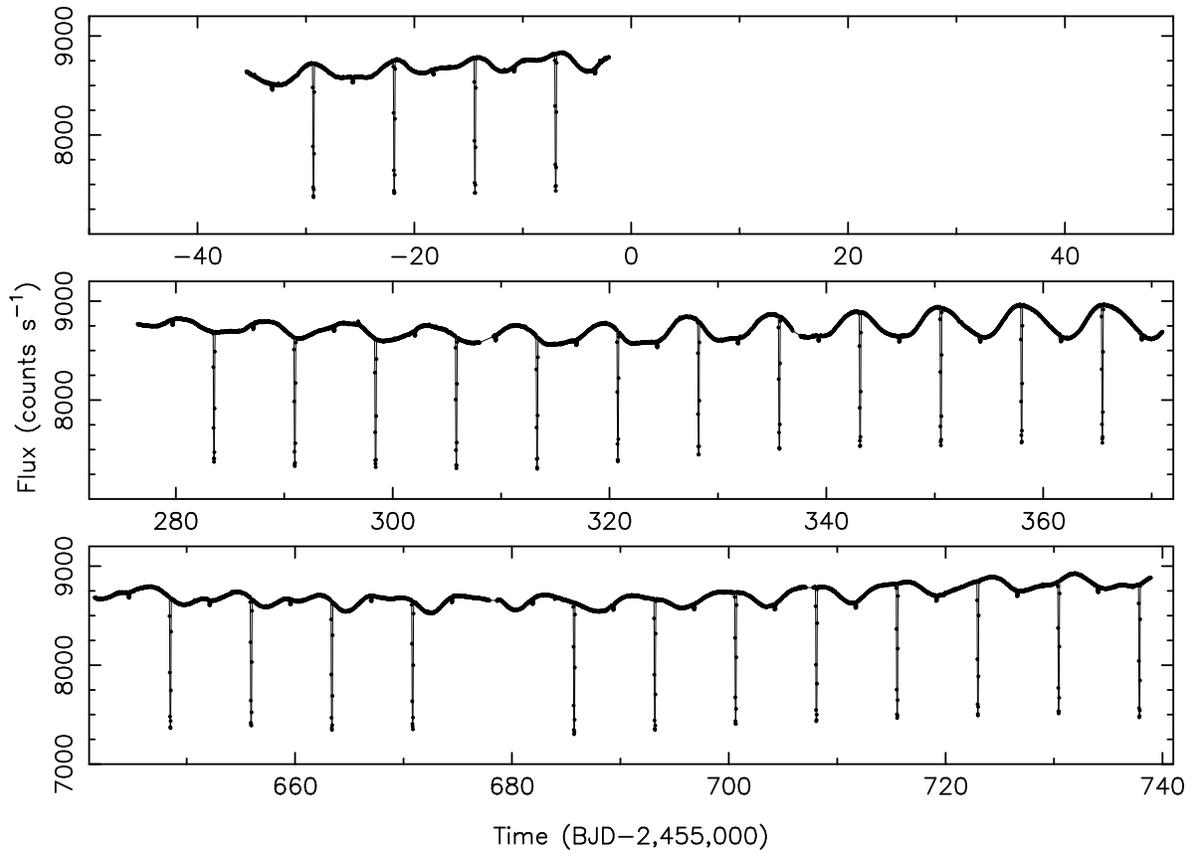}
       \end{center}
\caption{{\bf Light curves showing spot modulations.}
The SAP light curves of Kepler-47 from Q1 (top), Q5
(middle), and Q9 (bottom) are shown.  The target appeared on the
same detector module during these Quarters.
A modulation in the out-of-eclipse regions due to star spots is evident.
\label{plotmultiQ159}}
\end{figure} 

\renewcommand{\thefigure}{S\arabic{figure}}
 \begin{figure} 
    \begin{center}
      \includegraphics[angle = 0,clip=yes,width=1.1\textwidth]{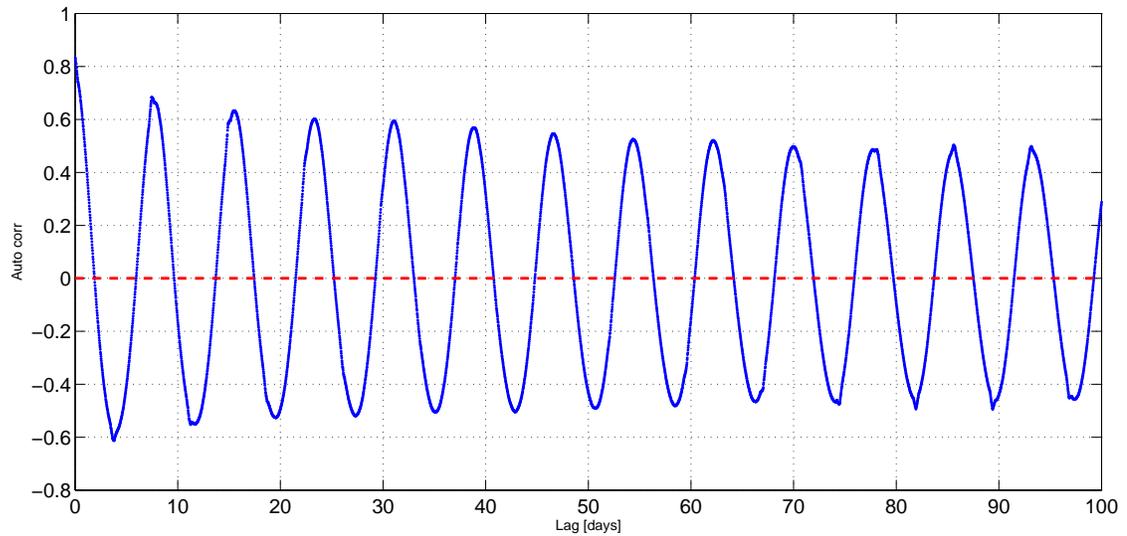}
       \end{center}
\caption{{\bf The 
autocorrelation function of the cleaned light curve with the eclipses removed.}
\label{autocorr}}
\end{figure} 

\renewcommand{\thefigure}{S\arabic{figure}}
 \begin{figure} 
    \begin{center}
      \includegraphics[angle = 0,clip=yes,width=1.1\textwidth]{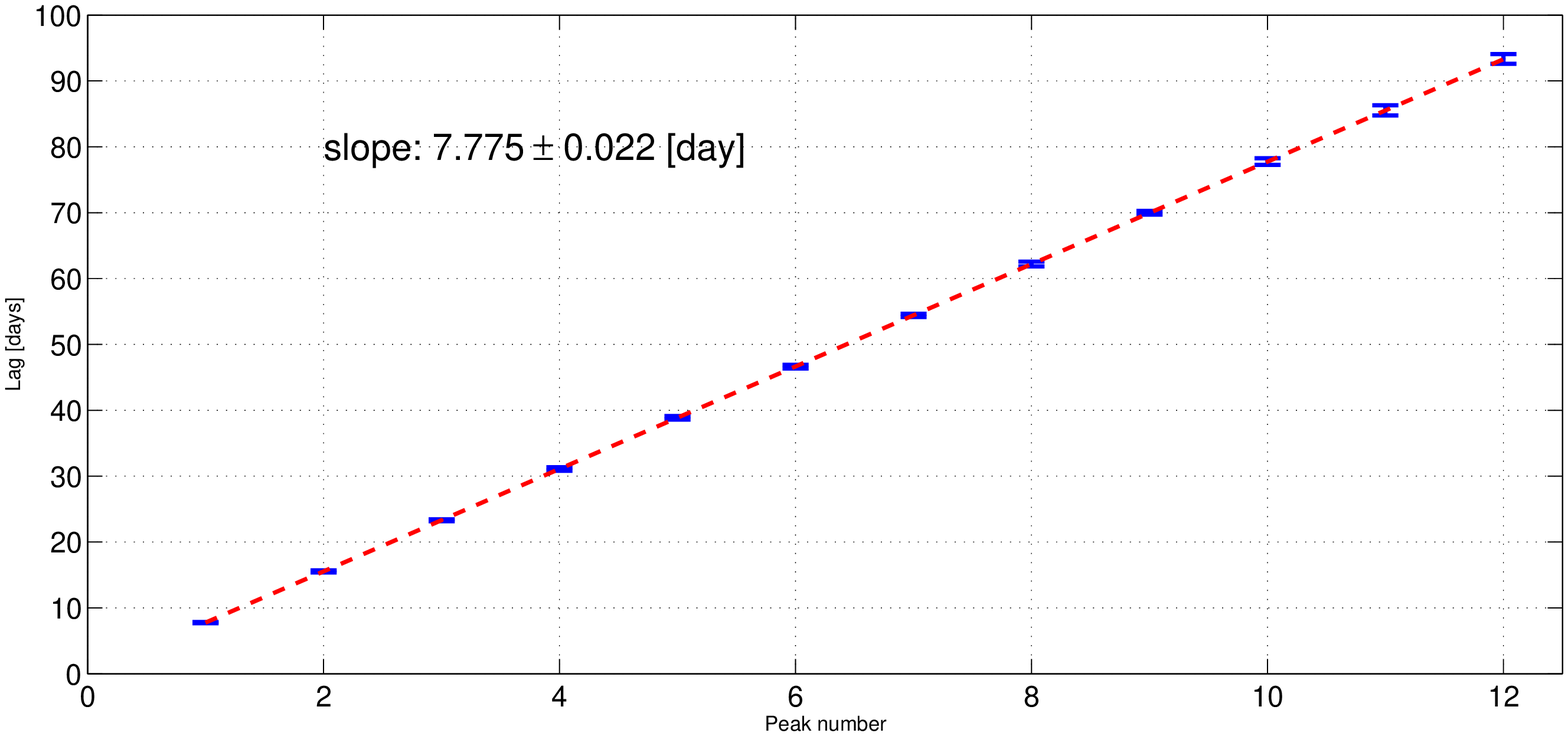}
       \end{center}
\caption{{\bf The measured lag versus 
the peak number in the autocorrelation function.}
The measured lag of the peaks in the autocorrelation
function displayed in Fig.\ \protect\ref{autocorr} is shown.
The dashed line is a linear fit to these points.  The slope of 
$7.775\pm 0.022$ days is taken to be the rotation 
period of the primary star.
\label{lagline}}
\end{figure}

\renewcommand{\thefigure}{S\arabic{figure}}
 \begin{figure} 
    \begin{center}
      \includegraphics[angle = -90,clip=yes,width=0.98\textwidth]{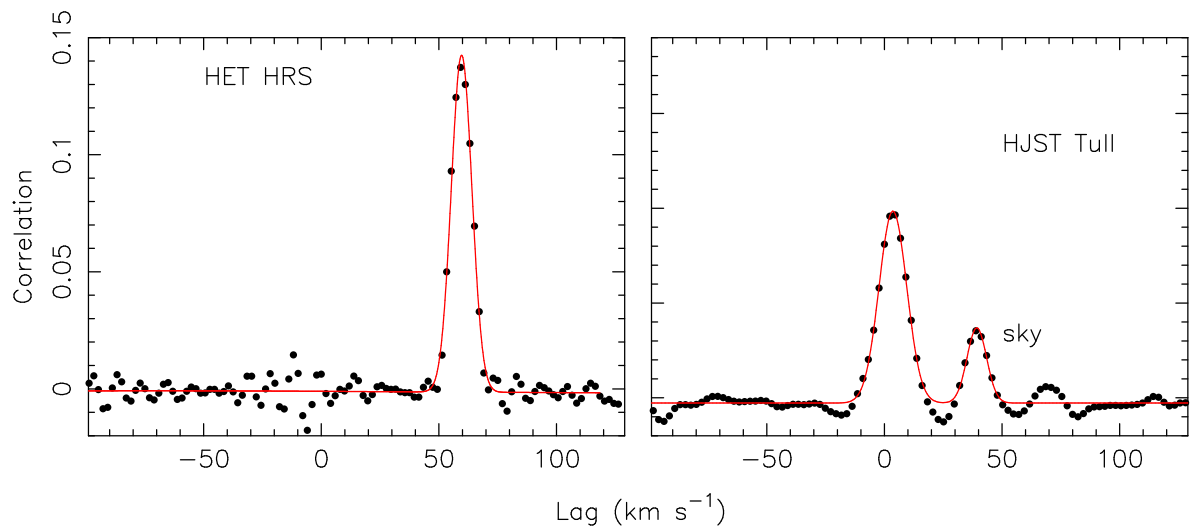}
       \end{center}
\caption{{\bf Representative broadening functions.} 
Broadening functions (BFs)
from the HET+HRS (left) and
the HJST+Tull spectrograph (right) are shown.  The solid lines are the 
best-fitting Gaussians.
The smaller peak in the HJST BF is due to the sky background.
In all cases, the BF peak due to the sky was resolved from the object BF peak.
\label{plotBF}}
\end{figure}

\renewcommand{\thefigure}{S\arabic{figure}}
 \begin{figure} 
    \begin{center}
  \includegraphics[angle = -90,clip=yes,width=0.98\textwidth]{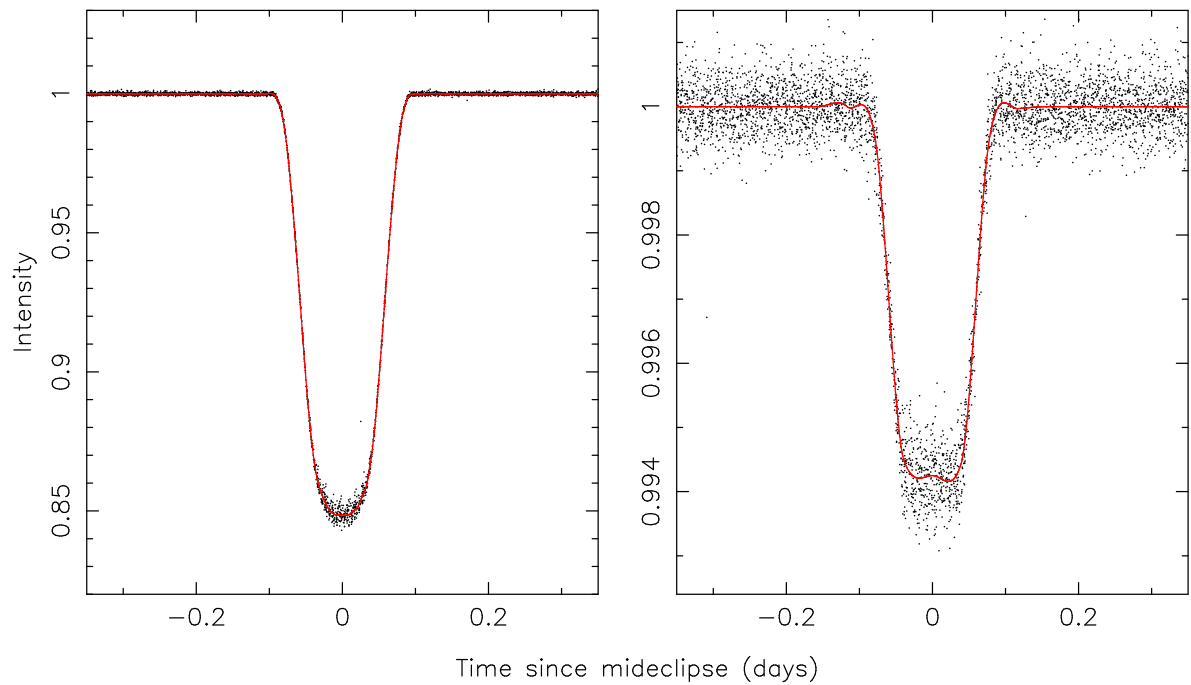}
       \end{center}
\caption{{\bf Mean primary and secondary eclipse profiles.}
The observed profiles for the primary eclipse (dots,
left panel) and the secondary eclipse (dots, right panel) arrived at
after an iterative process.   The Piecewise Cubic Hermit Spline
(PCHS) models are shown as the solid lines.  The increased scatter
in the middle of the primary eclipse is most likely due to the effects
of star spots.
\label{plot3154PCHS}}
\end{figure}

\renewcommand{\thefigure}{S\arabic{figure}}
 \begin{figure} 
    \begin{center}
  \includegraphics[angle = 0,clip=yes,width=0.85\textwidth]{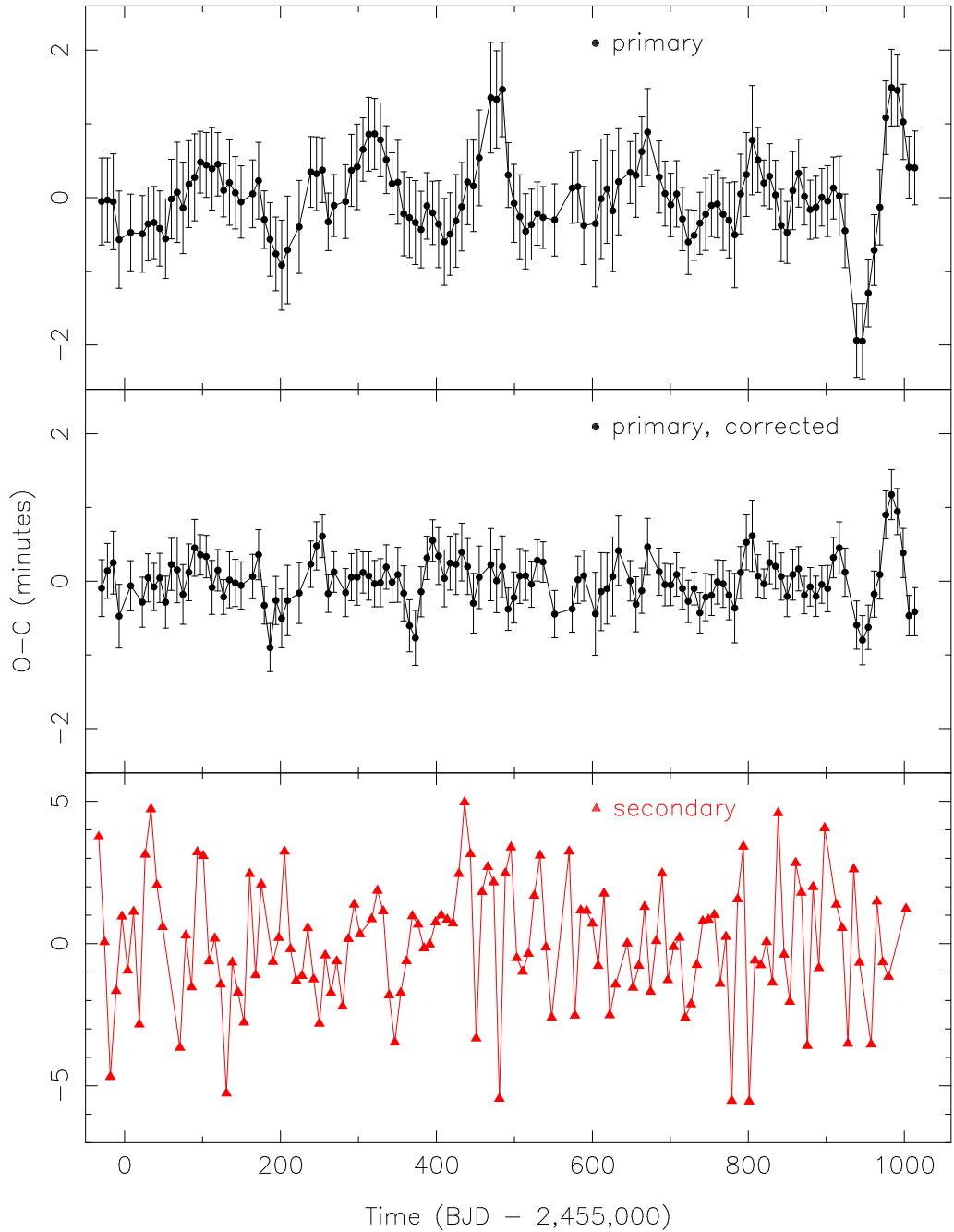}
       \end{center}
\caption{{\bf Observed minus computed curves for the stellar eclipses.}
Top:  The Observed minus Computed (O-C) residual times of the primary eclipses.
Coherent deviations of nearly two minutes are seen, with a quasiperiod of
$\approx 178$ days.
Middle:  The O-C times of the primary after correction for the effects
of star spots.  No periodicities or trends are evident.
Bottom:  The O-C times for the secondary star.  Note the change
in the vertical scale.  The error bars are not shown for clarity.
The scatter is much larger, and no periodicities or trends are seen.
\label{plotmultiOC}}
\end{figure}

\renewcommand{\thefigure}{S\arabic{figure}}
 \begin{figure} 
  \includegraphics[angle = -90,clip=yes,width=0.98\textwidth]{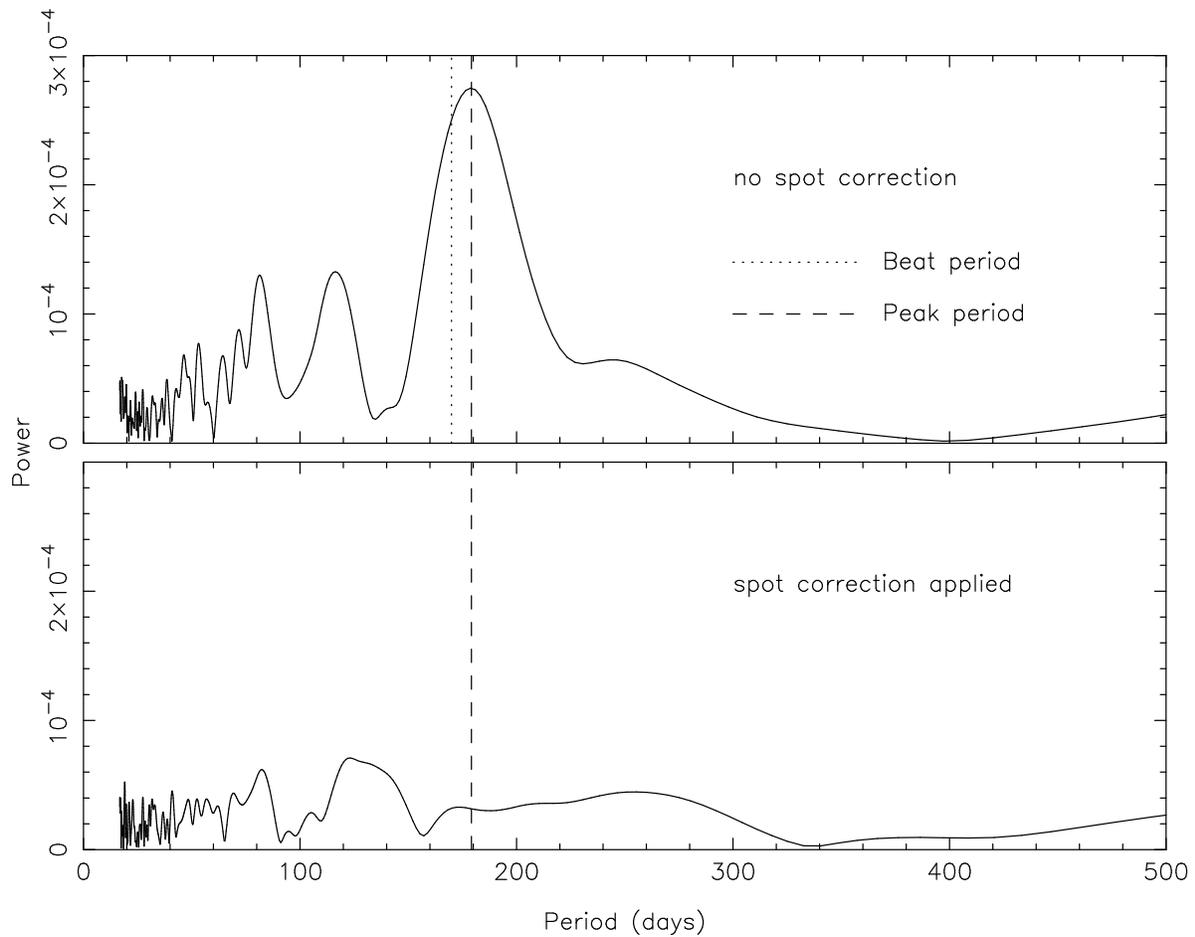}
\caption{{\bf Lomb-Scargle periodograms of O-C curves.}
Top:  A Lomb-Scargle periodogram of the O-Cs of the primary eclipses,
before any corrections for star spots have been applied.  The peak power
occurs at a period of 179.2 days (dashed line).  The expected beat period of
$\approx 170$ days is indicated by the dotted line.
Bottom:  A Lomb-Scargle periodogram of the primary eclipse O-Cs after 
a correction
for the effects of star spots has been applied.
There is no significant power at any period.
\label{plot3154lomb}}
\end{figure}

\renewcommand{\thefigure}{S\arabic{figure}}
 \begin{figure} 
    \begin{center}
  \includegraphics[angle = -90,clip=yes,width=0.98\textwidth]{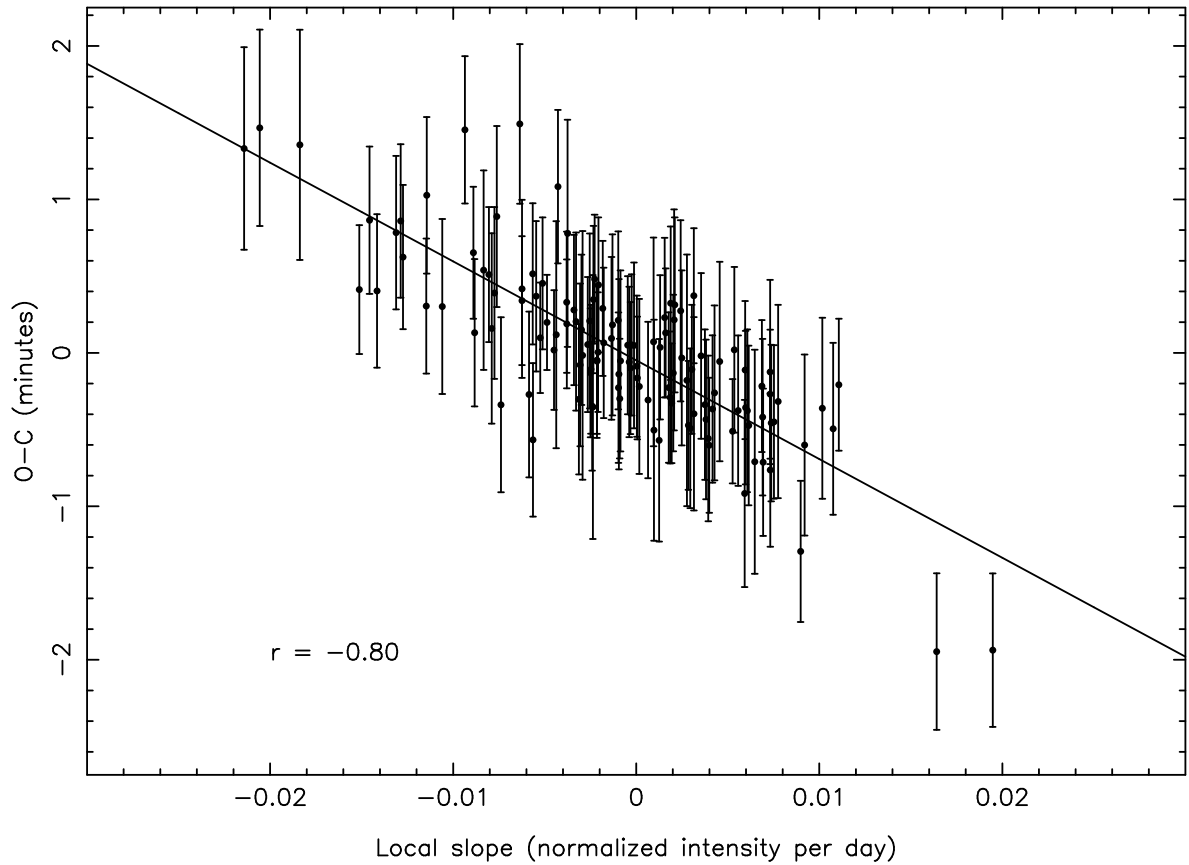}
       \end{center}
\caption{{\bf The correlation of residual O-C time and local slope 
near primary eclipse.}
The dependence of the primary eclipse O-C times on
the local SAP light curve slope is shown.  
A clear correlation is seen.  
The best-fitting line has a coefficient of
correlation of $r=-0.80$.
\label{slopevsoc}}
\end{figure} 

\clearpage

\renewcommand{\thefigure}{S\arabic{figure}}
 \begin{figure} 
    \begin{center}
      \includegraphics[angle = 0,clip=yes,width=0.99\textwidth]{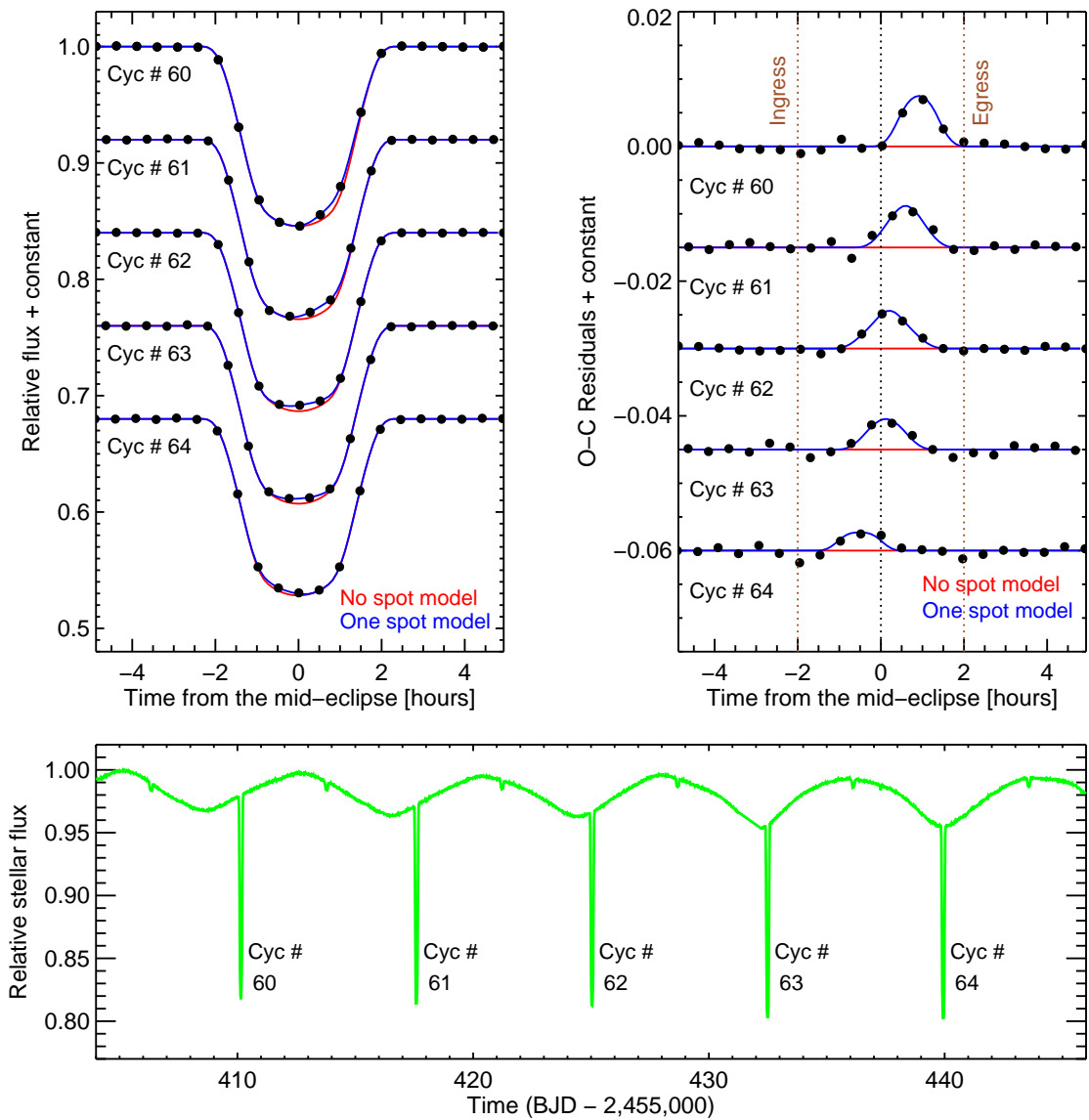}
       \end{center}
\caption{
{\bf 
The effect of star-spots on the primary eclipses.}
{Upper Left:} The observed eclipse light curves 
(black dots) for five consecutive primary eclipses
are shown. A model with no 
spots (red curves) does not fit the data well, whereas a model with 
a spot that is occulted by the secondary star fits much better (blue 
curves). 
{Upper Right:} The residuals for the same
eclipses are shown. As time passes (top to bottom) the residual feature from 
the no-spot model moves from the right side of the eclipse to the 
left.
{Lower:} A section of the light curve spanning
eclipse cycles 60--64 is shown. Notice how the local slope in the immediate
vicinity of the primary eclipse slowly changes from cycle to cycle.
\label{roberto1}}
\end{figure} 

\clearpage

\renewcommand{\thefigure}{S\arabic{figure}}
 \begin{figure} 
    \begin{center}
  \includegraphics[angle = 0,clip=yes,width=0.7\textwidth]{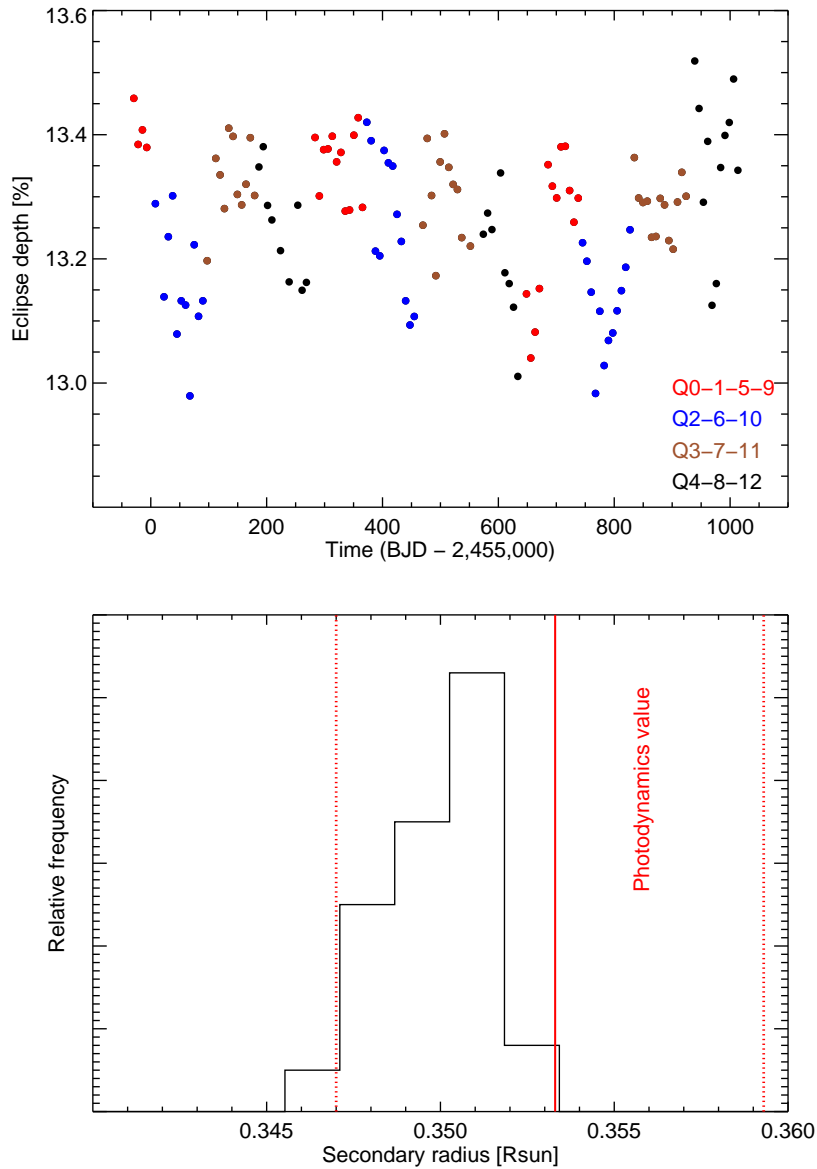}
       \end{center}
\caption{
{\bf 
Eclipse depth variation and its effect on the secondary star radius
estimate.}
{Top:} The individual depths for each primary 
eclipse calculated with a one-spot model are
shown (different color correspond
to different observing seasons).
The depth changes with time because the fraction of the star 
covered by spots changes with time. 
There is also a hint that the depths change with the
observing season (each season the star falls into a different CCD,
changing the level of contamination). 
{Bottom:} A histogram of the inferred radius of 
the secondary star (black line) for each eclipse
is shown. This demonstrates how 
the secondary star radius from the photometric-dynamical model 
(thick red line) is slightly underestimated (as expected), but 
the difference is not significant compared to the error bars on the 
measured radius (the 15.4\% and 84.6\% confidence levels are shown
with dotted red lines). 
\label{roberto2}}
\end{figure}

 \begin{figure} 
    \begin{center}
  \includegraphics[angle = -90,clip=yes,width=0.98\textwidth]{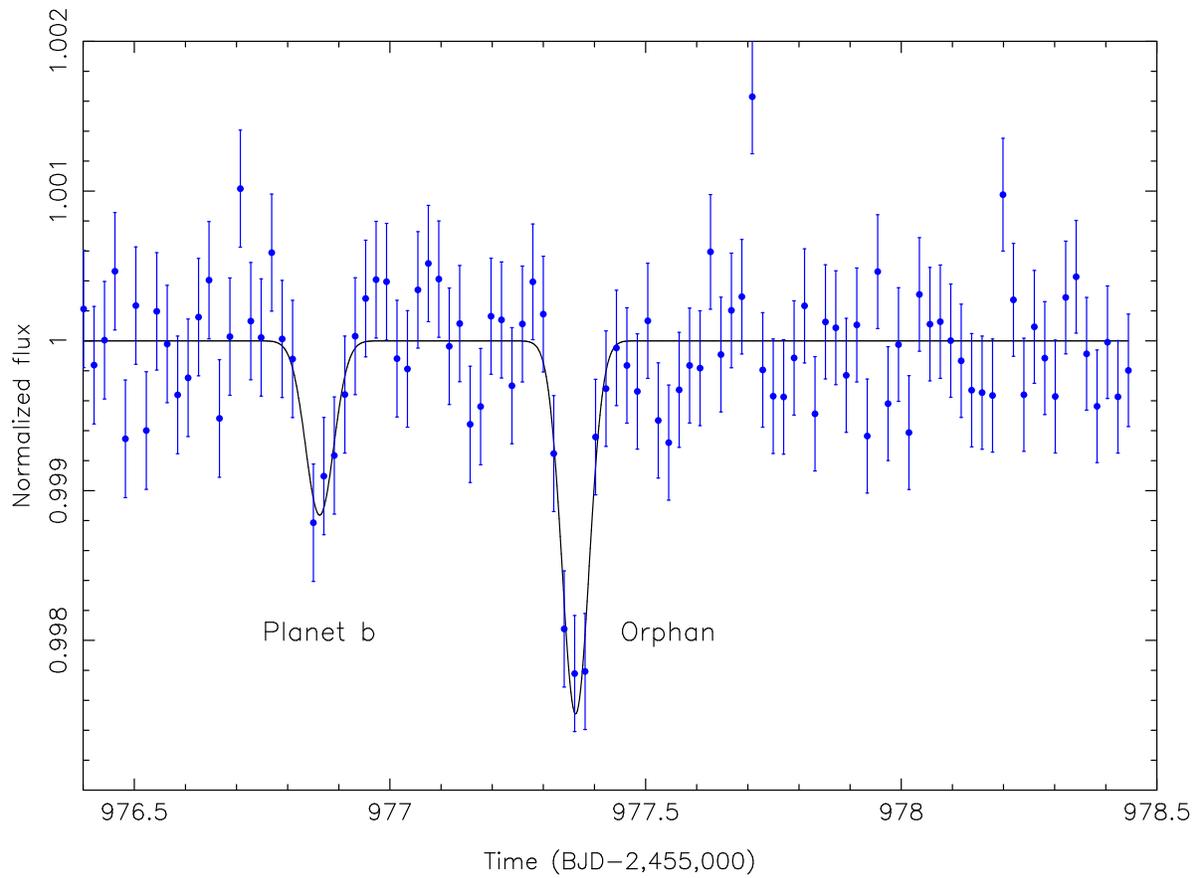}
       \end{center}
\caption{{\bf 
A segment of the Q12 light curve showing a transit of the inner planet
and an orphan transit.}
An ``orphan'' transit that cannot be accounted for by the inner or
outer planets appears near the middle of this data segment, about 12
hours after a transit of the inner planet (left). 
The solid line is a simple model consisting of two Gaussians
used to find the mid-transit time of the orphan and to evaluate the 
significance
of the event.
\label{plotorphan1}}
\end{figure}

 \begin{figure} 
    \begin{center}
  \includegraphics[angle = -90,clip=yes,width=0.98\textwidth]{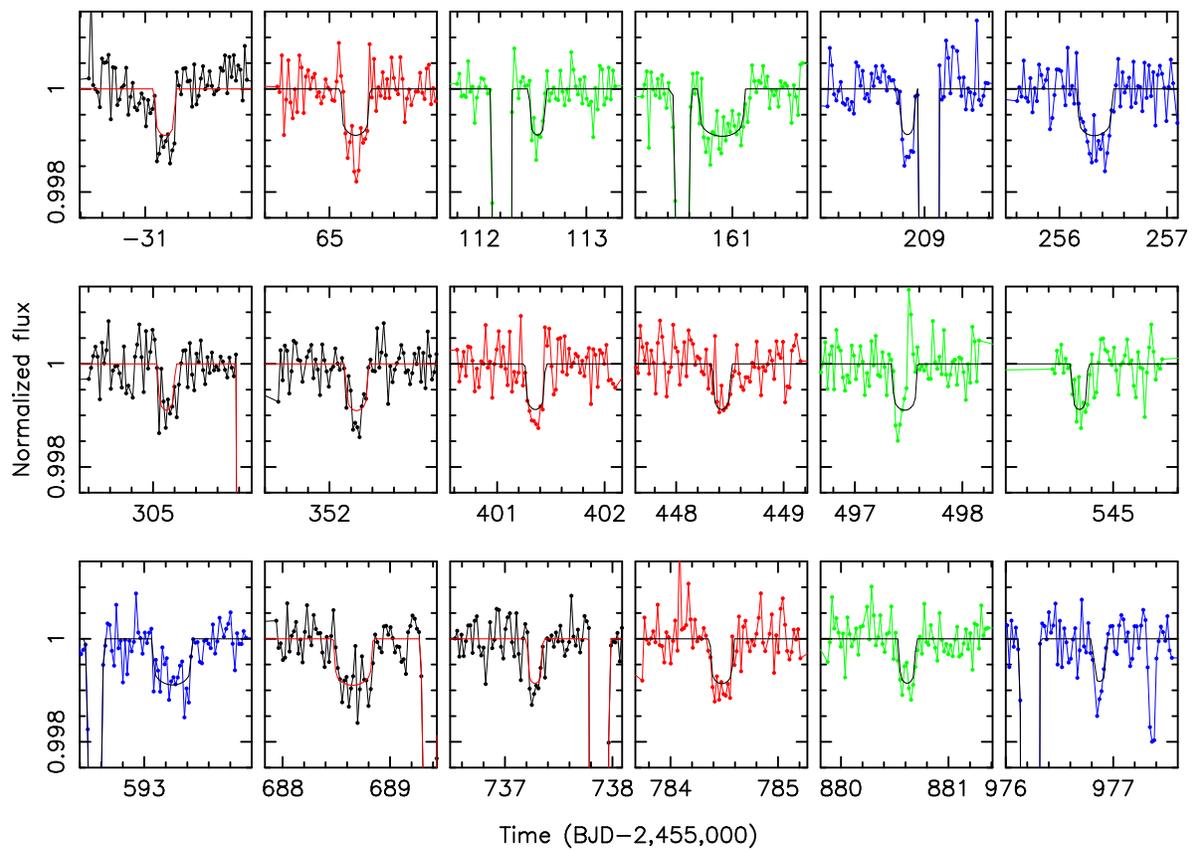}
       \end{center}
\caption{{\bf All observed transits of the inner planet.} 
The complete set of planet b transits
with the best-fitting model is shown.
The color coding is the same as in Fig.\ \protect\ref{plotrawmulti}.
\label{plot3154planet1}}
\end{figure}

 \begin{figure} 
    \begin{center}
  \includegraphics[angle = -90,clip=yes,width=0.98\textwidth]{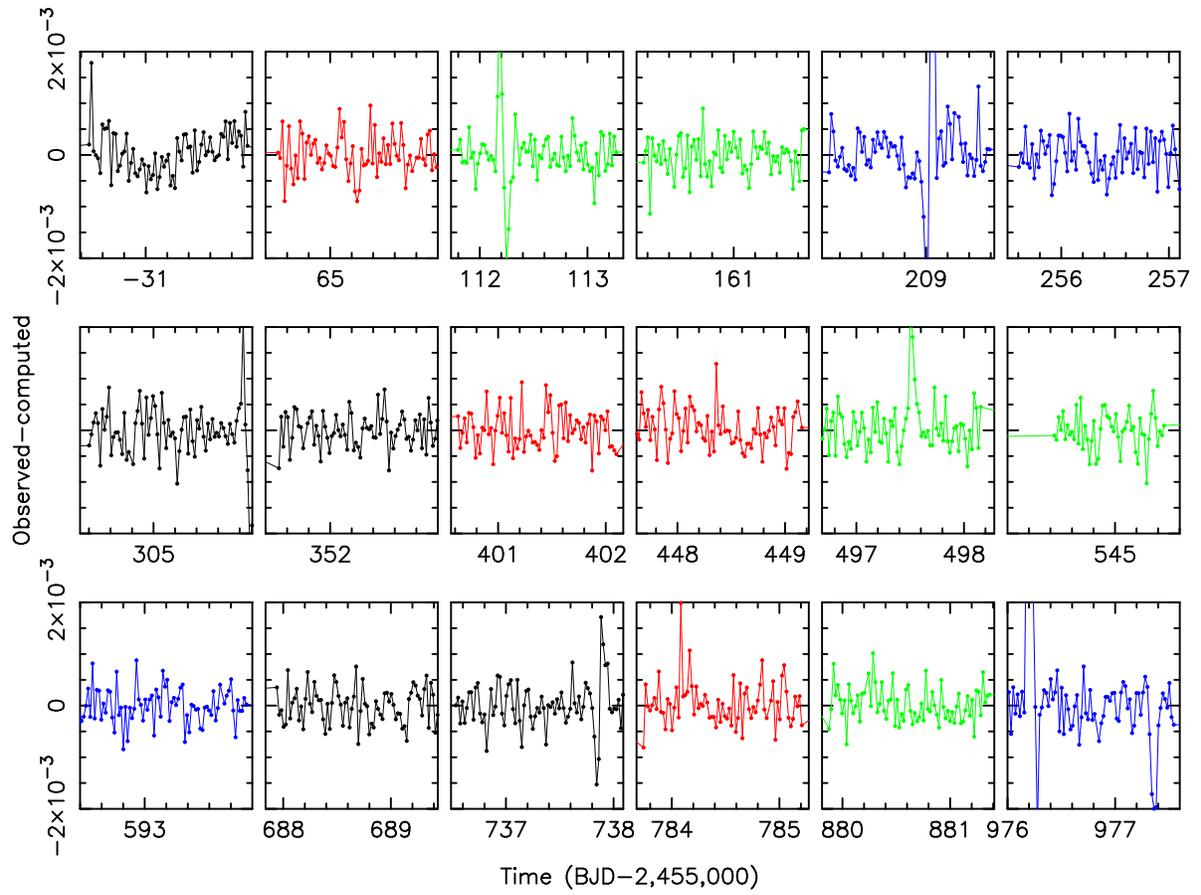}
       \end{center}
\caption{{\bf The residuals of the model fits of the inner planet transits.} 
The residuals of the model fits of the transits due to the inner planet
displayed
in Fig.\ \protect\ref{plot3154planet1} are shown.
The color coding is the same as in Fig.\ \protect\ref{plotrawmulti}.
\label{plot3154planet1res}}
\end{figure}

 \begin{figure} 
    \begin{center}
  \includegraphics[angle = -90,clip=yes,width=0.98\textwidth]{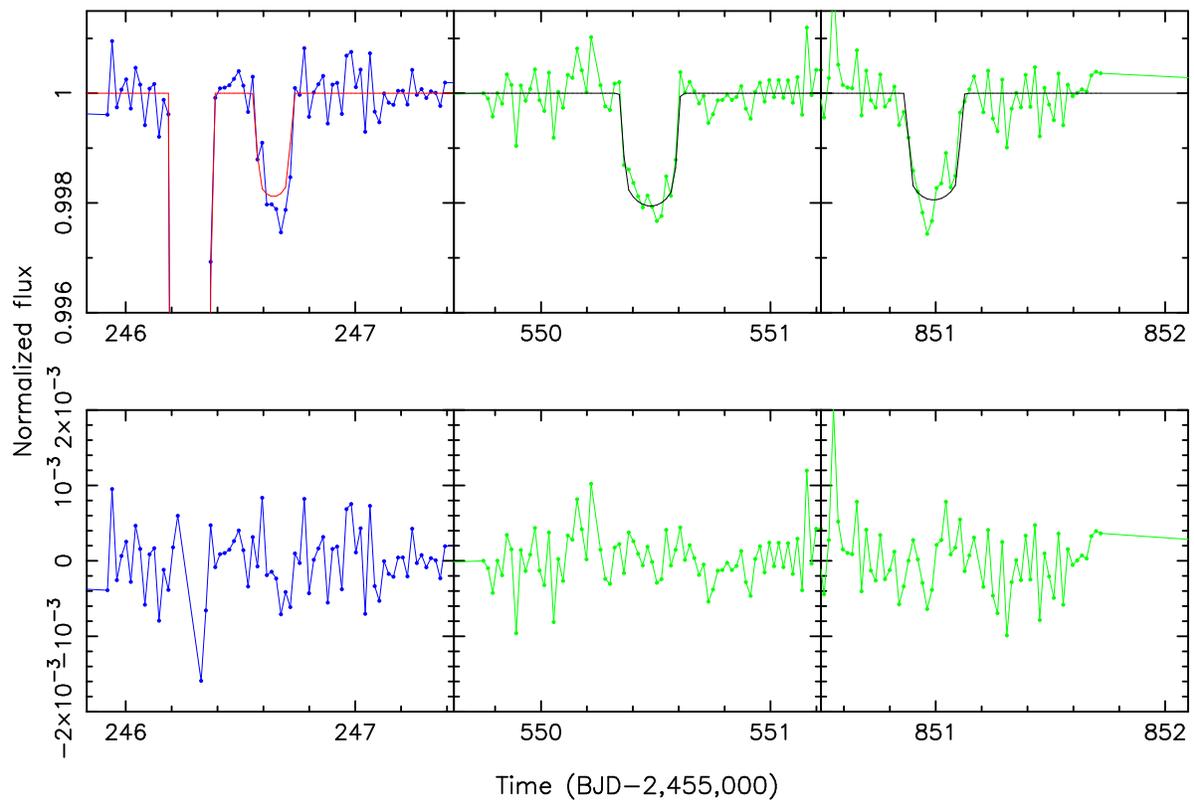}
       \end{center}
\caption{{\bf The model fits and residuals of the transits of the 
outer planet.}
The model fits to the transits of the outer planet are displayed 
in the top panels,
and the residuals are shown in the lower panels.
The color coding is the same as in Fig.\ \protect\ref{plotrawmulti}.
\label{plot3154planet2res}}
\end{figure}

 \begin{figure} 
 \hspace{-0.3in} \includegraphics[angle = 0,
     clip=yes,width=1.1\textwidth]{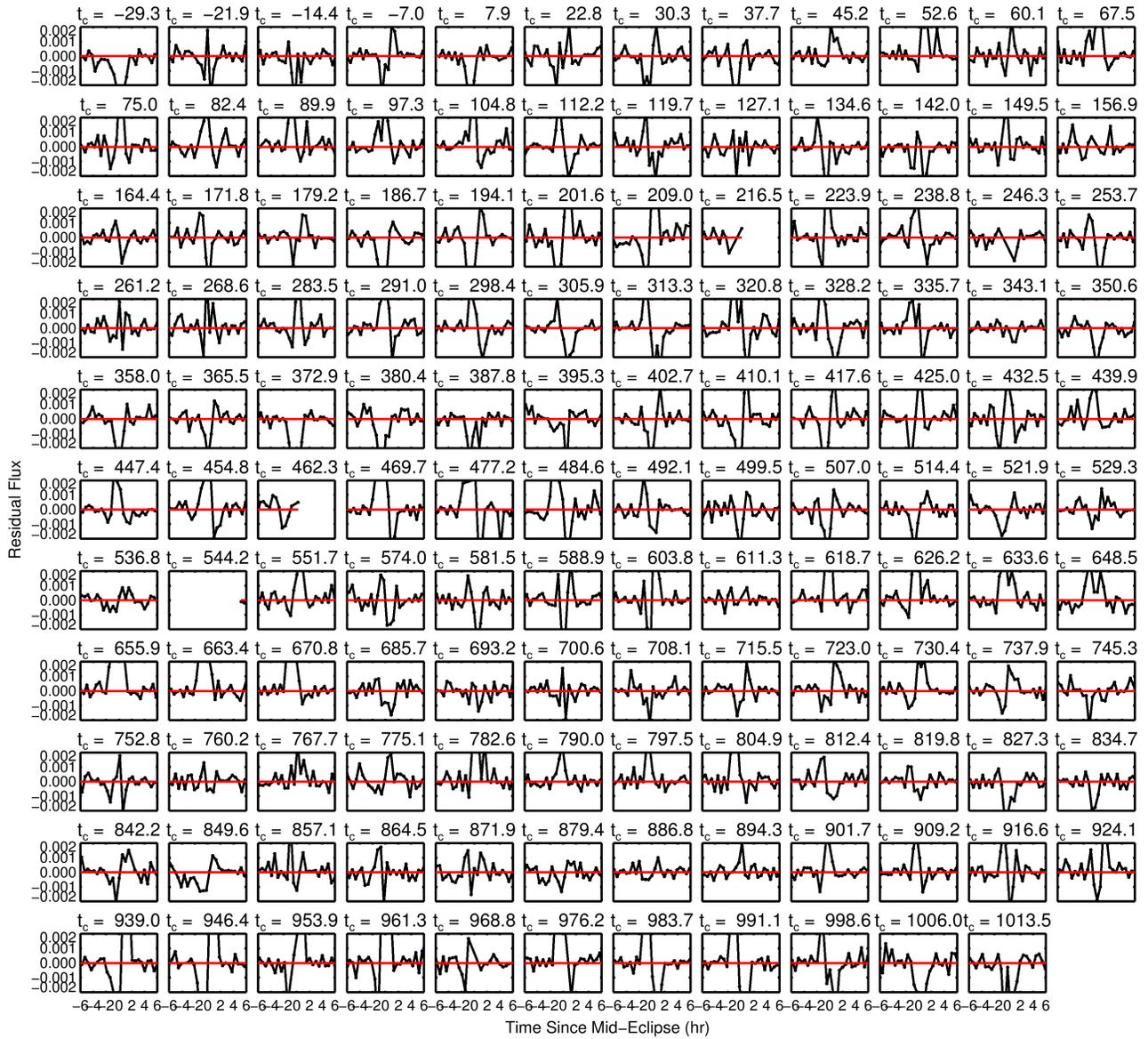}
\caption{{\bf The residuals of the fits to the primary eclipses.}
The residuals during each primary eclipse are displayed.  As expected, 
numerous  spot crossing events are seen.
\label{fitsFlux_100_Primaries_resid}}
\end{figure}

 \begin{figure} 
\hspace{-0.3in}\includegraphics[angle = 0,
     clip=yes,width=1.1\textwidth]{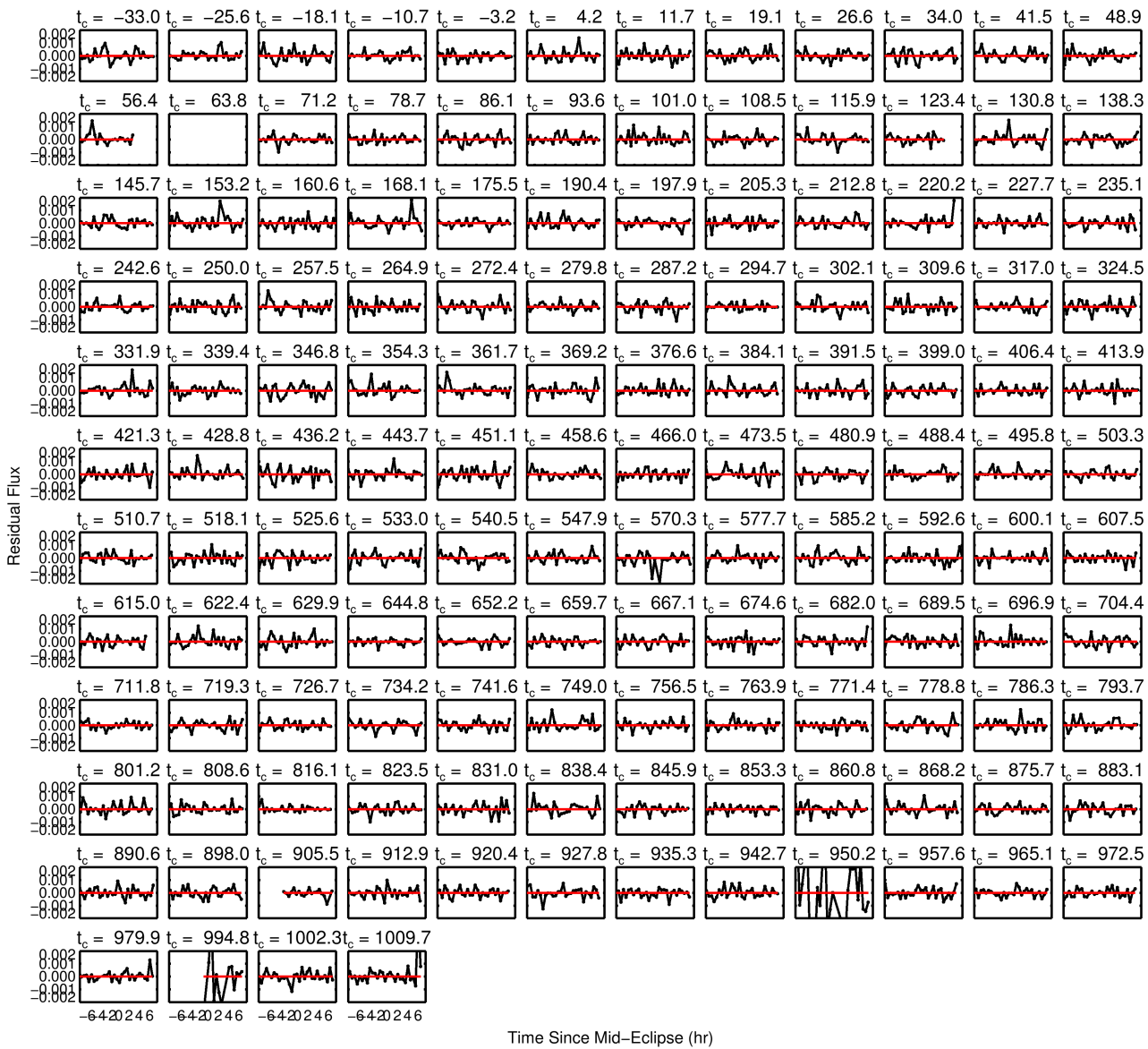}
\caption{{\bf The residuals of the fits to the secondary eclipses.}
The residuals during each secondary eclipse are shown.
As these eclipses are total there is much less structure seen in the
residuals, compared to the residuals for the primary eclipses
displayed in Fig.\ \protect\ref{fitsFlux_100_Primaries_resid}.
\label{fitsFlux_100_Secondaries_resid}}
\end{figure}

 \begin{figure} 
    \begin{center}
  \includegraphics[angle = 0,
     clip=yes,width=0.93\textwidth]{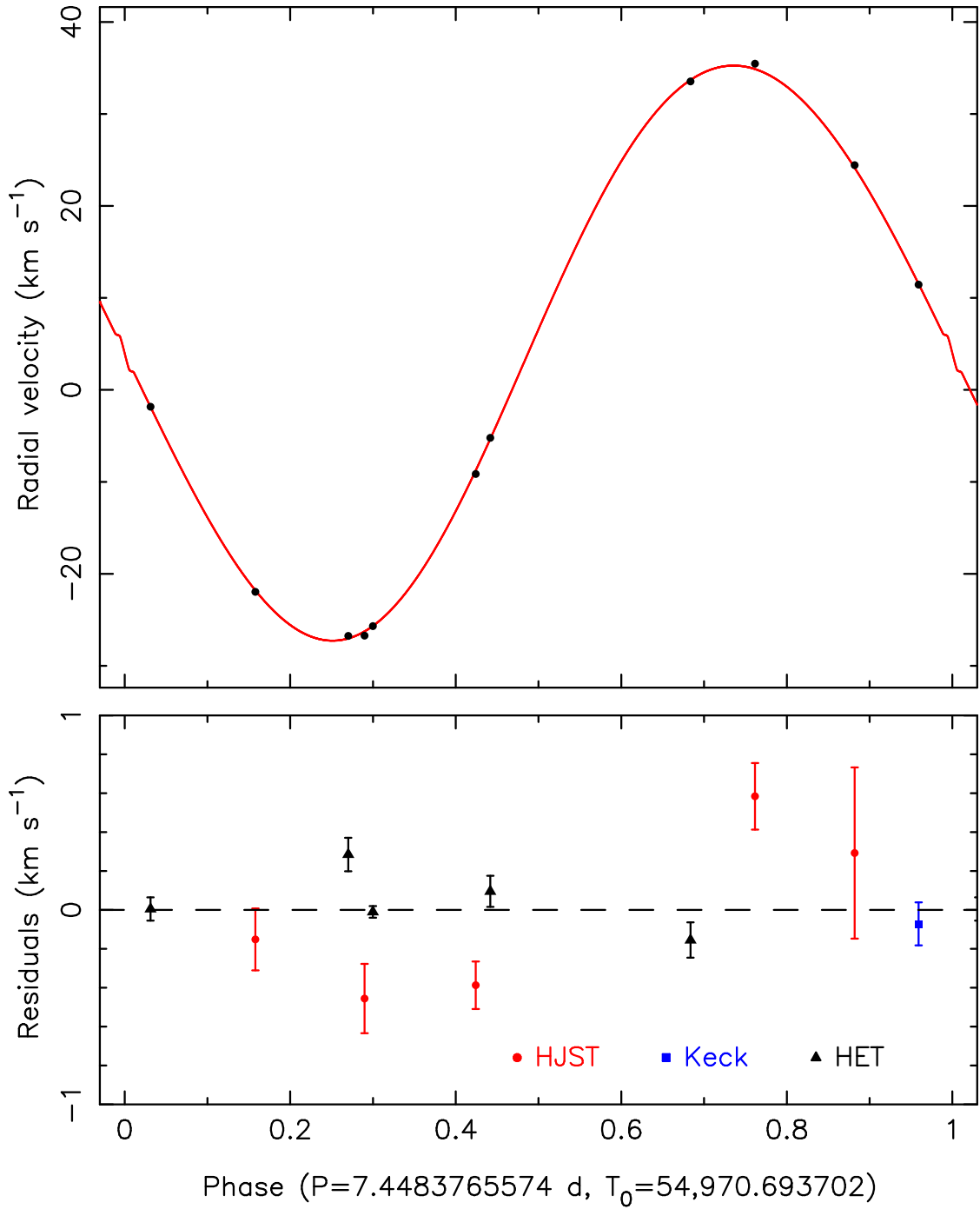}
       \end{center}
\caption{{\bf The observed and model radial velocity curve for the primary.}
Top:  The radial velocity measurements shown as a function of orbital
phase and the best-fitting model.  Bottom:  The residuals of the fit.  
Measurements from each telescope+instrument combination are denoted
with different symbols and colors.
\label{show3154RVres}}
\end{figure}

 \begin{figure} 
    \begin{center}
  \includegraphics[angle = 0,
     clip=yes,width=0.95\textwidth]{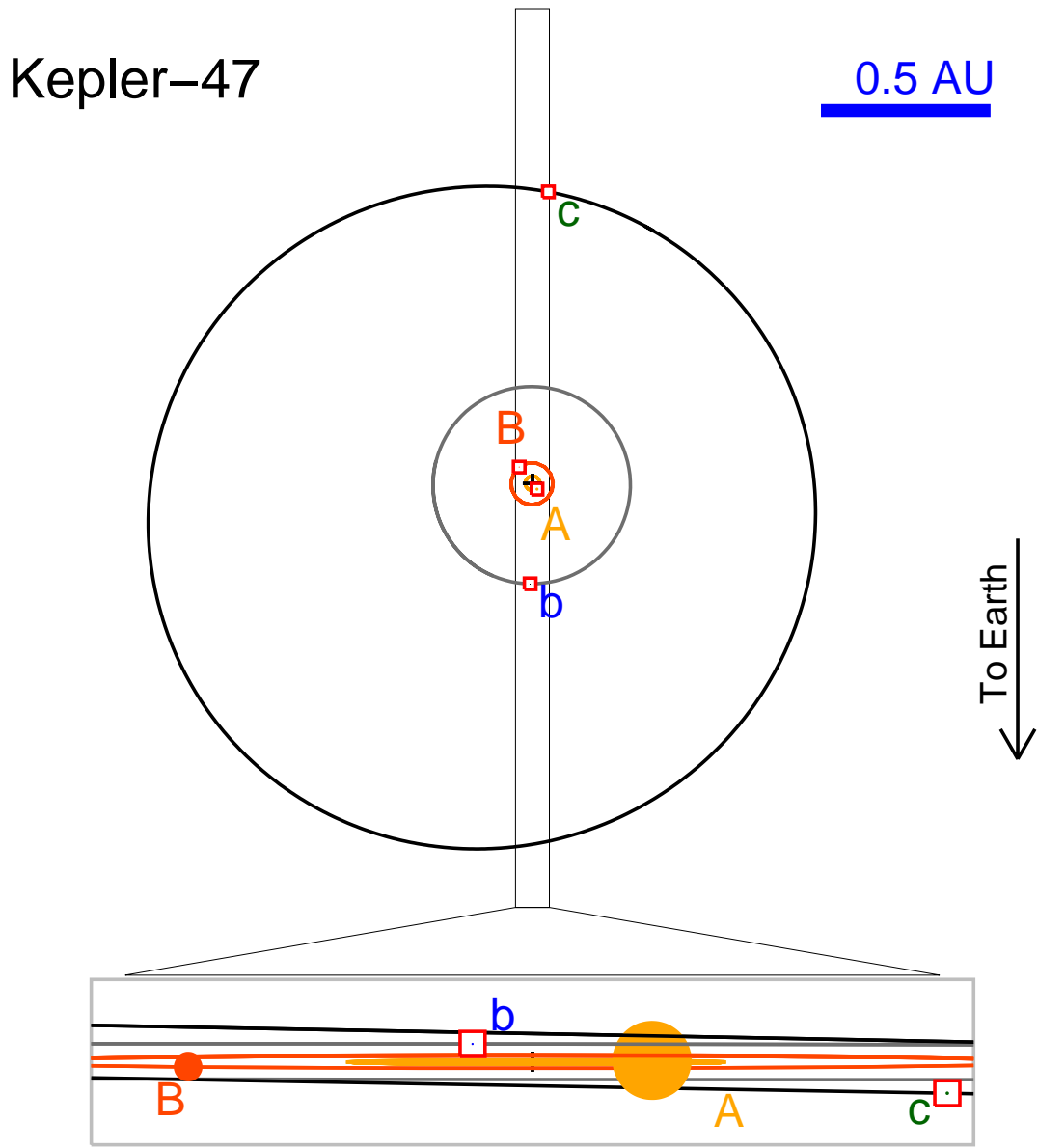}
       \end{center}
\caption{{\bf Schematic diagrams of the Kepler-47 orbits.}
Top:  A face-on view of the stellar and planetary orbits 
found from the best-fitting model
of the 
Kepler-47 system.  The center of mass of the system is marked with 
the cross.  The stars and the planets would not be seen at this scale, and
so their positions are marked with boxes.
Bottom:  The view of the system as seen from Earth on an expanded scale
is shown.  
The lines denote the
projected orbits of the various bodies.  Both planets can transit the
primary star (labeled A).  Transits of the secondary star (labeled
b) are narrowly
missed for the best-fitting orbital configuration.
\label{schem}}
\end{figure}

\begin{figure}
\centering
\includegraphics[width=6.2in]{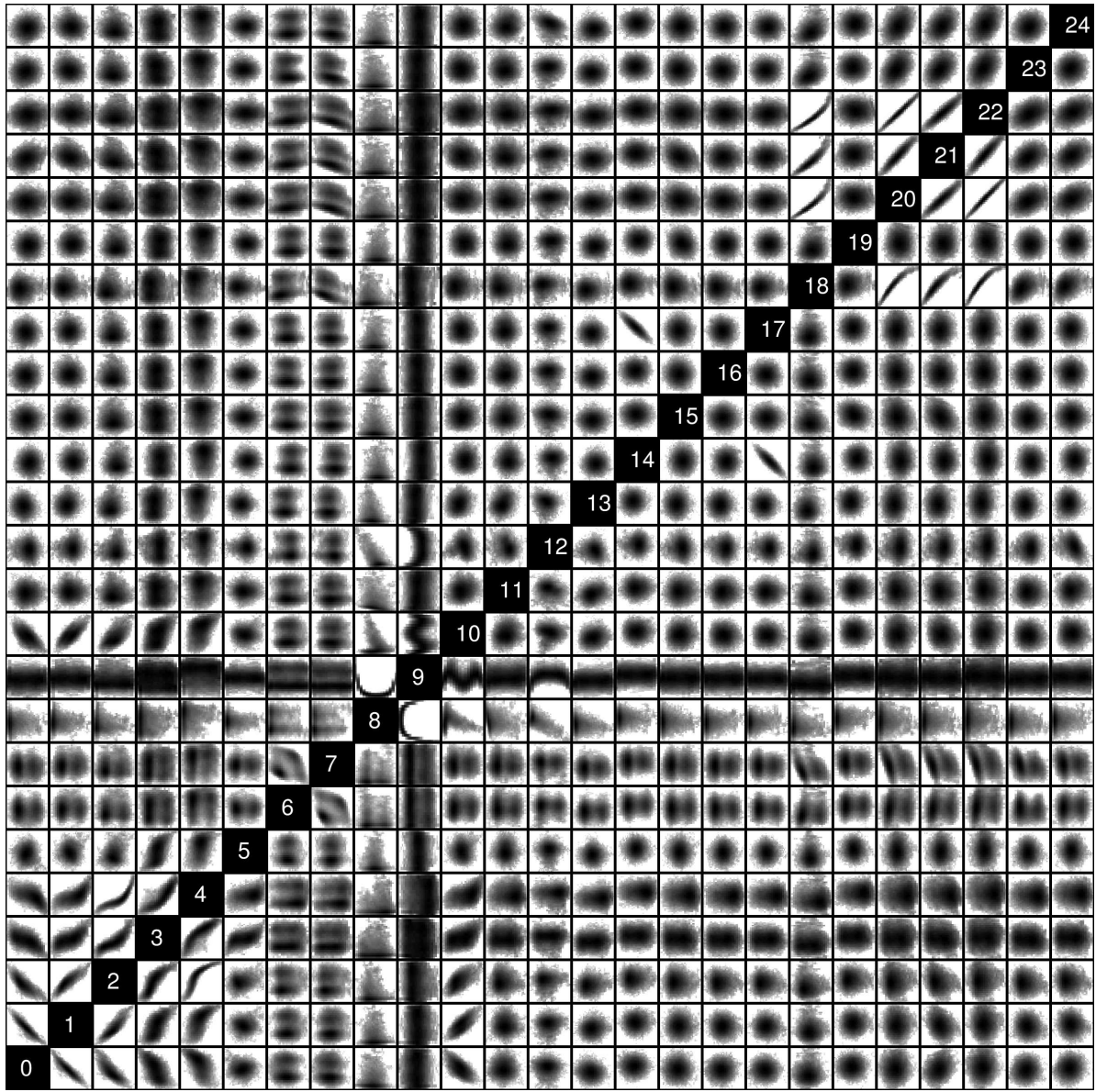}
\caption{{\bf Two-parameter joint posterior distributions of primary model
  parameters.}  The densities are plotted logarithmically in order to
  elucidate the nature of the parameter correlations.  The indices
  listed along the diagonal indicate which parameter is associated
  with the corresponding row and column.  The parameter name
  corresponding to a given index is indicated in Table \ref{tab:tab1}
  in the first column. \label{corr100} }
\end{figure}

\begin{figure}
\centering
\includegraphics[width=6.2in]{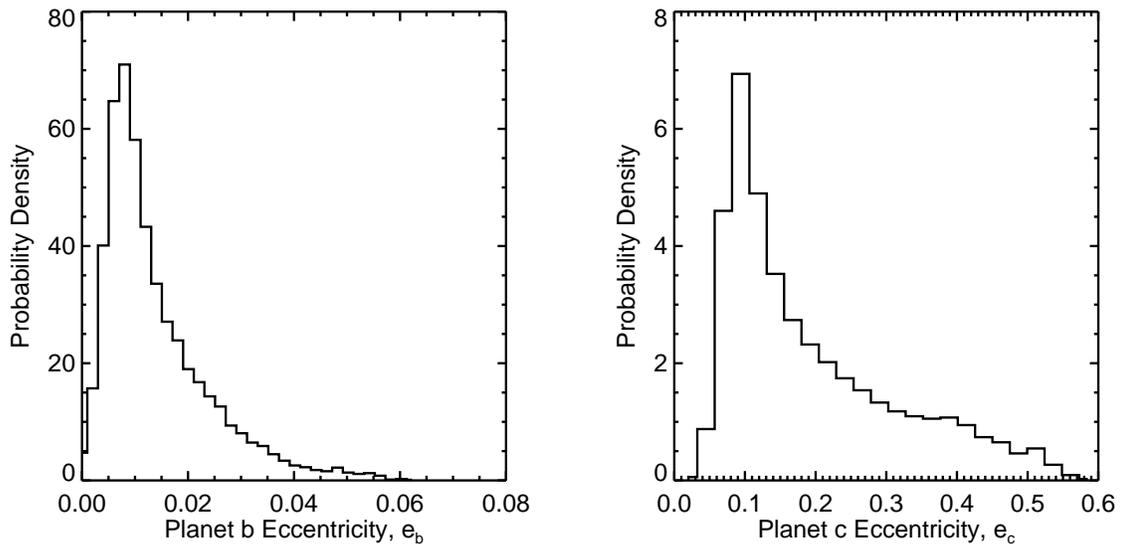}
\caption{{\bf Posterior distributions in eccentricity.}\label{fig:ecc} }
\end{figure}

\begin{figure}
\centering
\includegraphics[width=5in]{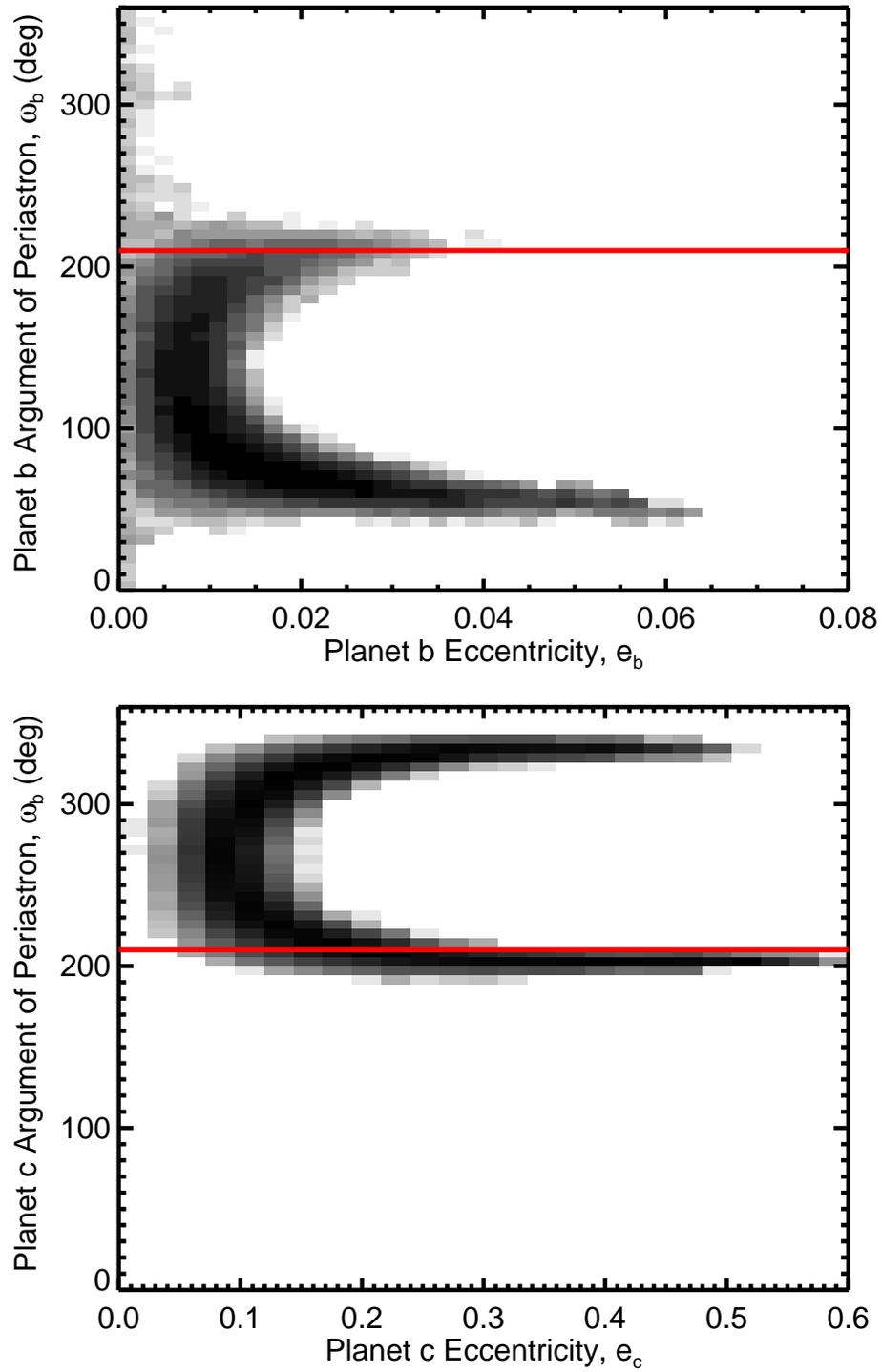}
\caption{{\bf Posterior distributions in the eccentricity 
and argument of pericenter planes.}\label{fig:omega} }
\end{figure}

\begin{figure}
\centering
\includegraphics[width=6.5in]{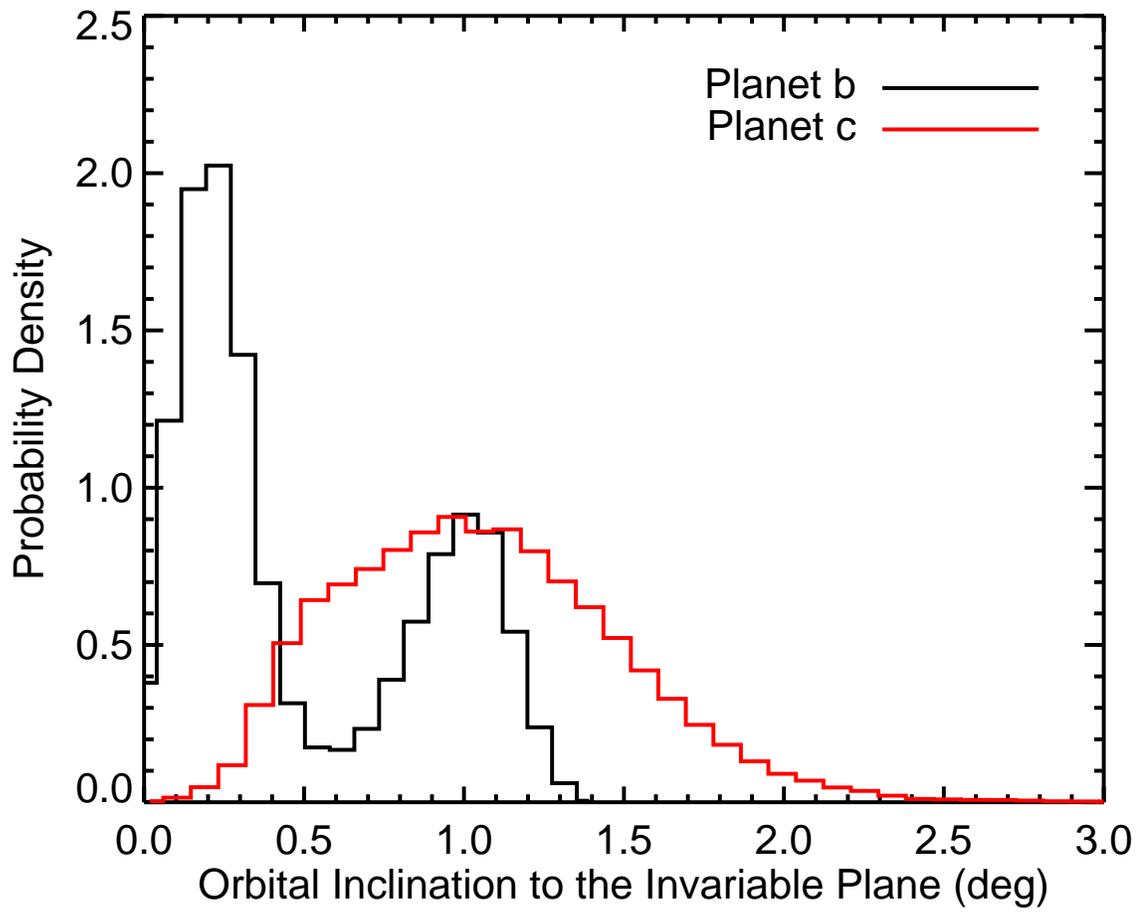}
\caption{{\bf Posterior distributions 
in the inclination of the planetary orbits relative to the 
invariable plane.}\label{fig:inclination_angle} }
\end{figure}

 \begin{figure} 
      \includegraphics[angle = 0,clip=yes,width=1.1\textwidth]{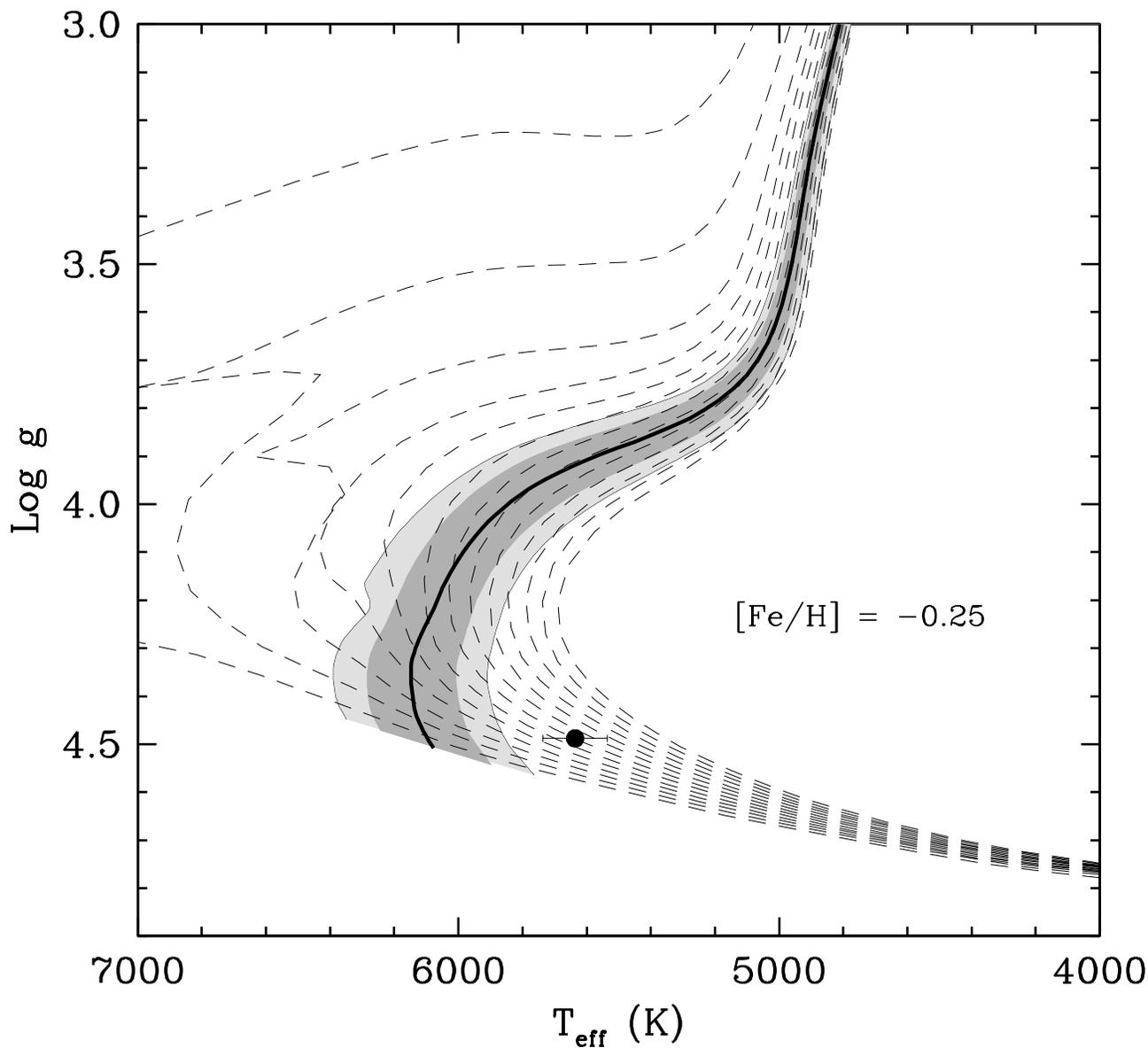}
\caption{{\bf 
Comparison of the absolute dimensions of the primary of
Kepler-47 against stellar evolution theory.} The thick solid line shows
an evolutionary track from the Yonsei-Yale series \cite{Yi:01,
Demarque:04} interpolated to the measured mass of that star
and its measured metallicity. The $1\sigma$ uncertainty in the
location of the track due to the mass error is indicated with the
darker shaded area.  The wider light shaded area includes the
additional $1\sigma$ contribution from the uncertainty in [Fe/H].
Isochrones from 1 to 13 Gyr (left to right) are shown with dashed
lines.
\label{plottrackg}}
\end{figure}

 \begin{figure} 
    \begin{center}
      \includegraphics[angle = 0,clip=yes,width=0.98\textwidth]{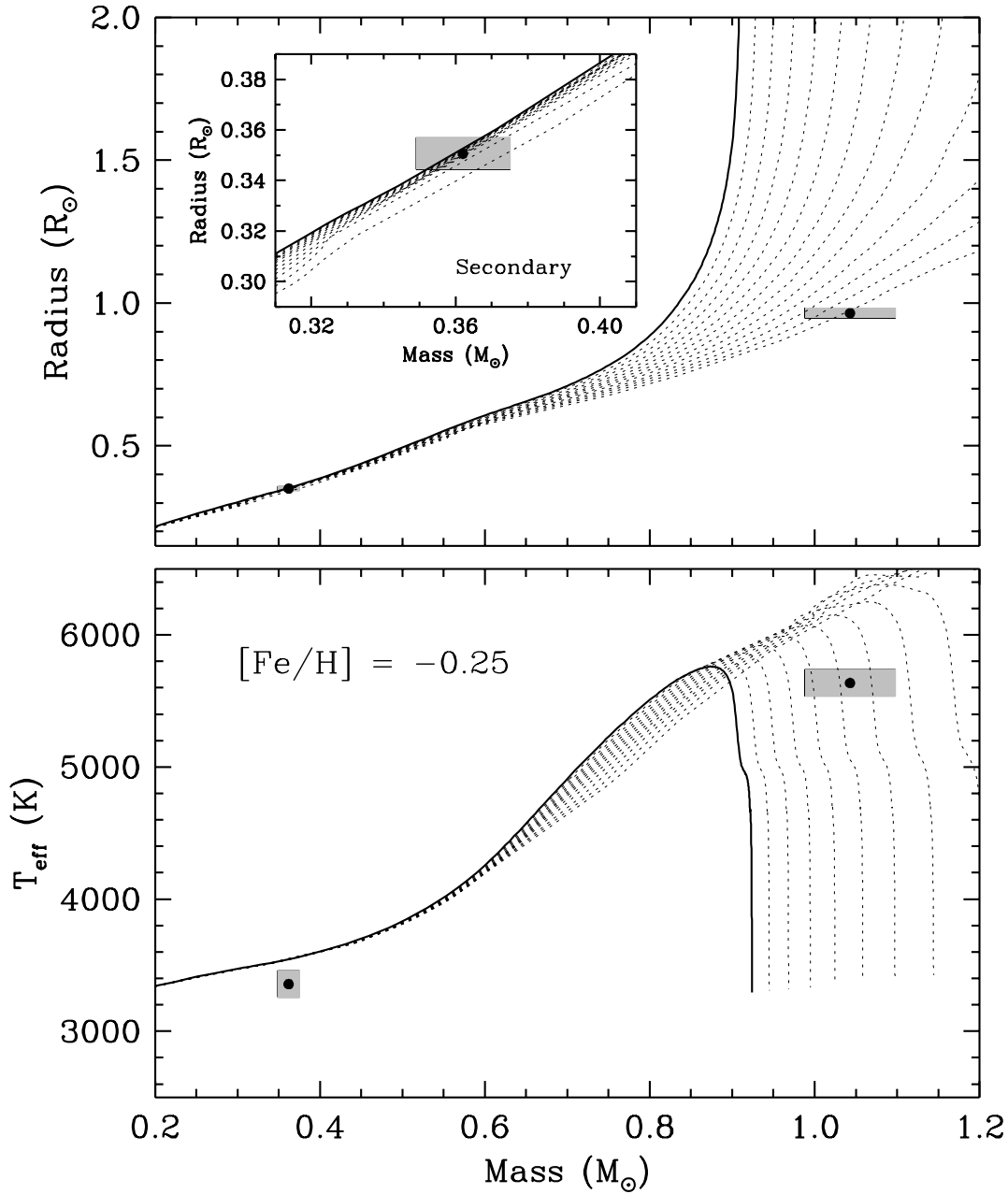}
       \end{center}
\caption{{\bf Isochrones in the mass-radius and mass-temperature
planes.}
Isochrones from the Dartmouth models \cite{Dotter:08}
corresponding to ages from 1 to 13 Gyr, compared against the measured
masses, radii, and temperatures of the stars in Kepler-47. The oldest
isochrone is indicated with a solid line, and the metallicity has been
set to the spectroscopically determined value of [Fe/H]$ = -0.25$.  The
error bars for the measurements are represented with the shaded boxes.
Top: The mass-radius diagram. The inset shows an enlargement around the
secondary, which appears to agree with the models. Bottom:
The mass-temperature diagram, showing the secondary to be cooler than
predicted.
\label{plotmrt2a}}
\end{figure} 

\clearpage

\renewcommand{\thetable}{S\arabic{table}}

\begin{table}
\caption{{\bf Radial velocities for Kepler-47.}}\label{tabRV}
\begin{center}
{\small
\begin{tabular}{lllrr}
\hline
Date       & UT Time & BJD & RV$_{A}$    & telescope \\
YYYY-MM-DD &         & (2,455,000+) & km s$^{-1}$ &  \\
\hline
\hline
2012-04-10 &  13:25:48.68  & 1028.05942 & $    11.442  \pm     0.011 $ &   Keck \\
2012-04-23 &  09:11:27.36  & 1040.90325 & $    33.534  \pm     0.091 $ &   HET \\
2012-05-01 &  09:52:55.08  & 1048.93237 & $    35.458  \pm     0.171 $ &   HJST \\
2012-05-02 &  07:23:45.95  & 1049.82882 & $    24.430  \pm     0.440 $ &   HJST \\
2012-05-04 &  08:34:10.40  & 1051.88474 & $   -21.957  \pm     0.159 $ &   HJST \\
2012-05-05 &  08:08:26.83  & 1052.86692 & $   -26.719  \pm     0.178 $ &   HJST \\
2012-05-06 &  08:08:55.42  & 1053.86729 & $    -9.150  \pm     0.122 $ &   HJST \\
2012-05-18 &  07:35:21.15  & 1065.83749 & $    -1.843  \pm     0.060 $ &   HET \\
2012-05-20 &  07:37:07.78  & 1067.83880 & $   -25.681  \pm     0.030 $ &   HET \\
2012-06-05 &  06:29:24.15  & 1083.79236 & $    -5.223  \pm     0.080 $ &   HET \\
2012-06-26 &  08:03:52.39  & 1104.85862 & $   -26.743  \pm     0.086 $ &   HJST \\
\hline
\end{tabular}
}
\end{center}
\end{table}

\begin{table}
\caption{{\bf Spectroscopic parameters from SPC.}}
\begin{center}
\begin{tabular}{rr}
\hline
\hline
parameter   & value  \\
\hline
$T_{\rm eff}$ (K) & $5636\pm 100$ \\
$\log g$ (cgs dex) & $4.42\pm 0.10$ \\
$[$m/H$]$ (dex) & $-0.25\pm 0.08$ \\
$V_{\rm rot}\sin i$ (km s$^{-1}$) & $4.1\pm 0.5$ \\
\hline
\end{tabular}
\end{center}
\end{table}

\begin{longtable}{rrrrrrr}
\caption{{\bf Times of stellar eclipses.}}\\
\hline\hline
cycle \# & primary   &
 corrected & uncertainty &
cycle \# & secondary  & uncertainty \\
  &    time\footnotemark &   time$^1$ & (min) & &
  time$^1$ & (min) \\
\hline  
\endfirsthead
\hline\hline
\caption[]{(continued)} \\
\hline  
\endhead
\hline
\endfoot
0.0\footnotetext{BJD-2,455,000}   &       ...    &           ...     &     ...   &   0.4873910   &   -33.12216  &    2.18    \\ 
1.0   &   -29.30630  &       -29.30631  &    0.39    &  1.4873910   &   -25.67634  &    2.58    \\ 
2.0   &   -21.85791  &       -21.85777  &    0.37    &  2.4873910   &   -18.23125  &    2.18    \\ 
3.0   &   -14.40955  &       -14.40931  &    0.42    &  3.4873910   &   -10.78077  &    2.18    \\ 
4.0   &    -6.96153  &        -6.96144  &    0.43    &  4.4873910   &    -3.33057  &    2.18    \\ 
5.0   &       ...    &           ...     &     ...   &   5.4873910   &     4.11649  &    2.18    \\ 
6.0   &     7.93529  &         7.93560  &    0.34    &  6.4873910   &    11.56631  &    2.48    \\ 
7.0   &       ...    &           ...     &     ...   &   7.4873910   &    19.01194  &    2.28    \\ 
8.0   &    22.83203  &        22.83220  &    0.34    &  8.4873910   &    26.46447  &    2.38    \\ 
9.0   &    30.28050  &        30.28080  &    0.33    &  9.4873910   &    33.91396  &    2.08    \\ 
10.0   &    37.72889  &        37.72909  &    0.32    & 10.4873910   &    41.36049  &    2.28    \\ 
11.0   &    45.17721  &        45.17755  &    0.33    & 11.4873910   &    48.80785  &    2.18    \\ 
12.0   &    52.62549  &        52.62570  &    0.35    & 12.4873910   &       ...    &       ...  \\ 
13.0   &    60.07424  &        60.07443  &    0.35    & 13.4873910   &       ...    &       ...  \\ 
14.0   &    67.52268  &        67.52276  &    0.44    & 14.4873910   &    71.15005  &    2.18    \\ 
15.0   &    74.97091  &        74.97090  &    0.41    & 15.4873910   &    78.60117  &    2.38    \\ 
16.0   &    82.41951  &        82.41949  &    0.39    & 16.4873910   &    86.04829  &    2.18    \\ 
17.0   &    89.86795  &        89.86809  &    0.39    & 17.4873910   &    93.49997  &    2.18    \\ 
18.0   &    97.31647  &        97.31640  &    0.27    & 18.4873910   &   100.94826  &    2.38    \\ 
19.0   &   104.76482  &       104.76476  &    0.29    & 19.4873910   &   108.39407  &    2.18    \\ 
20.0   &   112.21316  &       112.21285  &    0.37    & 20.4873910   &   115.84301  &    2.18    \\ 
21.0   &   119.66158  &       119.66138  &    0.28    & 21.4873910   &   123.29027  &    2.28    \\ 
22.0   &   127.10971  &       127.10951  &    0.24    & 22.4873910   &   130.73599  &    2.18    \\ 
23.0   &   134.55816  &       134.55805  &    0.38    & 23.4873910   &   138.18757  &    2.28    \\ 
24.0   &   142.00644  &       142.00639  &    0.32    & 24.4873910   &   145.63522  &    2.38    \\ 
25.0   &   149.45473  &       149.45475  &    0.32    & 25.4873910   &   153.08287  &    2.28    \\ 
26.0   &       ...    &           ...     &     ...   &  26.4873910   &   160.53488  &    2.48    \\ 
27.0   &   164.35156  &       164.35158  &    0.30    & 27.4873910   &   167.98079  &    2.08    \\ 
28.0   &   171.80006  &       171.80016  &    0.34    & 28.4873910   &   175.43139  &    2.18    \\ 
29.0   &   179.24807  &       179.24806  &    0.25    & 29.4873910   &       ...    &       ...  \\ 
30.0   &   186.69626  &       186.69604  &    0.33    & 30.4873910   &   190.32626  &    2.18    \\ 
31.0   &   194.14450  &       194.14486  &    0.33    & 31.4873910   &   197.77523  &    2.28    \\ 
32.0   &   201.59277  &       201.59307  &    0.40    & 32.4873910   &   205.22572  &    2.18    \\ 
33.0   &   209.04129  &       209.04161  &    0.48    & 33.4873910   &   212.67172  &    2.18    \\ 
34.0   &       ...    &           ...     &     ...   &  34.4873910   &   220.11933  &    2.18    \\ 
35.0   &   223.93826  &       223.93843  &    0.41    & 35.4873910   &   227.56783  &    2.28    \\ 
36.0   &       ...    &           ...     &     ...   &  36.4873910   &   235.01738  &    2.18    \\ 
37.0   &   238.83553  &       238.83546  &    0.31    & 37.4873910   &   242.46451  &    2.18    \\ 
38.0   &   246.28389  &       246.28401  &    0.33    & 38.4873910   &   249.91181  &    2.18    \\ 
39.0   &   253.73230  &       253.73247  &    0.29    & 39.4873910   &   257.36186  &    2.58    \\ 
40.0   &   261.18019  &       261.18031  &    0.25    & 40.4873910   &   264.80933  &    2.18    \\ 
41.0   &   268.62872  &       268.62889  &    0.27    & 41.4873910   &   272.25848  &    2.18    \\ 
42.0   &       ...    &           ...     &     ...   &  42.4873910   &   279.70576  &    2.18    \\ 
43.0   &   283.52551  &       283.52545  &    0.33    & 43.4873910   &   287.15579  &    2.58    \\ 
44.0   &   290.97418  &       290.97397  &    0.32    & 44.4873910   &   294.60501  &    2.18    \\ 
45.0   &   298.42259  &       298.42234  &    0.38    & 45.4873910   &   302.05267  &    2.18    \\ 
46.0   &   305.87113  &       305.87077  &    0.28    & 46.4873910   &       ...    &       ...  \\ 
47.0   &   313.31965  &       313.31911  &    0.33    & 47.4873910   &   316.94980  &    2.18    \\ 
48.0   &   320.76803  &       320.76741  &    0.31    & 48.4873910   &   324.39888  &    2.18    \\ 
49.0   &   328.21635  &       328.21580  &    0.33    & 49.4873910   &   331.84676  &    2.08    \\ 
50.0   &   335.66454  &       335.66432  &    0.30    & 50.4873910   &   339.29309  &    2.28    \\ 
51.0   &   343.11269  &       343.11255  &    0.27    & 51.4873910   &   346.74032  &    2.18    \\ 
52.0   &   350.56108  &       350.56100  &    0.37    & 52.4873910   &   354.18991  &    2.28    \\ 
53.0   &   358.00916  &       358.00920  &    0.37    & 53.4873910   &   361.63907  &    2.18    \\ 
54.0   &   365.45750  &       365.45727  &    0.35    & 54.4873910   &   369.08855  &    2.18    \\ 
55.0   &   372.90583  &       372.90553  &    0.37    & 55.4873910   &   376.53673  &    2.38    \\ 
56.0   &   380.35414  &       380.35434  &    0.34    & 56.4873910   &   383.98453  &    2.38    \\ 
57.0   &   387.80274  &       387.80304  &    0.29    & 57.4873910   &   391.43301  &    2.38    \\ 
58.0   &   395.25105  &       395.25158  &    0.28    & 58.4873910   &   398.88193  &    2.28    \\ 
59.0   &   402.69932  &       402.69981  &    0.39    & 59.4873910   &   406.33048  &    2.28    \\ 
60.0   &   410.14753  &       410.14797  &    0.39    & 60.4873910   &   413.77876  &    2.48    \\ 
61.0   &   417.59598  &       417.59649  &    0.37    & 61.4873910   &   421.22705  &    2.28    \\ 
62.0   &   425.04448  &       425.04486  &    0.41    & 62.4873910   &   428.67664  &    2.18    \\ 
63.0   &   432.49299  &       432.49335  &    0.39    & 63.4873910   &   436.12677  &    2.48    \\ 
64.0   &   439.94160  &       439.94159  &    0.38    & 64.4873910   &   443.57389  &    2.28    \\ 
65.0   &   447.38994  &       447.38962  &    0.41    & 65.4873910   &   451.01777  &    2.28    \\ 
66.0   &   454.83858  &       454.83824  &    0.42    & 66.4873910   &   458.46973  &    2.28    \\ 
67.0   &       ...    &           ...     &     ...   &  67.4873910   &   465.91872  &    2.18    \\ 
68.0   &   469.73590  &       469.73511  &    0.49    & 68.4873910   &   473.36673  &    2.18    \\ 
69.0   &   477.18426  &       477.18334  &    0.43    & 69.4873910   &   480.80983  &    2.18    \\ 
70.0   &   484.63273  &       484.63184  &    0.42    & 70.4873910   &   488.26371  &    2.18    \\ 
71.0   &   492.08030  &       492.07982  &    0.29    & 71.4873910   &   495.71273  &    2.08    \\ 
72.0   &   499.52841  &       499.52831  &    0.35    & 72.4873910   &   503.15841  &    2.28    \\ 
73.0   &   506.97666  &       506.97688  &    0.37    & 73.4873910   &   510.60646  &    2.28    \\ 
74.0   &   514.42490  &       514.42526  &    0.33    & 74.4873910   &   518.05528  &    2.18    \\ 
75.0   &   521.87334  &       521.87356  &    0.31    & 75.4873910   &   525.50508  &    2.28    \\ 
76.0   &   529.32182  &       529.32216  &    0.28    & 76.4873910   &   532.95444  &    2.38    \\ 
77.0   &   536.77016  &       536.77052  &    0.27    & 77.4873910   &   540.40058  &    2.28    \\ 
78.0   &       ...    &           ...     &     ...   &  78.4873910   &   547.84725  &    2.38    \\ 
79.0   &   551.66689  &       551.66678  &    0.32    & 79.4873910   &       ...    &       ...  \\ 
80.0   &       ...    &           ...     &     ...   &  80.4873910   &       ...    &       ...  \\ 
81.0   &       ...    &           ...     &     ...   &  81.4873910   &   570.19645  &    2.88    \\ 
82.0   &   574.01232  &       574.01196  &    0.31    & 82.4873910   &   577.64083  &    2.38    \\ 
83.0   &   581.46071  &       581.46061  &    0.32    & 83.4873910   &   585.09178  &    2.38    \\ 
84.0   &   588.90872  &       588.90902  &    0.35    & 84.4873910   &   592.54015  &    2.18    \\ 
85.0   &       ...    &           ...     &     ...   &  85.4873910   &   599.98822  &    2.08    \\ 
86.0   &   603.80549  &       603.80542  &    0.56    & 86.4873910   &   607.43557  &    2.28    \\ 
87.0   &   611.25410  &       611.25400  &    0.53    & 87.4873910   &   614.88572  &    2.08    \\ 
88.0   &   618.70257  &       618.70241  &    0.48    & 88.4873910   &   622.33113  &    2.38    \\ 
89.0   &   626.15074  &       626.15090  &    0.54    & 89.4873910   &   629.78026  &    2.18    \\ 
90.0   &   633.59939  &       633.59952  &    0.47    & 90.4873910   &       ...    &       ...  \\ 
91.0   &       ...    &           ...     &     ...   &  91.4873910   &   644.67803  &    2.18    \\ 
92.0   &   648.49623  &       648.49598  &    0.27    & 92.4873910   &   652.12533  &    2.18    \\ 
93.0   &   655.94458  &       655.94414  &    0.37    & 93.4873910   &   659.57425  &    2.18    \\ 
94.0   &   663.39318  &       663.39264  &    0.31    & 94.4873910   &   667.02407  &    1.98    \\ 
95.0   &   670.84174  &       670.84143  &    0.39    & 95.4873910   &   674.47038  &    2.18    \\ 
96.0   &       ...    &           ...     &     ...   &  96.4873910   &   681.92000  &    2.28    \\ 
97.0   &   685.73807  &       685.73795  &    0.32    & 97.4873910   &   689.37003  &    2.28    \\ 
98.0   &   693.18629  &       693.18620  &    0.29    & 98.4873910   &   696.81581  &    2.08    \\ 
99.0   &   700.63456  &       700.63458  &    0.28    & 99.4873910   &   704.26500  &    2.08    \\ 
100.0   &   708.08304  &       708.08305  &    0.29    &100.4873910   &   711.71361  &    2.08    \\ 
101.0   &   715.53118  &       715.53130  &    0.28    &101.4873910   &   719.16004  &    2.18    \\ 
102.0   &   722.97934  &       722.97955  &    0.29    &102.4873910   &   726.60875  &    2.18    \\ 
103.0   &   730.42778  &       730.42805  &    0.22    &103.4873910   &   734.05809  &    2.18    \\ 
104.0   &   737.87627  &       737.87620  &    0.27    &104.4873910   &   741.50754  &    2.18    \\ 
105.0   &   745.32473  &       745.32472  &    0.32    &105.4873910   &   748.95596  &    2.18    \\ 
106.0   &   752.77319  &       752.77311  &    0.27    &106.4873910   &   756.40446  &    2.28    \\ 
107.0   &   760.22158  &       760.22162  &    0.30    &107.4873910   &   763.85116  &    2.28    \\ 
108.0   &   767.66986  &       767.66997  &    0.32    &108.4873910   &   771.30069  &    2.28    \\ 
109.0   &   775.11818  &       775.11824  &    0.33    &109.4873910   &   778.74507  &    2.28    \\ 
110.0   &   782.56642  &       782.56650  &    0.47    &110.4873910   &   786.19837  &    2.48    \\ 
111.0   &   790.01518  &       790.01521  &    0.35    &111.4873910   &   793.64804  &    2.08    \\ 
112.0   &   797.46374  &       797.46387  &    0.37    &112.4873910   &   801.09020  &    2.28    \\ 
113.0   &   804.91244  &       804.91231  &    0.48    &113.4873910   &   808.54202  &    2.18    \\ 
114.0   &   812.36063  &       812.36030  &    0.29    &114.4873910   &   815.99029  &    2.38    \\ 
115.0   &   819.80879  &       819.80861  &    0.20    &115.4873910   &   823.43924  &    2.18    \\ 
116.0   &   827.25723  &       827.25718  &    0.29    &116.4873910   &   830.88663  &    2.28    \\ 
117.0   &   834.70543  &       834.70552  &    0.31    &117.4873910   &   838.33915  &    2.18    \\ 
118.0   &   842.15352  &       842.15380  &    0.32    &118.4873910   &   845.78408  &    2.18    \\ 
119.0   &   849.60183  &       849.60199  &    0.27    &119.4873910   &   853.23131  &    2.08    \\ 
120.0   &   857.05060  &       857.05057  &    0.35    &120.4873910   &   860.68308  &    2.18    \\ 
121.0   &   864.49914  &       864.49900  &    0.30    &121.4873910   &   868.13074  &    2.08    \\ 
122.0   &   871.94730  &       871.94713  &    0.25    &122.4873910   &   875.57538  &    2.18    \\ 
123.0   &   879.39555  &       879.39559  &    0.26    &123.4873910   &   883.02764  &    2.18    \\ 
124.0   &   886.84395  &       886.84387  &    0.27    &124.4873910   &   890.47404  &    2.08    \\ 
125.0   &   894.29242  &       894.29236  &    0.25    &125.4873910   &   897.92584  &    2.08    \\ 
126.0   &   901.74076  &       901.74070  &    0.31    &126.4873910   &       ...    &       ...  \\ 
127.0   &   909.18926  &       909.18937  &    0.27    &127.4873910   &   912.82074  &    2.28    \\ 
128.0   &   916.63756  &       916.63783  &    0.35    &128.4873910   &   920.26855  &    2.18    \\ 
129.0   &   924.08561  &       924.08598  &    0.33    &129.4873910   &   927.71411  &    2.18    \\ 
130.0   &       ...    &           ...     &     ...   & 130.4873910   &   935.16675  &    2.18    \\ 
131.0   &   938.98133  &       938.98223  &    0.33    &131.4873910   &   942.61285  &    2.18    \\ 
132.0   &   946.42970  &       946.43047  &    0.33    &132.4873910   &       ...    &       ...  \\ 
133.0   &   953.87853  &       953.87896  &    0.30    &133.4873910   &   957.50762  &    2.28    \\ 
134.0   &   961.32731  &       961.32765  &    0.31    &134.4873910   &   964.95949  &    2.68    \\ 
135.0   &   968.77609  &       968.77621  &    0.33    &135.4873910   &   972.40639  &    2.18    \\ 
136.0   &   976.22531  &       976.22515  &    0.33    &136.4873910   &   979.85442  &    2.28    \\ 
137.0   &   983.67397  &       983.67372  &    0.34    &137.4873910   &       ...    &       ...  \\ 
138.0   &   991.12232  &       991.12194  &    0.31    &138.4873910   &       ...    &       ...  \\ 
139.0   &   998.57040  &       998.56992  &    0.33    &139.4873910   &  1002.20122  &    1.98    \\ 
140.0   &  1006.01835  &      1006.01771  &    0.27    &140.4873910   &       ...    &       ...  \\ 
141.0   &  1013.46672  &      1013.46612  &    0.33    &141.4873910   &       ...    &       ...  \\ 
\hline
\end{longtable}

\clearpage

\begin{longtable}{rrrrrrr}
\caption{{\bf Times of planetary transits.}}\\
\hline\hline
cycle \# & measured   &
 uncertainty & duration &
model  & \quad model duration  & note \\
  &    time\footnotemark &   (minute) & (hour) & 
  time$^2$ & (hour) & \\
\hline  
\endfirsthead
\hline\hline
\caption[]{(continued)} \\
\hline  
\endhead
\hline
\endfoot
\multicolumn{7}{c}{Planet b}   \\
\hline
1.0\footnotetext{BJD-2,455,000}   
  &   -30.79061 &   11.98 &   4.80 &  -30.81466 &   4.50 & \\
 2.0 &         ... &     ... &    ... &   16.27142 &   3.81 & data gap\\
 3.0 &    65.24000 &   30.00 &   6.72 &   65.24426 &   6.23 & \\
 4.0 &   112.53000 &   30.00 &   3.60 &  112.54562 &   3.53 & \\
 5.0 &   160.94000 &   90.00 &  11.00 &  160.90698 &  10.29 & \\
 6.0 &   208.84245 &    5.30 &   3.60 &  208.84102 &   3.50 & \\
 7.0 &   256.35001 &   60.00 &   8.16 &  256.32285 &   7.80 & \\
 8.0 &   305.13831 &    0.70 &   3.84 &  305.12396 &   3.71 & \\
 9.0 &   352.26001 &   30.00 &   5.28 &  352.25223 &   5.00 & \\
10.0 &   401.35165 &    8.06 &   4.56 &  401.35574 &   4.24 & \\
11.0 &   448.43933 &   19.12 &   4.08 &  448.42355 &   3.94 & \\
12.0 &   497.42072 &   60.00 &   5.76 &  497.46506 &   5.50 & \\
13.0 &   544.74023 &   10.37 &   3.60 &  544.68713 &   3.48 & \\
14.0 &   593.25055 &    2.30 &   9.12 &  593.26556 &   8.77 & \\
15.0 &        ...  &    ...  &    ... &  640.98407 &   3.29 & data gap\\
16.0 &   688.61578 &   26.96 &   9.36 &  688.65100 &   8.65 & \\
17.0 &   737.27374 &   10.37 &   3.60 &  737.27942 &   3.31 & \\
18.0 &   784.40002 &   30.00 &   5.52 &  784.47644 &   5.13 & \\
19.0 &         ... &   ...   &    ... &  833.53937 &   3.58 & data gap\\
20.0 &   880.63666 &   12.21 &   3.84 &  880.61505 &   3.73 & \\
21.0 &         ... &     ... &    ... &  929.71033 &   4.30 & corrupted data\\
22.0 &   976.86499 &   60.00 &   4.08 &  976.87207 &   3.06 & \\
\hline
\multicolumn{7}{c}{Planet c}\\
\hline
1.0  &   246.64867  &  5.07  & 5.76 & 246.64379 & 4.02 & \\
2.0  &   550.47591  &  5.23  & 8.16 & 550.47833 & 6.12 & \\
3.0  &   850.99483  &  5.30  & 6.96 & 850.99053 & 6.00 & \\
\hline
\multicolumn{7}{c}{Orphan} \\
\hline
1.0  &    977.363  &  5.76  &  4.15 & ...       & ...  & \\
\end{longtable}

\newpage

\thispagestyle{empty}
\begin{table}[t]
\caption{{\bf Model fitting parameters for the photometric-dynamical
model.}  See the text for definitions of the terms. The numbers in
boldface refer to the parameters shown in Fig.\ \protect\ref{corr100}.
\label{tab:tab1}}
\vspace{0.99em}
\footnotesize{
\begin{tabular}{|l|l|lll|}
\hline
Parameter Name & Best-fit & 50\% & 15.8\% & 84.2\% \\ \hline
\quad{\it Mass parameters} & & & & \\
{\bf 0.} Mass of Star A, $M_A$ ($M_\odot$) & $1.043$ & $1.049 $ & $-0.055$ & $+0.054$\\
{\bf 1.} Mass ratio, Star B, $M_B/M_A$ & $0.3473$ & $0.3462 $ & $-0.0063$ & $+0.0069$\\
\hline
\quad{\it Planet b Orbit} (Epoch BJD 2,454,969.216) & & & &\\
{\bf 2.} Orbital Period, $P_b$ (day) & $  49.514$ & $  49.532 $ & $-   0.027$ & $+   0.040$\\
{\bf 3.} Eccentricity Parameter, $\sqrt{e_b} \cos(\omega_b)$ & $-0.094$ & $ 0.000 $ & $- 0.075$ & $+ 0.067$\\
{\bf 4.} Eccentricity Parameter, $\sqrt{e_b} \sin(\omega_b)$ & $ 0.003$ & $ 0.098 $ & $- 0.067$ & $+ 0.042$\\
{\bf 5.} Time of Barycentric Transit,  & & & & \\\
~~~$t_b$ (BJD - 2,455,000) & $ -31.367$ & $ -31.353 $ & $-   0.010$ & $+   0.011$\\
{\bf 6.} Orbital Inclination, $i_b$ (deg) & $ 89.59$ & $ 89.70 $ & $-  0.16$ & $+  0.50$\\
{\bf 7.} Relative Nodal Longitude, $\Delta \Omega_b$ (deg) & $  0.10$ & $  0.23 $ & $-  0.21$ & $+  0.58$\\
\hline
\quad{\it Planet c Orbit}  (Epoch BJD - 2,455,246.6545) & & & &\\
{\bf 8.} Orbital Period, $P_c$ (day) & $ 303.158$ & $ 303.137 $ & $-   0.020$ & $+   0.072$\\
{\bf 9.} Eccentricity Parameter, $\sqrt{e_c} \cos(\omega_c)$ & $ -0.35$ & $ -0.04 $ & $-  0.40$ & $+  0.41$\\
{\bf 10.} Eccentricity Parameter, $\sqrt{e_c} \sin(\omega_c)$ & $-0.237$ & $-0.257 $ & $- 0.041$ & $+ 0.039$\\
{\bf 11.} Time of Barycentric Transit,  & & & & \\
~~~~~$t_c$ (BJD - 2,455,000) & $ 246.985$ & $ 246.997 $ & $-   0.012$ & $+   0.016$\\
{\bf 12.} Orbital Inclination, $i_c$ (deg) & $89.826$ & $89.825 $ & $- 0.010$ & $+ 0.009$\\
{\bf 13.} Relative Nodal Longitude, $\Delta \Omega_c$ (deg) & $  1.06$ & $  0.99 $ & $-  0.50$ & $+  0.49$\\
\hline
\quad{\it Stellar Orbit} & & & &\\
{\bf 14.} Orbital Period, $P_{AB}$ (day) & $  7.44837695$ & $  7.44837703 $ & $-  0.00000021$ & $+  0.00000021$\\
{\bf 15.} Eccentricity Parameter, $e_{EB} \cos(\omega_{EB})$ & $-0.019778$ & $-0.019797 $ & $- 0.000045$ & $+ 0.000044$\\
{\bf 16.} Eccentricity Parameter, $e_{EB} \sin(\omega_{EB})$ & $  -0.0125$ & $  -0.0112 $ & $-   0.0019$ & $+   0.0019$\\
{\bf 17.} Time of Primary Eclipse,  & & & & \\
~~~~~$t_{EB}$ (BJD - 2455000) & $-29.306346$ & $-29.306342 $ & $-  0.000018$ & $+  0.000018$\\
{\bf 18.}  Orbital Inclination, $i_{EB}$ (deg) & $ 89.34$ & $ 89.40 $ & $-  0.10$ & $+  0.12$\\
\hline
\quad{\it Radius/Light Parameters} & & & &\\
{\bf 19.} Linear Limb Darkening Parameter for & & & & \\
~~~~~Star A, $u$ & $ 0.4151$ & $ 0.4137 $ & $- 0.0044$ & $+ 0.0044$\\
{\bf 20.} Stellar Flux Ratio, $F_B/F_A$ ($\times 100$) & $  0.568$ & $  0.579 $ & $-  0.017$ & $+  0.017$\\
{\bf 21.} Density of Star A, $\rho_A$ (g cm$^{-3}$) & $1.163$ & $1.176 $ & $-0.025$ & $+0.024$\\
{\bf 22.} Radius Ratio, Star B, $R_B/R_A$ & $ 0.3636$ & $ 0.3671 $ & $- 0.0047$ & $+ 0.0047$\\
{\bf 23.} Planetary Radius Ratio, $R_b/R_A$ & $ 0.0283$ & $ 0.0289 $ & $- 0.0011$ & $+ 0.0011$\\
{\bf 24.} Planetary Radius Ratio, $R_c/R_A$ & $ 0.0439$ & $ 0.0440 $ & $- 0.0018$ & $+ 0.0017$\\
\hline
\quad{\it Relative Contamination}, & & & & \\
\quad$100\times(F_{\rm cont}/F_A)$  & & & &\\
Season 0 & $ -2.9$ & $ -0.9 $ & $-  2.8$ & $+  2.8$\\
Season 1 & $ -1.5$ & $  0.5 $ & $-  2.8$ & $+  2.9$\\
Season 2 & $ -2.8$ & $ -0.8 $ & $-  2.8$ & $+  2.8$\\
Season 3 & $ -1.9$ & $  0.1 $ & $-  2.8$ & $+  2.9$\\
\hline
\quad{\it Noise Parameter} & & & & \\
Long Cadence Relative Width, $\sigma_{\rm LC}$ ($\times 10^5$) & $62.95$ & $62.75 $ & $- 0.42$ & $+ 0.43$\\
\hline
\quad{\it Radial Velocity Parameters} & & & & \\
RV Offset, $\gamma$ (km s$^{-1}$) & $  4.67$ & $  4.60 $ & $-  0.23$ & $+  0.22$\\
Zero-level Diff., $\Delta \gamma$ (km s$^{-1}$) & $ -0.12$ & $ -0.01 $ & $-  0.31$ & $+  0.30$\\
RV Jitter, $\sigma_{RV}$ (km s$^{-1}$) & $     0.31$ & $     0.43 $ & $-     0.12$ & $+     0.19$\\
\hline
\end{tabular}
}
\end{table}

\clearpage

\begin{table}
\caption {{\bf Derived parameters from the photometric-dynamical model.} \label{tab:tab2}}
\vspace{1em}
\centering
\begin{tabular}{|l|l|lll|}
\hline
~~Parameter & Best-fit & 50\% & 15.8\% & 84.2\% \\ \hline
~{\it Bulk Properties} & & & & \\
~~Mass of Star A, $M_A$ ($M_\odot$) & $1.043$ & $1.049 $ & $-0.055$ & $+0.054$\\
~~Mass of Star B, $M_B$ ($M_\odot$) & $0.362$ & $0.363 $ & $-0.013$ & $+0.012$\\
~~Radius of Star A, $R_A$ ($R_\odot$) & $0.964$ & $0.963 $ & $-0.017$ & $+0.017$\\
~~Radius of Star B, $R_B$ ($R_\odot$) & $0.3506$ & $0.3533 $ & $-0.0063$ & $+0.0060$\\
~~Radius of Planet b, $R_b$ ($R_\oplus$) & $2.98$ & $3.03 $ & $-0.12$ & $+0.12$\\
~~Radius of Planet c, $R_c$ ($R_\oplus$) & $4.61$ & $4.62 $ & $-0.20$ & $+0.20$\\
~~Density of Star A, $\rho_A$ (g cm$^{-3}$) & $ 1.163$ & $ 1.176 $ & $- 0.025$ & $+ 0.024$\\
~~Density of Star B, $\rho_B$ (g cm$^{-3}$) & $  8.41$ & $  8.24 $ & $-  0.20$ & $+  0.21$\\
~~Gravity of Star A, $\log g_A$ (cgs) & $  4.488$ & $  4.492 $ & $-  0.011$ & $+  0.010$\\
~~Gravity of Star B, $\log g_B$ (cgs) & $ 4.9073$ & $ 4.9017 $ & $- 0.0067$ & $+ 0.0067$\\
~{\it Orbital Properties} & & & & \\
~~Semimajor Axis of Stellar Orbit, $a_{AB}$ (AU) & $0.0836$ & $0.0838 $ & $-0.0014$ & $+0.0013$\\
~~Semimajor Axis of Planet b, $a_b$ (AU) & $0.2956$ & $0.2962 $ & $-0.0047$ & $+0.0044$\\
~~Semimajor Axis of Planet c, $a_c$ (AU) & $ 0.989$ & $ 0.991 $ & $- 0.016$ & $+ 0.015$\\
~~Eccentricity of Stellar Orbit, $e_{AB}$ & $0.0234$ & $0.0228 $ & $-0.0009$ & $+0.0010$\\
~~Argument of Periapse Stellar Orbit, $\omega_{AB}$ (Degrees) & $   212.3$ & $   209.5 $ & $-     4.4$ & $+     4.1$\\
~~Mutual Orbital Inclination, $I_b$ (deg) & $ 0.27$ & $ 0.43 $ & $- 0.24$ & $+ 0.66$\\
~~Mutual Orbital Inclination, $I_c$ (deg) & $ 1.16$ & $ 1.08 $ & $- 0.42$ & $+ 0.46$\\
~{\it Eccentricities Constraints} & & & & \\
~~Eccentricity of Planet b Orbit (95\% conf.), $e_b$ & $ < 0.035$ & & & \\
~~Eccentricity of Planet c Orbit (95\% conf.), $e_c$ & $ < 0.411$ & & & \\
\hline
\end{tabular}
\end{table}

\clearpage

\begin{longtable}{rrrrr}
\caption{{\bf Predicted transit times for planet b.}}\\
\hline\hline
Epoch & $T_{0}$-2,455,000 BJD & Impact Parameter & Transit Velocity ($R_A$/day) & Duration (hr) \\ 
\hline  
\endfirsthead
\hline\hline
\caption[]{(continued)} \\
\hline  
\endhead
\hline
\endfoot
22 & $ 1025.630\pm    0.035$ & $0.527\pm0.237$ & $ 6.007\pm 0.094$ & $7.069\pm0.963$\\
23 & $ 1073.156\pm    0.022$ & $0.602\pm0.265$ & $12.496\pm 0.136$ & $3.209\pm0.605$\\
24 & $ 1121.106\pm    0.075$ & $0.528\pm0.236$ & $ 4.522\pm 0.086$ & $9.353\pm1.339$\\
25 & $ 1169.457\pm    0.033$ & $0.642\pm0.261$ & $12.282\pm 0.111$ & $3.170\pm0.681$\\
26 & $ 1216.768\pm    0.072$ & $0.577\pm0.246$ & $ 7.221\pm 0.260$ & $5.674\pm0.960$\\
27 & $ 1265.733\pm    0.050$ & $0.691\pm0.261$ & $11.318\pm 0.129$ & $3.280\pm0.809$\\
28 & $ 1312.836\pm    0.073$ & $0.654\pm0.255$ & $ 9.880\pm 0.268$ & $3.885\pm0.847$\\
29 & $ 1361.940\pm    0.081$ & $0.742\pm0.266$ & $ 9.580\pm 0.264$ & $3.667\pm1.041$\\
30 & $ 1409.062\pm    0.083$ & $0.745\pm0.269$ & $11.584\pm 0.213$ & $3.016\pm0.852$\\
31 & $ 1457.985\pm    0.149$ & $0.774\pm0.275$ & $ 7.030\pm 0.490$ & $4.897\pm1.514$\\
32 & $ 1505.349\pm    0.099$ & $0.838\pm0.288$ & $12.403\pm 0.124$ & $2.596\pm0.840$\\
33 & $ 1553.638\pm    0.299$ & $0.783\pm0.285$ & $ 4.552\pm 0.261$ & $7.406\pm2.197$\\
34 & $ 1601.654\pm    0.123$ & $0.917\pm0.312$ & $12.433\pm 0.137$ & $2.454\pm0.872$\\
35 & $ 1649.136\pm    0.281$ & $0.835\pm0.301$ & $ 6.085\pm 0.750$ & $5.389\pm1.845$\\
36 & $ 1697.947\pm    0.159$ & $0.963\pm0.342$ & $11.752\pm 0.363$ & $2.532\pm0.949$\\
37 & $ 1745.100\pm    0.235$ & $0.917\pm0.331$ & $ 9.008\pm 0.752$ & $3.496\pm1.235$\\
38 & $ 1794.188\pm    0.221$ & $0.994\pm0.374$ & $10.313\pm 0.726$ & $2.827\pm1.076$\\
39 & $ 1841.284\pm    0.229$ & $0.993\pm0.368$ & $11.053\pm 0.532$ & $2.678\pm1.018$\\
40 & $ 1890.314\pm    0.359$ & $1.000\pm0.403$ & $ 8.063\pm 1.181$ & $3.692\pm1.560$\\
41 & $ 1937.556\pm    0.243$ & $1.053\pm0.409$ & $12.181\pm 0.256$ & $2.402\pm0.904$\\
42 & $ 1986.153\pm    0.619$ & $0.989\pm0.425$ & $ 5.333\pm 0.940$ & $5.488\pm2.230$\\
43 & $ 2033.860\pm    0.275$ & $1.090\pm0.447$ & $12.458\pm 0.233$ & $2.257\pm0.897$\\
44 & $ 2081.589\pm    0.657$ & $1.009\pm0.445$ & $ 5.096\pm 1.232$ & $5.669\pm2.273$\\
45 & $ 2130.158\pm    0.328$ & $1.106\pm0.478$ & $12.073\pm 0.675$ & $2.389\pm0.956$\\
46 & $ 2177.380\pm    0.519$ & $1.058\pm0.472$ & $ 7.999\pm 1.540$ & $3.815\pm1.701$\\
47 & $ 2226.418\pm    0.429$ & $1.093\pm0.498$ & $10.920\pm 1.298$ & $2.676\pm1.117$\\
48 & $ 2273.503\pm    0.452$ & $1.095\pm0.498$ & $10.406\pm 1.157$ & $2.930\pm1.211$\\
49 & $ 2322.594\pm    0.652$ & $1.059\pm0.509$ & $ 8.952\pm 1.842$ & $3.420\pm1.708$\\
50 & $ 2369.745\pm    0.442$ & $1.110\pm0.516$ & $11.826\pm 0.654$ & $2.554\pm1.039$\\
51 & $ 2418.551\pm    0.948$ & $1.027\pm0.513$ & $ 6.479\pm 1.596$ & $4.824\pm2.190$\\
52 & $ 2466.039\pm    0.467$ & $1.100\pm0.522$ & $12.314\pm 0.439$ & $2.437\pm0.942$\\
53 & $ 2514.061\pm    1.048$ & $1.007\pm0.515$ & $ 5.318\pm 1.565$ & $5.741\pm2.277$\\
54 & $ 2562.341\pm    0.533$ & $1.066\pm0.516$ & $12.245\pm 1.000$ & $2.523\pm0.980$\\
55 & $ 2609.673\pm    0.888$ & $1.007\pm0.514$ & $ 7.041\pm 2.193$ & $4.574\pm2.176$\\
56 & $ 2658.612\pm    0.680$ & $1.006\pm0.499$ & $11.336\pm 1.781$ & $2.798\pm1.204$\\
57 & $ 2705.720\pm    0.731$ & $1.003\pm0.504$ & $ 9.729\pm 1.868$ & $3.343\pm1.556$\\
58 & $ 2754.815\pm    0.941$ & $0.933\pm0.477$ & $ 9.607\pm 2.221$ & $3.583\pm1.858$\\
59 & $ 2801.933\pm    0.665$ & $0.977\pm0.482$ & $11.443\pm 1.205$ & $2.886\pm1.109$\\
60 & $ 2850.864\pm    1.217$ & $0.872\pm0.453$ & $ 7.453\pm 2.025$ & $4.872\pm2.182$\\
61 & $ 2898.207\pm    0.665$ & $0.931\pm0.450$ & $12.091\pm 0.809$ & $2.810\pm0.897$\\
62 & $ 2946.504\pm    1.336$ & $0.826\pm0.428$ & $ 6.071\pm 1.878$ & $5.869\pm2.223$\\
63 & $ 2994.499\pm    0.731$ & $0.874\pm0.411$ & $12.231\pm 1.283$ & $2.896\pm0.922$\\
64 & $ 3041.964\pm    1.214$ & $0.796\pm0.400$ & $ 6.417\pm 2.481$ & $5.578\pm2.385$\\
65 & $ 3090.777\pm    0.898$ & $0.799\pm0.370$ & $11.615\pm 1.982$ & $3.255\pm1.367$\\
66 & $ 3137.912\pm    0.997$ & $0.778\pm0.366$ & $ 9.062\pm 2.402$ & $4.034\pm1.866$\\
67 & $ 3187.001\pm    1.142$ & $0.713\pm0.332$ & $10.124\pm 2.354$ & $3.862\pm1.931$\\
68 & $ 3234.085\pm    0.863$ & $0.728\pm0.329$ & $11.034\pm 1.705$ & $3.445\pm1.234$\\
69 & $ 3283.097\pm    1.385$ & $0.623\pm0.298$ & $ 8.157\pm 2.238$ & $4.894\pm2.199$\\
70 & $ 3330.344\pm    0.823$ & $0.663\pm0.292$ & $11.893\pm 1.143$ & $3.340\pm0.832$\\
71 & $ 3378.853\pm    1.505$ & $0.548\pm0.267$ & $ 6.752\pm 2.073$ & $6.129\pm2.179$\\
72 & $ 3426.633\pm    0.871$ & $0.579\pm0.265$ & $12.139\pm 1.403$ & $3.476\pm1.016$\\
\end{longtable}

\clearpage

\begin{longtable}{rrrrr}
\caption{{\bf Predicted transit times for planet c.}}\\
\hline\hline
Epoch & $T_{0}$-2,455,000 BJD & Impact Parameter & Transit Velocity ($R_A$/day) 
        & Duration (hr) \\ 
\hline  
\endfirsthead
\hline\hline
\caption[]{(continued)} \\
\hline  
\endhead
\hline
\endfoot
3 & $ 1154.756\pm    0.011$ & $0.430\pm0.056$ & $ 8.131\pm 0.133$ & $ 4.735\pm 0.260$\\
4 & $ 1458.197\pm    0.264$ & $0.397\pm1.258$ & $ 3.369\pm 0.217$ & $12.967\pm 1.000$\\
5 & $ 1758.963\pm    0.024$ & $0.446\pm0.057$ & $ 7.600\pm 0.135$ & $ 5.086\pm 0.278$\\
6 & $ 2062.831\pm    0.033$ & $0.407\pm0.114$ & $ 7.249\pm 0.145$ & $ 5.953\pm 0.462$\\
7 & $ 2363.464\pm    0.076$ & $0.458\pm0.066$ & $ 4.141\pm 0.258$ & $ 9.823\pm 0.783$\\
8 & $ 2667.055\pm    0.045$ & $0.436\pm0.116$ & $ 8.195\pm 0.150$ & $ 5.217\pm 0.417$\\
9 & $ 2970.751\pm    0.085$ & $0.462\pm0.164$ & $ 5.117\pm 0.316$ & $ 9.026\pm 0.981$\\
10 & $ 3271.295\pm    0.083$ & $0.462\pm0.118$ & $ 6.923\pm 0.195$ & $ 6.232\pm 0.487$\\
11 & $ 3575.153\pm    0.080$ & $0.461\pm0.172$ & $ 7.787\pm 0.236$ & $ 5.922\pm 0.561$\\
\end{longtable}

\begin{table}[hb]
\caption {{\bf ELC model parameters.} \label{tab:ELC}}
\vspace{1em}
\centering
\begin{tabular}{cc}
\hline
\hline
Parameter & Best fit \\
\hline
$e$                 & $0.0306  \pm 0.0071$ \\
$\omega$ (deg)      & $226\pm 12$ \\
$R_A/a$             & $0.05322\pm       0.00068$ \\
$R_B/a$             & $0.01935\pm       0.00029$ \\
$T_{\rm eff, B}/T_{\rm eff, A}$           & $0.5958\pm        0.0035$  \\
$i$ (deg)           & $89.69   \pm    0.16$ \\
$x_A$               & $0.30\pm       0.13$ \\
$y_A$               & $0.38\pm       0.27$  \\
\hline
\end{tabular}
\end{table}

\clearpage

\end{document}